\documentclass[a4paper, 11pt]{article}

\usepackage[left=20mm, right=20mm, top=25mm, bottom=25mm]{geometry}
\usepackage{subfiles}

\usepackage[T1]{fontenc}
\usepackage{lmodern}

\usepackage{textcomp}
\usepackage{graphicx}
\graphicspath{{./figs/}}
\usepackage[svgnames]{xcolor}
\usepackage[hang, small, bf]{caption}
\usepackage[subrefformat=parens]{subcaption}
\usepackage[shortlabels]{enumitem}
\usepackage{exscale}
\usepackage[tbtags]{amsmath}
\allowdisplaybreaks[2]
\numberwithin{equation}{section}
\usepackage{amssymb, amsthm}
\usepackage{mathtools}
\usepackage{mathrsfs}
\usepackage{mleftright}
\usepackage{bm}
\usepackage{bbm}
\usepackage{physics2}
\usephysicsmodule{braket}
\usepackage{derivative}
\derivset{\pdv}[delims-eval=.|]
\usepackage{esvect}
\usepackage{extarrows}
\usepackage{interval}
\intervalconfig{soft open fences}
\usepackage{siunitx}
\usepackage{xifthen}

\theoremstyle{plain}
\newtheorem{theorem}{Theorem}[section]
\newtheorem{lemma}[theorem]{Lemma}

\newtheorem{proposition}[theorem]{Proposition}
\theoremstyle{definition}
\newtheorem{definition}[theorem]{Definition}
\newtheorem{assumption}[theorem]{Assumption}
\theoremstyle{remark}
\newtheorem{remark}[theorem]{Remark}

\NewDocumentCommand{\Napier}{}{\mathord{\mathrm{e}}}
\NewDocumentCommand{\imag}{}{\mathord{\mathrm{i}}}

\NewDocumentCommand{\evaluated}{s O{} m o}
  {%
    \IfBooleanTF{#1}
      {%
        \mleft[#3\mright]%
      }
      {%
        #2[#3#2]%
      }%
    \IfNoValueF{#4}{_{#4}}
  }

\DeclarePairedDelimiter{\abs}{\lvert}{\rvert}
\DeclarePairedDelimiter{\norm}{\lVert}{\rVert}

\DeclareMathOperator{\Tr}{Tr}
\DeclarePairedDelimiterX{\commutator}[2]\lbrack\rbrack{#1,#2}
\DeclarePairedDelimiterXPP{\order}[1]{\mathcal{O}}(){}{#1}

\ProvideDocumentCommand{\separator}{}{}
\NewDocumentCommand{\SetSymbol}{O{}}{%
  \nonscript#1:
  \allowbreak
  \nonscript
  \mathopen{}%
}
\DeclarePairedDelimiterX{\Set}[1]{\lbrace}{\rbrace}{%
  \RenewDocumentCommand{\separator}{}{\SetSymbol}  
  #1
}

\NewDocumentCommand{\ft}{m}{\hat{#1}}             

\NewDocumentCommand{\TField}{}{q}
\DeclarePairedDelimiter{\Expect}\langle\rangle
\DeclareMathOperator{\sgExpectOp}{\mathbb{E}}
\DeclarePairedDelimiterXPP{\sgExpect}[1]{\sgExpectOp}\lbrack\rbrack{}{#1}  

\usepackage{framed}
\usepackage{centernot}

\definecolor{shadecolor}{gray}{0.85}

\newcommand{\betac}{\beta_\mathrm{c}}

\newcommand{\bvec}{\boldsymbol{b}}

\newcommand{\Ccal}{\mathcal{C}}
\newcommand{\compl}{\mathrm{c}}

\newcommand{\dc}{d_\mathrm{c}}
\newcommand{\diff}{\mathrm{d}}

\newcommand{\Ebb}{\mathbb{E}}

\newcommand{\im}{\mathrm{i}}

\newcommand{\Jbb}{\mathbb{J}}
\newcommand{\Jc}{J_\mathrm{c}}

\newcommand{\lace}{\mathsf{L}}

\newcommand{\mvec}{\boldsymbol{m}}

\newcommand{\N}{\mathbb{N}}

\newcommand{\nn}{\nonumber}

\newcommand{\Ocal}{\mathcal{O}}
\newcommand{\ovec}{\boldsymbol{o}}

\newcommand{\Pcal}{\mathcal{P}}

\newcommand{\piv}{\texttt{piv}}
\newcommand{\Prob}{\mathbb{P}}

\newcommand{\psivec}{\boldsymbol{\psi}}

\newcommand{\R}{\mathbb{R}}

\newcommand{\Scal}{\mathcal{S}}

\newcommand{\Tbb}{\mathbb{T}}

\newcommand{\ttl}{\texttt{l}}

\newcommand{\ttr}{\texttt{r}}

\newcommand{\uvec}{\boldsymbol{u}}

\newcommand{\vep}{\varepsilon}
\newcommand{\vno}{\varnothing}

\newcommand{\vtri}{\vartriangle}
\newcommand{\vvec}{\boldsymbol{v}}

\newcommand{\wvec}{\boldsymbol{w}}

\newcommand{\xvec}{\boldsymbol{x}}
\newcommand{\xivec}{\boldsymbol{\xi}}

\newcommand{\yvec}{\boldsymbol{y}}

\newcommand{\Z}{\mathbb{Z}}
\newcommand{\Zd}{\mathbb{Z}^d}
\newcommand{\zerovec}{\boldsymbol{0}}

\newcommand{\zvec}{\boldsymbol{z}}

\newcommand{\lbeq}[1]{\label{eq:#1}}
\newcommand{\Refeq}[1]{(\ref{eq:#1})}
\newcommand{\sss}{\scriptscriptstyle}

\newcommand{\cn}[2]{\underset{#1}{\overset{#2}{\longleftrightarrow}}}
\newcommand{\db}[2]{\underset{#1}{\overset{#2}\Longleftrightarrow}}
\newcommand{\ncn}[2]{\underset{#1}{\overset{#2}{\centernot\longleftrightarrow}}}
\newcommand{\ocn}[2]{\underset{#1}{\overset{#2}{\longrightarrow}}}

\newcommand{\ind}[1]{\mathbbm{1}{\raisebox{-2pt}{$\scriptstyle \{#1\}$}}}
\newcommand{\indic}[1]{\mathbbm{1}{\raisebox{-2pt}{$\scriptstyle #1$}}}

\usepackage[backend=biber, style=numeric, sorting=nyt, sortcites, giveninits=true, alldates=iso, seconds=true, maxnames=8, minnames=3, maxbibnames=99]{biblatex}
\usepackage{ifdraft}
\ifoptiondraft
  {
    \usepackage{fancyhdr}
    \geometry{right=30mm}
    \setlength{\marginparwidth}{1.5\marginparwidth}
    \usepackage[english]{datetime2}
    \rhead{\textit{\DTMnow}}
    \chead{\fbox{DRAFT}}
    \usepackage[mathlines]{lineno}
    \pagewiselinenumbers
  }
  {}
\usepackage[obeyDraft]{todonotes}
\usepackage[final]{changes}  
\definechangesauthor[name={Yoshinori Kamijima}, color=red]{YK}
\definechangesauthor[name={Akira Sakai}, color=red]{AS}
\setauthormarkup{}

\ExplSyntaxOn
\ifoptiondraft
  {
    \usepackage[color]{showkeys}
    \definecolor{refkey}{rgb}{0, 1, 0}
    \definecolor{labelkey}{rgb}{0, 0, 1}
    \usepackage{seqsplit}

    \RenewDocumentCommand{\showkeyslabelformat}{+m}
    {
        \tl_set:Nn \l_tmpa_tl { \(\{\)#1\(\}\) }
        \tl_replace_all:Nnn \l_tmpa_tl { \textvisiblespace } { \nobreakspace }
        \breslend_seqsplit:V \l_tmpa_tl
    }

    \cs_new_protected:Nn \breslend_seqsplit:n
    {
      \parbox[t]{0.75\marginparwidth}{\raggedright\normalfont\small\ttfamily\seqsplit{#1}}
    }
    \cs_generate_variant:Nn \breslend_seqsplit:n { V }
  }{}
\ExplSyntaxOff

\usepackage{tcolorbox}
\tcbuselibrary{breakable, skins, theorems}

\usepackage[pdfencoding=auto, pdfusetitle, setpagesize=false]{hyperref}

\title{Mean-field behavior of the quantum Ising susceptibility\\ and a new lace expansion for the classical Ising model}
\author{%
  Yoshinori Kamijima\thanks{Information Networking for Innovation and Design, Toyo University, Japan. \url{https://orcid.org/0000-0003-3037-8250}}%
  \and
  Akira Sakai\thanks{Department of Mathematics, Faculty of Science, Hokkaido University, Japan. \url{https://orcid.org/0000-0003-0943-7842}}
}
\date{\today}

\begin{document}

\maketitle

\begin{abstract}
  \noindent
  The transverse-field Ising model is widely studied as one of the simplest quantum spin systems.
  It is known that this model exhibits a phase transition at the critical inverse temperature 
  \added[id=AS]{$\betac$, which is determined by the spin-spin couplings and the transverse field $\TField\ge0$}.
  Bj\"ornberg~\cite{b2013infrared} investigated the divergence rate of the susceptibility for the nearest-neighbor model as the critical point is approached by simultaneously changing the spin-spin coupling $J\ge0$ and $\TField$ in a proper manner, with fixed temperature. 
  In this paper, we \added[id=AS]{fix $J$ and $\TField$ and show} that the susceptibility diverges as $(\added[id=AS]{\betac}-\beta)^{-1}$ as $\beta\uparrow\added[id=AS]{\betac}$ for $d>4$ assuming an infrared bound on the space-time two-point function.
 One of the key elements is a stochastic-geometric representation in 
 Bj\"ornberg \& Grimmett~\cite{bg2009phase} 
 and Crawford \& Ioffe~\cite{ci2010random}. 
 As a byproduct, we derive a new lace expansion for the classical Ising model (i.e., $q=0$).
\end{abstract}

\tableofcontents

\section{Introduction and the main results}

\subsection{Introduction}
The Ising model is one of the most-studied models of ferromagnetism.  
It was invented by Wilhelm Lenz in 1920~\cite{l1920beitrage}, but is named after Ernst Ising who proved absence of a phase transition on a 1-dimensional lattice \cite{i1925beitrag}.
It is formally defined by the infinite-volume limit of the finite-volume Gibbs distribution $\propto \Napier^{-\beta \mathcal{H}(\vv{\sigma})}$, where $\beta$ represents the inverse temperature and 
$\mathcal{H}(\vv{\sigma})$ represents the energy function, called Hamiltonian, for a spin configuration $\vv{\sigma} =\{\sigma_x\}_{x \in \Lambda}\in\{\pm1\}^\Lambda$ on a finite graph $(\Lambda, \mathbb{J}_\Lambda)$:
\begin{align}\lbeq{CIH}
\mathcal{H}_\Lambda(\vv{\sigma}) = -\sum_{\{x, y\} \in \mathbb{J}_\Lambda} J_{x, y}\, \sigma_x\, \sigma_y.
\end{align}
Unless otherwise stated, we assume all spin-spin couplings $J_{x, y}$ are positive (i.e., \added[id=AS]{the Ising model is} ferromagnetic). 

It is \added[id=AS]{now} well-known that the Ising model exhibits a sharp phase transition on locally-finite transitive graphs in dimensions $d\geq2$: 
there is a critical point $\betac\in(0,\infty)$ such that the susceptibility $\chi_\beta$, which is the sum of the infinite-volume limit $\Lambda\uparrow\Zd$ of the two-spin expectation, 
becomes finite as soon as $\beta<\betac$~\added[id=AS]{\cite{a1982geometric,ag1983renormalized}}, 
while the spontaneous magnetization $m_\beta$, which is the infinite-volume limit of the single-spin expectation under the plus-boundary condition (i.e., all spins outside $\Lambda$ are fixed at $+1$), becomes positive as soon as $\beta>\betac$ \added[id=AS]{\cite{abf1987phase,af1986critical,dt2016new,ads2015random}}.
Moreover, it is generally believed that those order parameters exhibit power-law \added[id=AS]{singularity, called critical behavior,} in the vicinity of the critical point, e.g., $\chi_\beta\asymp(\betac - \beta)^{-\gamma}$ as $\beta\uparrow\betac$.  
\added[id=AS]{Identifying the values of those critical exponents, such as $\gamma$, is one of the most important problems in statistical physics and probability, 
as they are considered} to be universal in the sense that they depend only on the symmetry and the dimension $d$ of the underlying lattice, but not on the microscopic details of the concerned models.  
\added[id=AS]{At the same time, it is a difficult and challenging problem, since the critical behavior is a result of interaction among infinitely many components, such as spins.  
In mean-field theory, the two-point function is replaced by Green's function of the underlying random walk generated by the step-distribution $\propto J_{o,x}$, which yields an estimate $\gamma=1$.} 
For models, such as the nearest-neighbor model, which satisfy a stronger symmetry condition, called \added[id=AS]{\textit{reflection positivity}},
it is known \added[id=AS]{(see, e.g., \cite{a1982geometric,ag1983renormalized})} that $\gamma=1$ for all dimensions above the upper-critical dimension $\dc=4$; other critical exponents also take on their mean-field values in dimensions $d>\dc$. 
In two dimensions, on the other hand, Fisher's scaling or the exact result on the critical two-point function by Wu, McCoy, Tracy and Barouch~\cite{wmtb1976spinspin} implies 
$\gamma=7/4$.  In three dimensions, only numerical results are available so far, although there are some interesting predictions from the so-called conformal bootstrap by 
El-Showk, Paulos, Poland, Rychkov, Simmons-Duffin and Vichi~\cite{epprsv2012solving}. 

One of the key elements to show the aforementioned mean-field behavior \added[id=AS]{$\gamma=1$ in dimensions $d>4$} is an infrared bound on the two-point function, 
which is a bound in the infrared region in terms of \added[id=AS]{Green's function mentioned above}.
It was first proven for the nearest-neighbor model \added[id=AS]{in dimensions $d>2$}~\cite{fss1976infrared}, and then extended to a wider class of models that satisfy reflection positivity~\cite{fils1978phase}.  
For a more detailed review, see also~\cite{b2009reflection}.  
It is used to verify the so-called bubble condition, 
which is a sufficient condition for the mean-field behavior~\cite{a1982geometric,af1986critical,ag1983renormalized} 
and is the square summability of the two-point function up to 
the critical point. 
Although the class of reflection-positive models is large enough to include the nearest-neighbor model, it does not necessarily cover all important models, 
such as the next-nearest-neighbor models with relatively equal weight and the uniformly spread-out models over the support.

Another way to prove an infrared bound is the \added[id=AS]{\textit{lace expansion}}.  
It has been successful in various models, such as self-avoiding walk~\cite{bs1985selfavoiding,hs1992selfavoiding}, 
oriented/unoriented percolation~\cite{hs1990meanfield,ny1993triangle}, 
lattice trees and lattice animals~\cite{hs1990upper}, 
the contact process~\cite{s2001meanfield}, 
the Ising model~\cite{s2007lace,s2022correct}, 
the $\varphi^4$ model~\cite{s2015application,bhh2021continuoustime,s2022correct}, 
the random-connection model~\cite{hhlm2023lace} 
and self-repellent Brownian bridges~\cite{bkm2024selfrepellent}. 
Since the lace expansion yields a renewal equation for the two-point 
function, the infrared asymptotics at the critical point can be derived by 
deconvolution~\cite{ls2024gaussian}, without assuming reflection positivity.
However, because of its perturbative nature from the underlying random walk, 
the dimension $d$ must usually be way higher than the critical dimension $\dc$.


In this paper, we investigate the quantum Ising model under transverse field~\cite{lsm1961two}, which is a toy model for quantum spin systems.  
It has also become popular in the field of computer science, due to its 
application to combinatorial optimization problems using quantum 
annealing (e.g., \cite{mn2006convergence,mn2007convergence,mn2008mathematical}).
The model is defined by replacing each spin $\sigma_x$ in the classical 
Hamiltonian $\mathcal{H}_\Lambda(\vv{\sigma})$ in \Refeq{CIH} 
with $S^{\sss(3)}_x$, which is a tensor product of the z-axis Pauli matrix, 
and perturbing the Hamiltonian by the transverse field 
$-\TField\sum_{x\in\Lambda}S^{\sss(1)}_x$ in the x-axis direction. 
Since \added[id=AS]{$S_x^{\sss(1)}$ and $S_x^{\sss(3)}$} do not commute, 
the model becomes \added[id=AS]{more disordered} as \added[id=AS]{soon} as $\TField>0$, 
which may result in \added[id=AS]{a smaller susceptibility, hence a larger} critical point $\betac(\TField)$; 
existence of a phase transition for the transverse-field Ising model 
and other quantum models (e.g., anisotropic Heisenberg models) was 
first proved by Ginibre~\cite{g1969existence}. 
We are interested in the rate of divergence of the susceptibility $\chi$
as $\beta\uparrow\betac(\TField)$ and finding out how it is affected 
by quantum effects.

In \cite{b2013infrared}, Bj\"ornberg investigated the quantum Ising 
susceptibility $\chi$ with the nearest-neighbor spin-spin coupling 
$J_{x,y}=J\ind{\|y-x\|_1=1}$.  Since $\chi$ is increasing in $J$ 
when $\beta$ and $\TField$ are fixed, there must be a critical point 
$\Jc(q)$ such that $\chi$ converges if $J<\Jc(q)$ and diverges if 
$J>\Jc(\TField)$.  He showed that, if $\beta>0$ is fixed and 
$(J,\TField)$ approaches $(\Jc(\TField_0),\TField_0)$ within 
a certain region, then $\chi$ diverges 
as $\|(J,\TField)-(\Jc(\TField_0),\TField_0)\|_2^{-1}$ for $d>4$ 
(\added[id=AS]{see} Section~\ref{ss:main}).  
This appears as if it shows the mean-field behavior $\gamma=1$, 
but in fact it does not, since the region 
mentioned above does not cover the ray 
$\beta\uparrow\betac(\TField)$ with $J$ and $q$ fixed 
(\added[id=AS]{see} Figure~\ref{fig:phasediag}).  
The restriction to the nearest-neighbor model for $d>4$ is for 
the use of reflection positivity (so that the two-point function 
obeys an infrared bound) and a quantum version of the bubble 
condition.

We show that $\chi$ for the nearest-neighbor model in dimensions 
$d>4$ indeed exhibits the mean-field behavior $\gamma=1$ 
as $\beta\uparrow\betac(\TField)$ with $J$ and $q~(\ll1)$ fixed.  
This implies that the critical behavior is robust against small 
quantum perturbation as long as $d>4$.  
The proof is based on an inequality for $\pdv{\chi}/{\beta}$ 
that is valid for a wider class of models than that of 
reflection-positive models.
The inequality for $\pdv{\chi}/{\beta}$ is obtained by reorganizing 
two differential \added[id=AS]{inequalities} in Bj\"ornberg~\cite{b2013infrared}.  
Those differential \added[id=AS]{inequalities} were obtained by using a stochastic-geometric 
representation in space-time~\cite{bg2009phase,ci2010random}. 
By setting $\TField=0$ in this representation, 
we also derive a new lace expansion for the classical Ising model.  
In the \added[id=AS]{forthcoming paper}~\cite{ks2025lace}, we will extend the lace expansion 
to the $\TField>0$ case, in order to prove the aforementioned 
mean-field results \added[id=AS]{for the quantum Ising model} without assuming reflection positivity.

\subsection{Transverse-field Ising model}
For each pair of sites $b = \{x, y\} \subset \mathbb{Z}^d$, we denote its translation-invariant spin-spin coupling by
\begin{equation}\label{eq:Jdef}
  J_b = J_{x, y} = J_{y - x}.
\end{equation}
For a finite subset $\Lambda\subset\Zd$ containing the origin $o = (0, \dots, 0)$, we define its bond set by
\begin{align}
  \mathbb{J}_\Lambda = \Set[\big]{b=\{x,y\}\subset\Lambda \separator J_b \ne 0}.
\end{align}
Throughout this paper, we assume that \added[id=AS]{$\Lambda$ is a $d$-dimensional torus centered at $o$ (i.e., $\Lambda\approx(\Z/L\Z)^d$ for some $L<\infty$) and} that the spin-spin coupling satisfies the following conditions:

\begin{shaded}
\begin{assumption}  \label{asm:ferromagnetism}
\begin{enumerate}[(i)]
\item {\itshape $\Zd$-symmetric}:
$J_x=J_{T(x)}$, where $T(x)$ is the mirror reflection of $x$ with respect to a coordinate hyperplane, or the image of $x$ rotated by 90 degrees around the origin.
In addition, $J_{o} = 0$.
\item {\itshape (Strongly) summable}: 
$\exists\alpha > 2$ such that 
$\sum_{x\in\mathbb{Z}^d} \abs{x}^\alpha J_x < \infty$.
\item {\itshape Irreducible}:
$\forall x,y\in\Zd$, $\exists v_1,v_2,\dots,v_n\in\Zd$ such that 
$J_{x,v_1}J_{v_1,v_2}\cdots J_{v_n,y}>0$.
\item {\itshape Ferromagnetic}:
$J_x> 0$ for every $\{o,x\} \in \Jbb_\Lambda$.
\end{enumerate}
\end{assumption}
\end{shaded}

We note that the nearest-neighbor model, defined by
\begin{align}\lbeq{nearest-neighbor}
J_x=\begin{cases}
 J&\text{if }\|x\|_1=1,\\
 0&\text{otherwise},
\end{cases}
\end{align}
where $\|\cdot\|_p$ for $p\ge1$ is the $\ell^p$ norm,
satisfies the above assumptions.

Next we define the Hamiltonian $H_\Lambda$ as an operator acting on 
$\bigotimes_{x\in\Lambda}\mathbb{C}^2 = (\mathbb{C}^2)^{\otimes \Lambda}$ as follows.  Let 
\begin{align}
I=
  \begin{bmatrix}
    1 & 0\\
    0 & 1
  \end{bmatrix},&&
S^{\sss(1)}=
  \begin{bmatrix}
    0 & 1\\
    1 & 0
  \end{bmatrix},&&
S^{\sss(3)}=
  \begin{bmatrix}
    1 & 0\\
    0 & -1
  \end{bmatrix},
\end{align}
and define
\begin{align}
  H_\Lambda &= H_{\Lambda,0} + Q_\Lambda, &
  H_{\Lambda,0} &= -\sum_{\{x,y\}\in\mathbb{J}_\Lambda}J_{x,y}S_x^{\sss(3)} S_y^{\sss(3)}, &
  Q_\Lambda &= -\TField \sum_{x\in\Lambda}S_x^{\sss(1)},
\end{align}
where $q \ge 0$ is the strength of the transverse field,
and the subscript $x$ in $S^{\sss(j)}_x$ is the location of 
$S^{\sss(j)}$ in a tensor product of operators:
\begin{align}
  S^{\sss(j)}_x &=
    \underbrace{
      I \cdots \otimes I \otimes
      \underset{
        \substack{\uparrow\\ x}
      }{
        S^{\sss(j)}
      }
      \otimes I \otimes \cdots I
    }_\Lambda.
\end{align}
We use the bra-ket notation commonly used in physics to denote the eigenvectors of $S^{\sss(3)}$ by
\begin{align}
\bra{+1}=\begin{bmatrix}1 & 0\end{bmatrix},&&
\ket{+1}=\begin{bmatrix}1 \\ 0\end{bmatrix},&&
\bra{-1}=\begin{bmatrix}0 & 1\end{bmatrix},&&
\ket{-1}=\begin{bmatrix}0 \\ 1\end{bmatrix}.
\end{align}
For $\vv{\sigma}=\{\sigma_x\}_{x\in\Lambda}$, we define
\begin{align}
    \ket{\vv{\sigma}} =
    \bigotimes_{x\in\Lambda}
    \ket{\sigma_x},
\end{align}
and denote its transpose by $\bra{\vv{\sigma}}$.

Finally we define the expectation at the inverse temperature $\beta\ge0$ 
of an operator $A$ on $(\mathbb{C}^2)^{\otimes \Lambda}$ as
\begin{align}
  \Expect{A}_\Lambda =
    \frac{
      \Tr[A \Napier^{-\beta H_\Lambda}]
    }{
      \Tr[\Napier^{-\beta H_\Lambda}]
    },&&
  \Tr[\cdots]=
    \sum_{\vv{\sigma}\in\{\pm 1\}^\Lambda}
    \braket[3]{\vv{\sigma}}{\cdots}{\vv{\sigma}}.
\end{align}
In particular, we are interested in the susceptibility defined as
\begin{align}  \label{eq:chidef}
  \chi_\Lambda
  =
    \chi_\Lambda(\beta, \added[id=AS]{\{J_b\}}, \TField)
  =
    \int_\mathbb{T} \odif{t}
    \sum_{x\in\Lambda}
    \Expect*{S_o^{\sss(3)}(0) S_x^{\sss(3)}(t)}_\Lambda,
\end{align}
where \added[id=AS]{$\Tbb=\R/\Z$} (i.e., $\interval{0}{1}$ with the periodic-boundary condition) and
\begin{align}
  S_x^{\sss(3)}(t) = \Napier^{-t\beta H_\Lambda} S_x^{\sss(3)} \Napier^{t\beta H_\Lambda}
\end{align}
is interpreted in physics as an imaginary-time evolution operator.
Since $S^{\sss(3)}_x$ and $H_{\Lambda,0}$ commute, we have that, 
for any $t\in\mathbb{T}$,
\begin{align}
  \Tr[S_o^{\sss(3)}(0) S_x^{\sss(3)}(t) \Napier^{-\beta H_{\Lambda,0}}] =
    \sum_{\vv{\sigma}\in\{\pm1\}^\Lambda}
    \sigma_o \sigma_x
    \Napier^{-\beta \mathcal{H}_\Lambda(\vv{\sigma})}
    \underbrace{
      \braket{\vv{\sigma}}{\vv{\sigma}}
    }_{= 1},
\end{align}
where $\mathcal{H}_\Lambda(\vv{\sigma})$ is the classical Ising Hamiltonian \Refeq{CIH}, hence
\begin{align}
  \chi_{\Lambda}(\beta, \added[id=AS]{\{J_b\}}, 0)
  =
    \int_\mathbb{T} \odif{t}
    \sum_{x\in\Lambda}
    \Expect*{S_o^{\sss(3)}(0) S_x^{\sss(3)}(t)}_{\Lambda}
  =
    \sum_{x\in\Lambda}
    \frac{
      \sum_{\vv{\sigma}}
      \sigma_o
      \sigma_x
      \Napier^{-\beta \mathcal{H}_\Lambda(\vv{\sigma})}
    }{
      \sum_{\vv{\sigma}}
      \Napier^{-\beta \mathcal{H}_\Lambda(\vv{\sigma})}
    }.
\end{align}
Since this is identical to the susceptibility for the classical 
Ising model, $\chi_\Lambda$ in \eqref{eq:chidef} is a natural 
extension to the transverse-field Ising model.

\subsection{Main results}\label{ss:main}
\begin{figure}[t]
  \centering
  \includegraphics[scale=0.8]{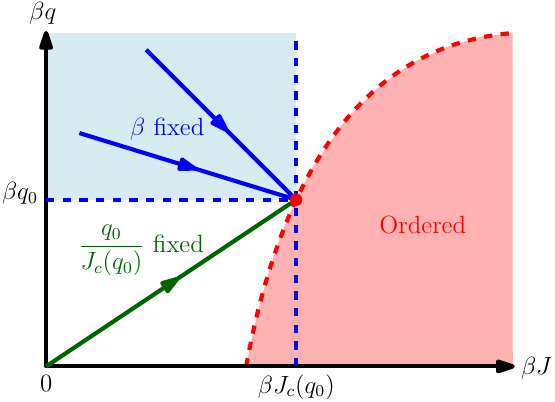}
  \caption{A phase diagram in the $(\beta J, \beta \TField)$ plane, with the imaginary 
  critical curve (slashed in red).  
  Bj\"ornberg~\cite[Theorem~1.3]{b2013infrared} covers the blue-shaded region 
  with fixed $\beta$, while Theorem~\ref{thm:MF_susceptibility} covers rays coming out of 
  the origin \added[id=AS]{with fixed angle (depicted} in green).}
  \label{fig:phasediag}
\end{figure}

For now, let us restrict ourselves to the nearest-neighbor model \Refeq{nearest-neighbor}.  
Although $\chi_\Lambda$ is not increasing in $\beta$ in general due to the quantum effect, 
it is increasing in $J$, so there is a critical value \added[id=AS]{$\Jc(\beta,\TField)\in(0,\infty)$ for all $d\ge2$ \cite[(1.5)]{bg2009phase}} such that 
$\added[id=AS]{\chi(\beta,J,\TField)}\equiv\lim_{\Lambda\uparrow\Zd}\chi_\Lambda$ 
is finite\footnote{\added[id=AS]{In this regime, the infinite-volume limit is independent of the boundary condition (i.e., independent of whether $\Lambda$ is a torus or a finite box in a vacuum), thanks to analyticity of the the free energy.}} as long as $J<\added[id=AS]{\Jc(\beta,\TField)}$:
\begin{equation}
  \added[id=AS]{J_{\mathrm{c}}(\beta,\TField)} = \inf\Set*{J \geq 0 \separator \added[id=AS]{\chi(\beta,J,\TField)} = \infty}.
\end{equation}
\added[id=AS]{It is known \cite{bg2009phase} that $\Jc(\beta,\TField)$ is non-decreasing in $\TField$, due to the quantum effect mentioned earlier, but not much else is known about the shape of the curve.}  
For $d>4$ with \added[id=AS]{$\beta\in(0,\infty)$} fixed, Bj\"ornberg~\cite[Theorem~1.3]{b2013infrared}\footnote{%
  Bj\"ornberg~\cite{b2013infrared} also proved that $\chi$ exhibits the same mean-field behavior for $d>3$ when the temperature is zero (i.e., $\beta=\infty$).
  He also proved that the upper bound is loosened with a logarithmic factor
  at the critical dimension $d=4$ for $\beta<\infty$ and $d=3$ for $\beta=\infty$: 
  \added[id=AS]{there are $c_1,c_2\in(0,\infty)$ such that, as $(J, \TField)\to(\Jc(\beta,\TField_0),\TField_0)$ along any ray strictly inside the region $\{(J, \TField): \TField>\TField_0,~0\le J<\added[id=AS]{\Jc(\beta,\TField_0)}\}$,
  \begin{align}
  \frac{c_1}{\|(J,\TField) - (J_{\mathrm{c}}(\beta,\TField_0),\TField_0)\|_2}
  \le\tilde\chi(J,q)\le
  \frac{c_2\log\|(J,\TField) - (J_{\mathrm{c}}(\beta,\TField_0),\TField_0)\|_2^{-1}}
  {\|(J,\TField) - (J_{\mathrm{c}}(\beta,\TField_0),\TField_0)\|_2}.
  \end{align}
  }
} proved that \added[id=AS]{$\tilde\chi(J,\TField
)\equiv\chi(\beta,J,\TField)$} 
exhibits the following mean-field behavior as $(J, \TField)\to(\added[id=AS]{\Jc(\beta,\TField_0)},\TField_0)$ along any ray strictly inside the region $\{(J, \TField): \TField>\TField_0,~0\le J<\added[id=AS]{\Jc(\beta,\TField_0)}\}$ (\added[id=AS]{see} Figure~\ref{fig:phasediag}):
\begin{equation}
  \added[id=AS]{\tilde\chi(J, \TField)}
  \asymp
    \norm*{(J, \TField) - \bigl(\added[id=AS]{J_{\mathrm{c}}(\beta,\TField_0)}, \TField_0\bigr)}_2^{-1},
\end{equation}
where ``$\asymp$'' means that the ratio of the left-hand side to the right-hand side is bounded away from zero and infinity in the prescribed limit.

However, this is a bit unsatisfactory from a physics point of view, 
since it does not cover the \added[id=AS]{isotherm} limit $\TField\downarrow \TField_0$ with  
$J=\added[id=AS]{J_{\mathrm{c}}(\beta,\TField_0)}$ fixed (unlike the temperature, $J$ is 
usually uncontrollable, as it is inherent in the concerned material).  
\added[id=AS]{Furthermore, it} does not cover the limit $\beta\uparrow\added[id=AS]{\betac}$ along 
the ray $\Set{(\beta J,\beta\TField):0\le\beta <\added[id=AS]{\betac}}$ with $J$ 
and $\TField$ fixed (\added[id=AS]{see} Figure~\ref{fig:phasediag}), where
\begin{equation}
  \added[id=AS]{\betac=\betac(J,\TField)} = \inf\Set*{\beta \geq 0 \separator \added[id=AS]{\chi(\beta,J,\TField)} = \infty}.
\end{equation}
The following theorem elucidates the behavior of $\chi$ for $d>4$ in the latter region.

\begin{shaded}
\begin{theorem}[Mean-field behavior of the quantum susceptibility]\label{thm:MF_susceptibility}
For the nearest-neighbor model on $\Z^{d>4}$ with $J>0$ fixed, 
there \added[id=AS]{is a $\TField_0>0$ such that the following holds for every $\TField \in \interval{0}{\TField_0}$: 
there is a $\beta_0<\betac$ such that} 
$\chi$ is increasing in $\beta\in[\beta_0,\added[id=AS]{\betac})$ 
and diverges as $\chi\asymp(\added[id=AS]{\betac}-\beta)^{-1}$ 
as $\beta\uparrow\added[id=AS]{\betac}$.
\end{theorem}
\end{shaded}

The restriction to the nearest-neighbor model in the above theorem is
inherited from the result of Bj\"ornberg~\cite{b2013infrared}, 
where he proved an infrared bound only for the nearest-neighbor model, 
though it is believed to be true for all reflection-positive models
that satisfy Assumption~\ref{asm:ferromagnetism}.  

The proof of the above theorem is based on a differential inequality
for $\chi$ that involves a space-time bubble diagram.  
The infrared bound mentioned above is used to show convergence of 
this space-time bubble diagram as long as $d>4$. 

To prove the aforementioned differential inequality for $\chi$, we use a 
stochastic-geometric representation for the transverse-field Ising model. 
As a byproduct of this representation, we derive the following new lace 
expansion for the classical Ising model, in which we use the notation
\begin{align}\lbeq{G-def}
\xvec=(t_{\xvec},x)\in\Tbb\times\Lambda,&&
G(\ovec,\xvec)=\Expect{S^{\sss(3)}_o(0)S^{\sss(3)}_x(t_{\xvec})}_\Lambda.
\end{align}

\begin{shaded}
\begin{theorem}[The lace expansion for $q=0$]\label{thm:LEintro}
There are model-dependent expansion coefficients 
$\pi^{\sss(j)}\colon (\mathbb{T}\times\Lambda)^2 \to[0,\infty)$ 
for $j\ge0$ such that, \added[id=AS]{by defining $\pi^{\sss(\le j)}(\bm{o}, \bm{x}) = \sum_{i=0}^{j} (-1)^i \pi^{\sss(i)}(\bm{o}, \bm{x})$, we have}
\begin{equation}\lbeq{LEintro}
G(\bm{o},\bm{x})=\pi^{\sss(\le j)}(\bm{o}, \bm{x})+\int_{\Tbb}
 \diff t\sum_{\substack{\yvec,\zvec\in\Tbb\times\Lambda:\\
 t_{\yvec}=t_{\zvec}=t}}\beta J_{y,z}\,\pi^{\sss(\le j)}(\ovec,
 \yvec)\,G(\bm{z},\bm{x})+(-1)^{j+1}R^{\sss(j+1)}(\bm{o},\bm{x}),
  \end{equation}
  where the remainder $R^{\sss(j+1)}(\bm{o}, \bm{x})$ is bounded as
\begin{equation}
0\le R^{\sss(j+1)}(\bm{o},\bm{x})\le\int_{\Tbb}\diff t
 \sum_{\substack{\yvec,\zvec\in\Tbb\times\Lambda:\\
 t_{\yvec}=t_{\zvec}=t}}\beta J_{y,z}\,\pi^{\sss(j)}
 (\ovec,\yvec)\,G(\bm{z},\bm{x}).
  \end{equation}
\end{theorem}
\end{shaded}

Another lace expansion for the classical Ising 
model~\cite{s2007lace,s2022correct} was obtained in a totally 
different way, based on the so-called random-current representation 
on the lattice (i.e., no time variable).  
As is roughly explained in the beginning of Section~\ref{s:LE}, 
the lace expansion yields an infrared bound on $G$ in a relatively 
easy way, without assuming reflection positivity.  
In the \added[id=AS]{forthcoming paper}~\cite{ks2025lace}, 
we will report on the extension to the $q>0$ case.

\subsection{Organization}
The remainder of the paper is organized as follows.
In Section~\ref{sec:diff-ineq}, we prove Theorem~\ref{thm:MF_susceptibility} 
by using a differential inequality for $\pdv{\chi_\Lambda}/{\beta}$ 
(\added[id=AS]{see} Proposition~\ref{prp:diff-ineq_susceptibility_temperature}) 
that is a result of two differential inequlities for 
$\pdv{\chi_\Lambda}/{J_b}$ and $\pdv{\chi_\Lambda}/{\TField}$ (\added[id=AS]{see}  
Lemma~\ref{lem:diff-ineq_susceptibility_coupling-coefficient_transverse-field}).
In Section~\ref{s:SGRSW}, we review the stochastic-geometric representation 
for the transverse-field Ising model~\cite{bg2009phase,ci2010random}.  
One of the key features of this representation is the so-called source 
switching (\added[id=AS]{see} Lemma~\ref{lmm:SST}).  
In Section~\ref{s:bjornberg}, we use this representation and the source 
switching to explain the aforementioned differential inequalities.  
Finally, in Section~\ref{s:LE}, we \added[id=AS]{derive the lace expansion \Refeq{LEintro}} in a heuristic way 
and give a precise definition of the expansion coefficients $\pi^{\sss(j)}(\ovec,\xvec)$ for $j\ge0$.  
\added[id=AS]{The lace expansion is useless unless the expansion coefficients obey good control. 
We will demonstrate how to derive diagrammatic bounds (in terms of two-point functions) on a part of the expansion coefficient $\pi^{\sss(0)}(\ovec,\xvec)$.}

\section{Mean-field behavior of the quantum Ising susceptibility}  \label{sec:diff-ineq}

In this section, we prove Theorem~\ref{thm:MF_susceptibility} by using a 
differential inequality for the susceptibility 
(\added[id=AS]{see} Proposition~\ref{eq:diff-ineq_susceptibility_temperature}), 
which is a result of combining two differential inequalities in Bj\"ornberg~\cite{b2013infrared}.
The differential inequality for the susceptibility involves the so-called 
bubble diagram (\added[id=AS]{see} \Refeq{bubble-def}), 
whose convergence for $d>4$ is ensured by an infrared bound on the two-point 
function (\added[id=AS]{see} Lemma~\ref{lem:infrared-bound}); 
\added[id=AS]{here we rely on the fact that the nearest-neighbor model satisfies reflection positivity.}

In Section~\ref{s:bjornberg}, we will explain the derivation of those differential inequalities in Bj\"ornberg~\cite{b2013infrared} 
by using expressions for the derivatives of the susceptibility in Section~\ref{ss:derivatives} and a stochastic-geometric representation in Section~\ref{s:SGRSW}.

\subsection{Proof of the mean-field behavior of the susceptibility}
For $k\in\frac\pi{L} \Lambda$ and $\omega\in2\pi\Z$, we define
\begin{align}
\hat J(k)=\sum_{x\in\added[id=YK]{\Lambda}}\Napier^{\imag k\cdot x}\added[id=AS]{J_x},&&
\hat G(\omega,k)=\int_{\Tbb}\diff t\sum_{\xvec:t_{\xvec}=t}
 \Napier^{\imag\omega t}\Napier^{\imag k\cdot x}G(\ovec,\xvec).
\end{align}
We also define the bubble diagram as
\begin{align}\lbeq{bubble-def}
  B
  = \limsup_{\beta\uparrow\betac}
    \limsup_{L\uparrow\infty}\int_{\Tbb}
    \diff t\sum_{\xvec:t_{\xvec}=t}G(\ovec,\xvec)^2
  = \limsup_{\beta\uparrow\betac}
    \limsup_{L\uparrow\infty}\sum_{(\omega,k)
    \in2\pi\Z\times\frac\pi{L}\Lambda}\hat G(\omega,k)^2,
\end{align}
where the second equality is due to Parseval's identity.
The following lemma \added[id=AS]{implies} that $B<\infty$ for all $d>4$.

\begin{shaded}
\begin{lemma}[{Infrared bound for the nearest-neighbor model 
\cite{b2013infrared}}]\label{lem:infrared-bound}
There is \added[id=AS]{an $L$-independent constant $c=c(\beta,\TField)>0$} 
such that, 
for any $\omega\in2\pi\Z$ and 
$k=(k_1,\dots,k_d)\in\frac\pi{L}\Lambda$,
\begin{align}\label{eq:infrared-bound}
    \abs[\big]{\ft G(\omega, k)}
    \leq
      \frac{
        48
      }{
        \added[id=AS]{c}\,\omega^2 + \added[id=AS]{2\beta}J \sum_{j=1}^{d} \bigl(1 - \cos(k_j)\bigr)
      }.
\end{align}
\end{lemma}
\end{shaded}

\begin{remark}
  As mentioned in \cite[Section~1.3]{b2013infrared},
  it should be straightforward to extend the above result 
  to other reflection-positive models, 
  such as those \added[id=AS]{defined by} Yukawa and power-law \added[id=AS]{pair} potentials~\added[id=AS]{\cite{b2009reflection}}.  
  However, if we have a lace expansion for the two-point function, then we do not \added[id=AS]{have to assume} reflection positivity 
  to obtain an infrared bound \added[id=AS]{above the upper-critical dimension $\dc=4$ (or $\dc=2(\alpha\wedge2)$ for the power-law long-range case $J_x\propto|x|^{-d-\alpha}$ 
  with $\alpha>0$~\cite{cs2015critical,cs2019critical})}, as briefly explained in the beginning of Section~\ref{s:LE}, 
  where \added[id=AS]{the constant $c$ in \Refeq{infrared-bound}} may be described by the second derivative with respect to $\omega$ 
  of the alternating series of the expansion coefficients (\added[id=AS]{see} \Refeq{IRbd}).  
  Notice that \added[id=AS]{$c=1/(2\beta\TField)$} in \cite[Theorem~1.2]{b2013infrared}, which does not make sense in the classical limit $\TField \downarrow 0$, 
  \added[id=AS]{as it implied $|\ft G(\omega, k)|\to0$ at any temperature $\beta<\infty$ (e.g., around the critical point for the classical Ising model), as long as $\omega\ne0$}.  
  We will report in the forthcoming paper~\cite{ks2025lace} that, by using the lace expansion, 
  we must have $c=c_0+O(\TField)$ with $c_0\in(0,\infty)$.
\end{remark}

Theorem~\ref{thm:MF_susceptibility} is an immediate consequence of the 
following differential inequality (with $\hat J(0)=2dJ$ for the 
nearest-neighbor model) and the fact that $B<\infty$ for $d>4$.

\begin{shaded}
\begin{proposition}  \label{prp:diff-ineq_susceptibility_temperature}
For any spin-spin coupling that satisfies Assumption~\ref{asm:ferromagnetism},
  \begin{equation}
    \label{eq:diff-ineq_susceptibility_temperature}
    \frac{
      \ft J(0) \chi_\Lambda^2
    }{
      1 + \beta \ft J(0) B
    }
    \left(
      1
      -
      \frac{
        2 B
      }{
        \chi_\Lambda
      }
      -
      2 \TField
      \frac{
        1+5 \beta \ft J(0) B
      }{
        \ft J(0)
      }
    \right)
    \leq
      \pdv{\chi_\Lambda}{\beta}
    \leq
      \ft J(0) \chi_\Lambda^2.
  \end{equation}
\end{proposition}
\end{shaded}

The differential inequality \eqref{eq:diff-ineq_susceptibility_temperature} is obtained by combining the following two differential inequalities for $\chi_\Lambda$.
\added[id=YK]{They are originally derived by Bj\"ornberg~\cite{b2013infrared} for the nearest-neighbor model, but we can easily extend them to the general spin-spin coupling coefficients that satisfy Assumption~\ref{asm:ferromagnetism}.
For convenience of the readers, we review their proofs in Section~\ref{s:bjornberg} without using inequalities as much as possible (i.e., using equalities where possible) and explicitly using  expressions of correlation functions.}

\begin{shaded}
\begin{lemma}[{Generalization of \cite[Lemma~3.1]{b2013infrared}}]
  \label{lem:diff-ineq_susceptibility_coupling-coefficient_transverse-field}
For any spin-spin coupling that satisfies Assumption~\ref{asm:ferromagnetism},
  \begin{equation}
    \label{eq:diff-ineq_susceptibility_coupling-coefficient}
    \left[\ft J(0) \chi_\Lambda^2
      - 2 \ft J(0) B \chi_\Lambda
      - \ft J(0) B \sum_{b\in\mathbb{J}_\Lambda} J_b \pdv{\chi_\Lambda}{J_b}
      - 4 \TField \ft J(0) B \left(-\pdv{\chi_\Lambda}{\TField}\right)
    \right]^{+}
    \leq \sum_{b\in\mathbb{J}_\Lambda} \frac{J_b}{\beta} \pdv{\chi_\Lambda}{J_b}
    \leq \ft J(0) \chi_\Lambda^2,
  \end{equation}
  and
  \begin{equation}
    \label{eq:diff-ineq_susceptibility_transverse-field}
    \left[
      2 \chi_\Lambda^2
        - 2 B \chi_\Lambda
        - B \sum_{b\in\mathbb{J}_\Lambda} J_b \pdv{\chi_\Lambda}{J_b}
        - 4 B \left(-\pdv{\chi_\Lambda}{\TField}\right)
    \right]^{+}
    \leq \added[id=AS]{\frac{-1}{\beta}}\pdv{\chi_\Lambda}{\TField}
    \leq 2 \chi_\Lambda^2\replaced[id=YK]{,}{.}
  \end{equation}
  \replaced[id=YK]{where $[t]^{+} = \max\{0, t\}$ for $t \in \mathbb{R}$}{During the course of their proof, we can also conclude
  $\pdv{\chi_\Lambda}/{J_b} \geq 0$ and $\pdv{\chi_\Lambda}/{\TField} \leq 0$}.
\end{lemma}
\end{shaded}

\deleted[id=YK]{Derivation of these inequalities is explained in Section~\ref{s:bjornberg}.}


\begin{proof}[Proof of Proposition~\ref{prp:diff-ineq_susceptibility_temperature}]
  \replaced[id=YK]{Since $\chi_\Lambda(\beta, \{J_b\}, \TField) = \chi_\Lambda(1, \{\beta J_b\}, \beta \TField)$ (recall \eqref{eq:chidef}), by}{By} the chain rule,
  \begin{equation}
    \label{eq:relation-among-derivatives}
    \pdv{\chi_\Lambda}{\beta}
    \added[id=YK]{=
      \sum_{b\in\mathbb{J}_\Lambda}
        \pdv{(\beta J_b)}{\beta}
        \pdv{\chi_\Lambda}{(\beta J_b)}
      +
        \pdv{(\beta \TField)}{\beta}
        \pdv{\chi_\Lambda}{(\beta \TField)}}
    =
      \sum_{b\in\mathbb{J}_\Lambda}
        \frac{J_b}{\beta}
        \pdv{\chi_\Lambda}{J_b}
      \added[id=AS]{+
        \frac{\TField}{\beta}
        \pdv{\chi_\Lambda}{\TField}}.
  \end{equation}
  Using the second inequality in 
  \eqref{eq:diff-ineq_susceptibility_coupling-coefficient}
  and $\pdv{\chi_\Lambda}/{\TField} \leq 0$, we obtain 
  the upper bound on $\pdv{\chi_\Lambda}/{\beta}$ in 
  \Refeq{diff-ineq_susceptibility_temperature}.  
  For the lower bound on $\pdv{\chi_\Lambda}/{\beta}$, 
  we use the first inequality in 
  \eqref{eq:diff-ineq_susceptibility_coupling-coefficient} to obtain
  \begin{align}
    \pdv{\chi_\Lambda}{\beta}
    \geq {} \MoveEqLeft[0]
      \ft J(0)
      \chi_\Lambda^2
      -
        2
        \ft J(0)
        B
        \chi_\Lambda 
      -
        \ft J(0)
        B
        \underbrace{
          \sum_{b\in\mathbb{J}_\Lambda}
          J_b
          \pdv{\chi_\Lambda}{J_b}
        }_{
            \beta
            \pdv{\chi_\Lambda}{\beta}
            + \TField
              \left(-\pdv{\chi_\Lambda}{\TField}\right)
        }
      -
        4
        \TField
        \ft J(0)
        B
        \left(-\pdv{\chi_\Lambda}{\TField}\right)
      -
        \added[id=AS]{\TField
        \left(
          \frac{-1}{\beta} \pdv{
            \chi_\Lambda
          }{
            \TField
          }
        \right)},
  \end{align}
  which is equivalent to
  \begin{equation*}
    \left(
      1
      +
        \beta
        \ft J(0)
        B
    \right)
    \pdv{\chi_\Lambda}{\beta}\\
    \geq
      \ft J(0)
      \chi_\Lambda^2
      -
        2
        \ft J(0)
        B
        \chi_\Lambda
      -
        \added[id=AS]{\TField}
        \left(
          1
          +
            5
            \beta
            \ft J(0)
            B
        \right)
        \left(\added[id=AS]{\frac{-1}{\beta}}\pdv{\chi_\Lambda}{\TField}\right).
  \end{equation*}
  Finally, by using the second inequality in 
  \eqref{eq:diff-ineq_susceptibility_transverse-field},
  we obtain
  \begin{align}
    \left(
      1
      +
        \beta
        \ft J(0)
        B
    \right)
    \pdv{\chi_\Lambda}{\beta}
    \geq
      \ft J(0)
      \chi_\Lambda^2
      \left(
        1
        -
          \frac{
            2
            B
          }{
            \chi_\Lambda
          }
        -
          2
          \TField
          \frac{
            1
            +
              5
              \beta
              \ft J(0)
              B
          }{
            \ft J(0)
          }
      \right),
  \end{align}
  as required.
\end{proof}

\begin{proof}[Proof of Theorem~\ref{thm:MF_susceptibility}]
\added[id=AS]{First we show that $\chi$ is increasing in $\beta\in[\beta_0,\betac)$, where $\beta_0>0$ is determined shortly.  
Recall that $B<\infty$ for the nearest-neighbor model for all $d>4$, thanks to the infrared bound \Refeq{infrared-bound}.  
Recall also \Refeq{diff-ineq_susceptibility_temperature}.  
For any $\vep\in(0,1)$, there is a $\TField_0>0$ such that $2\TField(1+5\beta\hat J(0)B)/\hat J(0)<1-\vep$ for all $q\in[0,\TField_0]$.  
Then, by \cite[Theorem~6.3]{bg2009phase}, there is a $\beta_0<\betac$ such that $\limsup_{L\uparrow\infty}2B/\chi_\Lambda<\vep$ for all $\beta\in[\beta_0,\betac)$. 
Therefore, the leftmost expression in \Refeq{diff-ineq_susceptibility_temperature} becomes positive, and $\chi_\Lambda$ is increasing in $\beta\in[\beta_0,\betac)$ for sufficiently large $\Lambda$.
This completes the proof of monotonicity of $\chi$ in $\beta\in[\beta_0,\betac)$.}

\added[id=AS]{The remaining task is staightforward. 
By \Refeq{infrared-bound}, we now know that $\pdv{\chi_\Lambda}{\beta}\asymp\chi_\Lambda^2$, or equivalently $\pdv{}{\beta}\frac{-1}{\chi_\Lambda}\asymp1$, for all $\TField\in[0,\TField_0]$ and $\beta\in[\beta_0,\betac)$. 
Integrating this from $\beta$ to $\betac$ yields
$\frac1{\chi_\Lambda(\beta)}-\frac1{\chi_\Lambda(\betac)}\asymp\betac-\beta$, hence the infinite-volume limit $\chi(\beta)\asymp(\betac-\beta)^{-1}$ as $\beta\uparrow\betac$. 
This completes the proof of Theorem~\ref{thm:MF_susceptibility}.
}
\end{proof}

\subsection{Derivatives of the susceptibility}\label{ss:derivatives}
To explain the differential inequalities 
\Refeq{diff-ineq_susceptibility_coupling-coefficient}--\Refeq{diff-ineq_susceptibility_transverse-field} in Section~\ref{s:bjornberg}, 
we use the following lemma to derive expressions for the derivatives 
$\pdv{\chi_\Lambda}/{J_{u,v}}$ depending on an edge $\{u, v\} \in \mathbb{J}_\Lambda$
(\added[id=AS]{see} \Refeq{pre_derivative_susceptibility_coupling-coefficient}) and 
$\pdv{\chi_\Lambda}/{q}$ 
(\added[id=AS]{see} \Refeq{pre_derivative_susceptibility_transverse-field}).

\begin{shaded}
\begin{lemma}[Special case of {\cite[Equation~(2.1)]{w1967exponential}}]  \label{lem:1st-derivative_exp}
  Let $A(\alpha)$ be a finite-dimensional matrix parameterized by 
  $\alpha$ in an open interval $D \subset \mathbb{R}$.
  If $A(\alpha)$ is differentiable and $\odv{A(\alpha)}/{\alpha}$ 
  is continuous, then, for any $\alpha \in D$,
  \begin{equation}
    \label{eq:1st-derivative_exp}
    \added[id=AS]{\odv*{\Napier^{A(\alpha)}}{\alpha}} =
      \int_{0}^{1}\diff t~
      \Napier^{t A(\alpha)}\,
      \odv{A(\alpha)}{\alpha}\,
      \Napier^{(1 - t) A(\alpha)}.
  \end{equation}
\end{lemma}
\end{shaded}

First we consider $\pdv{\chi_\Lambda}/{J_{u,v}}$, i.e.,
\begin{align}
\pdv{\chi_\Lambda}{J_{u, v}}
&=\int_{\Tbb}\diff t\sum_{x\in\Lambda}\pdv{}{J_{u, v}}\frac{\Tr
 [S_o^{\sss(3)}\Napier^{-t\beta H_\Lambda}S_x^{\sss(3)}
 \Napier^{-(1-t)\beta H_\Lambda}]}{\Tr[\Napier^{-\beta H_\Lambda}]}\nn\\
&=\int_{\Tbb}\diff t\sum_{x\in\Lambda}\frac1{\Tr[\Napier^{-\beta
 H_\Lambda}]}\Bigg(\Tr\mleft[S^{\sss(3)}_o\pdv{\Napier^{-t\beta
 H_\Lambda}}{J_{u, v}}S^{\sss(3)}_x\Napier^{-(1 - t)\beta H_\Lambda}
 \mright]+\Tr\mleft[
            S^{\sss(3)}_o
            \Napier^{-t\beta H_\Lambda}
            S^{\sss(3)}_x
            \pdv{\Napier^{-(1 - t)\beta  H_\Lambda}}{J_{u, v}}
            \mright]\nn\\
&\hskip10pc-\Expect{S^{\sss(3)}_o(0)\,S^{\sss(3)}_x(t)}_\Lambda
            \Tr\mleft[
            \pdv{\Napier^{-\beta H_\Lambda}}{J_{u, v}}\mright]\Bigg).
\end{align}
By \eqref{eq:1st-derivative_exp} and change of variables, we have
\begin{align}
\Tr\mleft[S^{\sss(3)}_o\pdv{\Napier^{- t\beta H_\Lambda}}{J_{u, v}}
 S^{\sss(3)}_x\Napier^{-(1 - t)\beta  H_\Lambda}\mright]
&=t\beta\int_{\Tbb}\diff s\,\Tr\bigg[S^{\sss(3)}_o\Napier^{-st\beta
 H_\Lambda}S^{\sss(3)}_uS^{\sss(3)}_v\Napier^{-(1-s)t\beta H_\Lambda}
 S^{\sss(3)}_x\Napier^{-(1 - t)\beta  H_\Lambda}\bigg]\nn\\
&=\beta\int_0^t\diff s\,\Tr\bigg[S^{\sss(3)}_o(0)\,S^{\sss(3)}_u(s)\,
 S^{\sss(3)}_v(s)\,S^{\sss(3)}_x(t)\,\Napier^{-\beta H_\Lambda}\bigg].
\end{align}
We can compute the other two terms in a similar way.  As a result,
\begin{align}\lbeq{pre_derivative_susceptibility_coupling-coefficient}
\pdv{\chi_\Lambda}{J_{u, v}}=\beta\int_{\Tbb}\diff t\sum_{x\in\Lambda}
      \Biggl(
        &\int_{0}^{t} \odif{s}\,
        \Expect*{S^{\sss(3)}_o(0)\,S^{\sss(3)}_u(s)\,S^{\sss(3)}_v(s)\, S^{\sss(3)}_x(t)}_\Lambda 
         +\int_{t}^{1} \odif{s}\,
        \Expect*{S^{\sss(3)}_o(0)\,S^{\sss(3)}_x(t)\,S^{\sss(3)}_u(s)\,
        S^{\sss(3)}_v(s)}_\Lambda \nn\\
        &-\int_{0}^{1} \odif{s}\,
        \Expect[\big]{S^{\sss(3)}_o(0)\,S^{\sss(3)}_x(t)}_\Lambda\,
        \Expect[\big]{S^{\sss(3)}_u(s)\,S^{\sss(3)}_v(s)}_\Lambda
      \Biggr).
\end{align}

Next we consider $\pdv{\chi_\Lambda}/{q}$, i.e.,
\begin{align}
\pdv{\chi_\Lambda}{\TField}
&=\int_{\Tbb}\diff t\sum_{x\in\Lambda}\frac1{\Tr[\Napier^{-\beta
 H_\Lambda}]}\Bigg(\Tr\mleft[S^{\sss(3)}_o\pdv{\Napier^{-t\beta
 H_\Lambda}}{\TField}S^{\sss(3)}_x\Napier^{-(1 - t)\beta
 H_\Lambda}\mright]+\Tr\mleft[S^{\sss(3)}_o\Napier^{-t\beta
 H_\Lambda}S^{\sss(3)}_x\pdv{\Napier^{-(1 - t)\beta 
 H_\Lambda}}{\TField}\mright]\nn\\
&\hskip10pc-\Expect{S^{\sss(3)}_o(0)\,S^{\sss(3)}_x(t)}_\Lambda
 \Tr\mleft[\pdv{\Napier^{-\beta H_\Lambda}}{\TField}\mright]
 \Bigg).
\end{align}
Again, by \eqref{eq:1st-derivative_exp} and change of variables, we have
\begin{align}
\Tr\mleft[S^{\sss(3)}_o\pdv{\Napier^{- t\beta H_\Lambda}}{\TField}
 S^{\sss(3)}_x\Napier^{-(1 - t)\beta  H_\Lambda}\mright]
&=t\beta\int_{\Tbb}\diff s\sum_{y\in\Lambda}\Tr\bigg[S^{\sss(3)}_o
 \Napier^{-st\beta H_\Lambda}S^{\sss(1)}_y\Napier^{-(1-s)t\beta
 H_\Lambda}S^{\sss(3)}_x\Napier^{-(1 - t)\beta  H_\Lambda}\bigg]\nn\\
&=\beta\int_0^t\diff s\sum_{y\in\Lambda}\Tr\bigg[S^{\sss(3)}_o(0)\,
 S^{\sss(1)}_y(s)\,S^{\sss(3)}_x(t)\,\Napier^{-\beta H_\Lambda}\bigg].
\end{align}
The other two terms can be computed in a similar way.  As a result,
\begin{align}\lbeq{pre_derivative_susceptibility_transverse-field}
\pdv{\chi_\Lambda}{\TField}=\beta\int_{\Tbb}\diff t\sum_{x,y\in\Lambda}
 \Bigg( &\int_0^t\diff s\,\Expect*{S^{\sss(3)}_o(0)\,
        S^{\sss(1)}_y(s)\,S^{\sss(3)}_x(t)}_\Lambda
         +\int_t^1\diff s\,\Expect*{S^{\sss(3)}_o(0)\,
        S^{\sss(3)}_x(t)\,S^{\sss(1)}_y(s)}_\Lambda\nn\\
        &-\int_0^1\diff s\,\Expect[\big]{S^{\sss(3)}_o(0)\,
        S^{\sss(3)}_x(t)}_\Lambda\,\Expect[\big]{S^{\sss(1)}_y(s)}_\Lambda
 \Bigg).
\end{align}

\added[id=YK]{It remains to prove upper and lower bounds on the correlation functions $\Expect*{S^{\sss(3)}_o(0)\,S^{\sss(1)}_y(s)\,S^{\sss(3)}_x(t)}_\Lambda$ and $\Expect{S^{\sss(3)}_o(0)\,S^{\sss(3)}_u(s)\,S^{\sss(3)}_v(s)\, S^{\sss(3)}_x(t)}_\Lambda$ in the differential inequalities.
The stochastic-geometric representation below plays an important role.}

\section{Stochastic-geometric representation and the source switching}\label{s:SGRSW}
For further investigation of the expectations in 
\Refeq{pre_derivative_susceptibility_coupling-coefficient} and
\Refeq{pre_derivative_susceptibility_transverse-field}, 
the stochastic-geometric representation and the so-called source 
switching in \cite{bg2009phase,ci2010random} are quite useful.  
We review the stochastic-geometric representation in Section~\ref{ss:SGR} 
and the source switching in Section~\ref{s:SW}.

\subsection{Stochastic-geometric representation}\label{ss:SGR}
\begin{figure}[t]
  \centering
  \includegraphics[width=0.65\textwidth]{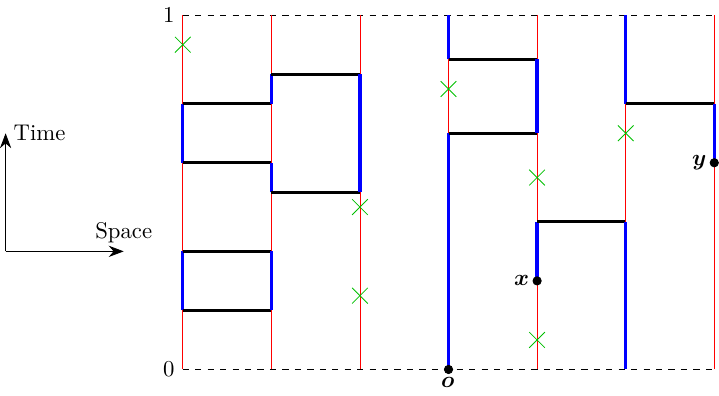}
  \caption{An example of a space-time spin configuration $\psivec\colon \Tbb^\Lambda \to \Set{\ttl, \ttr}$.
  The horizontal lines (thicker in black) represent elements of the bridge configuration $\xivec$, the cross marks (in green) represent elements of the mark configuration $\mvec$, and the vertical line at $z\in\Lambda$ represents $\psi_z\colon \Tbb \to \Set{\ttl, \ttr}$.
  The spin values $\ttl$ and $\ttr$ are shown in blue and red, respectively.
  The spin configuration $\psivec$ flips at every bridge in $\xivec$ and at the sources in $A = \Set{\bm{x}, \bm{y}}$.}
  \label{fig:example_space-time-spin-configuration}
\end{figure}

First we recall that, for $\ovec=(0,o)$ and $\xvec=(t,x)$,
\begin{align}
G(\ovec,\xvec)=\Expect{S_o^{\sss(3)}(0)\,S_x^{\sss(3)}(t)}_\Lambda=\frac{\mathrm{Tr}[S_o^{\sss(3)}
 e^{-t\beta H_\Lambda}S_x^{\sss(3)}e^{-(1-t)\beta H_\Lambda}]}{\mathrm{Tr}
 [e^{-\beta H_\Lambda}]}.
\end{align}
To compute the traces, we use the 1st-basis (the eigenvectors 
$\begin{bsmallmatrix}1 \\ 1\end{bsmallmatrix}$, 
$\begin{bsmallmatrix}1 \\ -1\end{bsmallmatrix}$ for $S^{\sss(1)}$), 
instead of using the 3rd-basis (the eigenvectors 
$\begin{bsmallmatrix}1 \\ 0\end{bsmallmatrix}$, 
$\begin{bsmallmatrix}0 \\ 1\end{bsmallmatrix}$ for $S^{\sss(3)}$) 
as in the previous section.  Also, we rewrite the 
Hamiltonian $H_\Lambda$ as
\begin{align}
H_\Lambda=-\sum_{\{x,y\}\in\Jbb_\Lambda}J_{x,y}S_x^{\sss(3)}S_y^{\sss
 (3)}-2q\sum_{z\in\Lambda}U_z^{\sss(1)}+q|\Lambda|,
\end{align}
where
\begin{align}\lbeq{U-def}
U^{\sss(1)}=\frac{S^{\sss(1)}+I}2
\end{align}
Let
\begin{align}
\ket{\ttr}=\begin{bmatrix}1 \\ 1\end{bmatrix},&&
\ket{\ttl}=\begin{bmatrix}1 \\ -1\end{bmatrix},
\end{align}
($\ttr$ for ``right'', $\ttl$ for ``left'') and denote their transpose by 
$\bra{\ttr}$ and $\bra{\ttl}$, respectively.  Notice that
\begin{align}\lbeq{ttrttl}
S^{\sss(3)}\ket{\ttr}=\ket{\ttl},&&
U^{\sss(1)}\ket{\ttr}=\ket{\ttr},&&
U^{\sss(1)}\ket{\ttl}=0.
\end{align}
Then, by the Lie-Trotter formula,
\begin{align}
e^{-\beta H_\Lambda}&=e^{-\beta q|\Lambda|}\lim_{\ell\uparrow\infty}\Bigg(\exp
 \bigg(\frac\beta\ell\sum_{\{x,y\}\in\Jbb_\Lambda}J_{x,y}S_x^{\sss(3)}
 S_y^{\sss(3)}\bigg)\exp\bigg(\frac{2\beta q}\ell\sum_{z\in\Lambda}U_z^{\sss(1)}
 \bigg)\Bigg)^\ell\nn\\
&=e^{\beta q|\Lambda|+\beta|J|}\lim_{\ell\uparrow
 \infty}\Bigg(\prod_{\{x,y\}\in\Jbb_\Lambda}e^{\frac{\beta J_{x,y}}\ell
 (S_x^{(3)}S_y^{(3)}-I)}\prod_{z\in\Lambda}e^{\frac{2\beta q}\ell(U_z^{(1)}-I)}
 \Bigg)^\ell,
\end{align}
where $|J|=\sum_{b\in\Jbb_\Lambda}\!J_b$.  Since
\begin{align}
e^{\frac{\beta J_{x,y}}\ell(S_x^{(3)}S_y^{(3)}-I)}&=I+\frac{\beta J_{x,y}}\ell
 (S_x^{(3)}S_y^{(3)}-I)+o(\ell^{-1})\nn\\
&=\Big(1-\frac{\beta J_{x,y}}\ell\Big)I+\frac{\beta J_{x,y}}\ell S_x^{(3)}
 S_y^{(3)}+o(\ell^{-1}),
\end{align}
and similarly
\begin{align}
e^{\frac{2\beta q}\ell(U_z^{(1)}-I)}=\Big(1-\frac{2\beta q}\ell\Big)I+\frac{2
 \beta q}\ell U_z^{(1)}+o(\ell^{-1}),
\end{align}
we obtain 
\begin{align}
e^{-\beta H_\Lambda}=e^{\beta q|\Lambda|+\beta|J|}\lim_{\ell\uparrow\infty}
 \Bigg(\prod_{\{x,y\}\in\Jbb_\Lambda}\bigg(\Big(1-\frac{\beta J_{x,y}}\ell
 \Big)I+\frac{\beta J_{x,y}}\ell S_x^{(3)}S_y^{(3)}\,\added[id=AS]{+o(\ell^{-1})}\bigg)\nn\\
\times\prod_{z\in\Lambda}\bigg(\Big(1-\frac{2\beta q}\ell\Big)I+\frac{2\beta q}
 \ell U_z^{(1)}\,\added[id=AS]{+o(\ell^{-1})}\bigg)&\Bigg)^\ell.
\end{align}
This yields a representation in terms of independent Poisson point processes 
$\xivec=\{\xi_b\}_{b\in\Jbb_\Lambda}$ and $\mvec=\{m_z\}_{z\in\Lambda}$, 
where each $\xi_b=(\xi_b^{\sss(j)}:j\in\N)$ (could be empty) has intensity 
$\beta J_b$ and each $m_z=(m_z^{\sss(j)}:j\in\N)$ (could be empty) has 
intensity $2\beta q$.  Let $\Prob$ be the joint probability measures of 
$\xivec\subset\Tbb^{\Jbb_\Lambda}$ and $\mvec\subset\Tbb^\Lambda$.  
Then, we obtain (\added[id=AS]{see} e.g., \cite[Equation (2.1)]{ci2010random})
\begin{align}\label{eq:crawford-ioffe}
\mathrm{Tr}[e^{-\beta H_\Lambda}]=e^{\beta q|\Lambda|+\beta|J|}\int\Prob
 (\diff\xivec,\diff\mvec)\sum_{\substack{\psivec=\{\psi_z\}_{z\in\Lambda}:\\
 \psi_z:\Tbb\to\{\ttr,\ttl\}}}\ind{\psivec\sim(\vno,\,\xivec,\mvec)},
\end{align}
where $\psivec$ is a time-dependent (but piecewise constant) spin 
configuration, and for a finite set $A\subset\Tbb^\Lambda$ of points, 
we mean by $\psivec\sim(A,\,\xivec,\mvec)$ that
\begin{enumerate}[(i)]
\item
$\psi_z$ flips at every $(t,z)\in A$ (we call $A$ the source set),
\item
$\psi_u$ and $\psi_v$ simultaneously flip at every $t\in\xi_{u,v}$ (we call 
$\xivec$ the bridge configuration),
\item
$\psi_z(t)=\ttr$ at every $t\in m_z$ (we call $\mvec$ the mark configuration).
\end{enumerate}
\added[id=YK]{We show an example of a spin configuration $\psivec$ in Figure~\ref{fig:example_space-time-spin-configuration}.}
Let
\begin{align}
\ind{\partial\psivec=A}(\xivec,\mvec)=\ind{\psivec\sim(A,\,\xivec,\mvec)},
\end{align}
where the notation $\partial\psivec$ for the source set is similar to the one 
used in the random-current representation for the classical Ising model (\added[id=AS]{see} e.g., 
\cite{a1982geometric})\footnote{In the random-current representation, each bond $b$ 
is assigned a nonnegative integer $n_b$, called current, which is equal to 
the number of bridges in the present setting, i.e., 
$\xi_b=(\xi_b^{\sss(1)},\dots,\xi_b^{\sss(n_b)})$, or equivalently  
$n_b={}^\#\xi_b\equiv\int_\Tbb\delta_{\xi_b}(t)\,\diff t$.}.  Denoting by $\Ebb$ 
the expectation against the measure $\Prob(\diff\xivec,\diff\mvec)$, we define 
$Z$ as
\begin{align}
Z=\int\Prob(\diff\xivec,\diff\mvec)\sum_{\substack{\psivec=\{\psi_z\}_{z\in
 \Lambda}:\\ \psi_z:\Tbb\to\{\ttr,\ttl\}}}\ind{\psivec\sim(\vno,\,\xivec,\mvec)}
 \equiv\Ebb\bigg[\sum_{\psivec}\ind{\partial\psivec=\vno}\bigg].
\end{align}

Similarly, for $\ovec=(0,o)$ and $\xvec=(t,x)$, we can rewrite the numerator of 
the two-point function as 
\begin{align}
\mathrm{Tr}[S_o^{\sss(3)}e^{-t\beta H_\Lambda}S_x^{\sss(3)}e^{-(1-t)\beta
 H_\Lambda}]=e^{\beta q|\Lambda|+\beta|J|}\,\Ebb\bigg[\sum_{\psivec}
 \ind{\partial\psivec=\ovec\vtri\xvec}\bigg],
\end{align}
where $\ovec\vtri\xvec$ is an abbreviation for the symmetric difference 
$\{\ovec\}\triangle\{\xvec\}$.  As a result,
\begin{align}\label{eq:crawford-ioffe2}
  G(\ovec,\xvec)=
    \frac{1}{Z}
    \sgExpect[\bigg]{
      \sum_{\bm{\psi}}
      \ind{\partial\bm{\psi} = \bm{o} \vtri \bm{x}}
    }.
\end{align}
Also,\todo{Added by Kamijima.} for $\bm{y} = (s, y)$, we can rewrite 
the one-point function $\Expect{S_y^{\sss(1)}(s)}_\Lambda$ as 
(\added[id=AS]{see} \Refeq{U-def}--\Refeq{ttrttl})
\begin{align}\lbeq{representation_x-correlation}
\Expect*{S_y^{\sss(1)}(s)}_\Lambda
 =\Expect*{2U_y^{\sss(1)}(s)-I}_\Lambda
 =\frac2Z\sgExpect[\Bigg]{\sum_{\partial\psivec=\vno}
 \ind{\psi_y(s)=\ttr}}-1.
\end{align}

We will also need a restricted version of the right-hand side of 
\eqref{eq:crawford-ioffe2} on the complement of a set $\Ccal\subset\Tbb^\Lambda$, 
as follows.  Let 
$\Ccal^\compl=\Tbb^\Lambda\setminus\Ccal=\bigcup_{z\in\Lambda}I_z$, where 
each $I_z\subset\Tbb$ is a union of finite number of ``intervals'' (an interval 
is a maximal connected component of \added[id=AS]{$\Tbb=\R/\Z$}\deleted[id=YK]{ oriented in the 
time-increasing direction}) and let $\Prob_{\Ccal^\compl}$ be the joint measure 
of independent Poisson point processes $\xivec=\{\xi_b\}_{b\in\Jbb_\Lambda}$ 
and $\mvec=\{m_z\}_{z\in\Lambda}$, where each 
$\xi_{u,v}=(\xi_{u,v}^{\sss(j)}:j\in\N)\subset I_u\cap I_v$ has intensity 
$\beta J_{u,v}$ and each $m_z=(m_z^{\sss(j)}:j\in\N)\subset I_z$ has intensity 
$2\beta q$.  Then, we let
\begin{align}
Z_{\Ccal^\compl}=\Ebb_{\Ccal^\compl}\bigg[\sum_{\psivec}\ind{\partial
 \psivec=\vno}\bigg]\equiv\int\Prob_{\Ccal^\compl}(\diff\xivec,\diff\mvec)
 \sum_{\substack{\psivec=\{\psi_z\}_{z\in\Lambda}:\\ \psi_z:I_z\to\{\ttr,
 \ttl\}}}\ind{\psivec\sim(\vno,\,\xivec,\mvec)},
\end{align}
where each $\psi_z$, if $I_z\ne\Tbb$, is under the ``free-boundary condition'' 
in the sense that $\psi_z(t)=\ttr$ for all $t$ at the ``boundaries'' of $I_z$ 
(unless there are sources or elements in $m_z$ at the boundaries of $I_z$, 
but the latter is unlikely to occur).  Finally, we define a restricted version 
of \eqref{eq:crawford-ioffe2} as
\begin{align}\label{eq:crawford-ioffe3}
G_{\Ccal^\compl}(\ovec,\xvec)=\frac1{Z_{\Ccal^\compl}}\Ebb_{\Ccal^\compl}
 \bigg[\sum_{\psivec}\ind{\partial\psivec=\ovec\vtri\xvec}\bigg].
\end{align}
If $\Ccal=\vno$, we simply omit the subscript and denote it by 
$G(\ovec,\xvec)$.

\subsection{Source switching}\label{s:SW}
In later sections, we will have to deal with the difference between 
\eqref{eq:crawford-ioffe2} and \eqref{eq:crawford-ioffe3}, that is
\begin{align}\lbeq{2pt-difference}
G(\ovec,\xvec)-G_{\Ccal^\compl}(\ovec,\xvec)
&=\frac1Z\Ebb\bigg[\sum_{\psivec}\ind{\partial\psivec=\ovec\vtri\xvec}\bigg]
 -\frac1{Z_{\Ccal^\compl}}\Ebb_{\Ccal^\compl}\bigg[\sum_{\psivec}\ind{\partial
 \psivec=\ovec\vtri\xvec}\bigg]\nn\\
&=\frac1{ZZ_{\Ccal^\compl}}\Bigg(\Ebb^1\Ebb_{\Ccal^\compl}^2\Bigg[\sum_{
 \substack{\partial\psivec^1=\ovec\vtri\xvec\\ \partial\psivec^2=\vno}}1\Bigg]
 -\Ebb^1\Ebb_{\Ccal^\compl}^2\Bigg[\sum_{\substack{\partial\psivec^1=\vno\\
 \partial\psivec^2=\ovec\vtri\xvec}}1\Bigg]\Bigg),
\end{align}
where $\Ebb^1\Ebb_{\Ccal^\compl}^2$ is the expectation against the product 
measure $\Prob^1(\diff\xivec^1,\diff\mvec^1)\,\Prob_{\Ccal^\compl}^2(\diff
\xivec^2,\diff\mvec^2)$, and each $\psivec^j$ is compatible with $\xivec^j$ and 
$\mvec^j$.  Let $\tilde\Prob$ be the joint measure of 
$\xivec=\xivec^1\cup\xivec^2$ and $\mvec=\mvec^1\cup\mvec^2$, i.e., the 
measure of independent Poisson point processes 
$\xivec=\{\xi_b\}_{b\in\Jbb_\Lambda}$ and $\mvec=\{m_z\}_{z\in\Lambda}$ 
whose intensities are doubled in the common region $\Ccal^\compl$, and denote 
its expectation by $\tilde\Ebb$.  Let $\xivec|_{\Ccal^\compl}$ (resp., 
$\mvec|_{\Ccal^\compl}$) be the restriction of $\xivec$ (resp., $\mvec$) to 
$\Ccal^\compl$, and denote its cardinality by ${}^\#\xivec|_{\Ccal^\compl}$ 
(resp., ${}^\#\mvec|_{\Ccal^\compl}$).  Then, we obtain the rewrite 
\begin{align}\label{eq:2pt-diff-rewr1}
\Ebb^1\Ebb_{\Ccal^\compl}^2\Bigg[\sum_{\substack{\partial\psivec^1=\ovec
 \vtri\xvec\\ \partial\psivec^2=\vno}}1\Bigg]&=\tilde\Ebb\Bigg[\bigg(\frac12
 \bigg)^{{}^\#\xivec|_{\Ccal^\compl}+{}^\#\mvec|_{\Ccal^\compl}}\hskip-5pt
 \sum_{\substack{\xivec^2\subset\xivec|_{\Ccal^\compl}\\ \mvec^2\subset
 \mvec|_{\Ccal^\compl}}}\sum_{\substack{\psivec^1\sim(\ovec\vtri\xvec,\,\xivec
 \setminus\xivec^2,\mvec\setminus\mvec^2)\\ \psivec^2\sim(\vno,\,\xivec^2,
 \mvec^2)}}1\Bigg]\nn\\
&\equiv\tilde\Ebb\Bigg[\bigg(\frac12\bigg)^{{}^\#\xivec|_{\Ccal^\compl}+{}^\#
 \mvec|_{\Ccal^\compl}}\hskip-1.5pc\sum_{\substack{(\partial\psivec^1,\partial
 \psivec^2)=(\ovec\vtri\xvec,\,\vno)\\ \psivec^2\text{ on }\Ccal^\compl}}1
 \Bigg].
\end{align}
Similarly, we have
\begin{align}\label{eq:2pt-diff-rewr2}
\Ebb^1\Ebb_{\Ccal^\compl}^2\Bigg[\sum_{\substack{\partial\psivec^1=\vno\\
 \partial\psivec^2=\ovec\vtri\xvec}}1\Bigg]=\tilde\Ebb\Bigg[\bigg(\frac12
 \bigg)^{{}^\#\xivec|_{\Ccal^\compl}+{}^\#\mvec|_{\Ccal^\compl}}\hskip-1.5pc
 \sum_{\substack{(\partial\psivec^1,\partial\psivec^2)=(\vno,\,\ovec\vtri\xvec)\\
 \psivec^2\text{ on }\Ccal^\compl}}1\Bigg].
\end{align}
We will swap the source constraints by using the so-called source switching 
(\added[id=AS]{see} Lemma~\ref{lmm:SST}) and simplify the expression 
\eqref{eq:2pt-difference}.  To explain this technique, we first introduce some 
notions and notation.
%

\begin{shaded}
\begin{definition}\label{def:connectivity}
\begin{enumerate}[(i)]
\item
A path from $\xvec$ to $\yvec$ in the bridge configuration $\xivec$ is \replaced[id=YK]{an interval between $\bm{x}$ and $\bm{y}$ or an ordered set $\mathcal{P} = (I^{\sss(1)}, I^{\sss(2)}, \dots, I^{\sss(n)})$ of disjoint intervals that satisfy the following property: there are $n-1$ bridges $\{\bm{u}_1, \bm{v}_1\}, \dots, \{\bm{u}_{n-1}, \bm{v}_{n-1}\} \in \xivec$ such that $I^{\sss(1)}$ is an interval between $\bm{x}$ and $\bm{u}_1$, $I^{\sss(2)}$ is an interval between $\bm{v}_1$ and $\bm{u}_2$, and so on, and $I^{\sss(n)}$ is an interval between $\bm{v}_{n-1}$ and $\bm{y}$}{a self-avoiding path with endpoints $\xvec,\yvec$, traversing time intervals and possibly bridges in $\xivec$}.
\item
Let $\ttr(\psivec)=\{(t,z):\psi_z(t)=\ttr\}$.  We say that a path \added[id=YK]{$\mathcal{P} = (I^{\sss(1)}, I^{\sss(2)}, \dots, I^{\sss(n)})$} in the bridge 
configuration $\xivec$ is $(\psivec^1,\psivec^2,\mvec)$-open (or simply say 
that a path is open) if it \added[id=AS]{does not include} sub-intervals of 
$\ttr(\psivec^1)\cap\ttr(\psivec^2)$ containing marks in $\mvec$.
\added[id=YK]{Notice that the endpoints of any bridge in $\xivec$ do not coincide with any mark in $\mvec$ with probability $1$, since two Poisson point processes do not intersect almost surely.}
\item
Given a set $\Ccal\subset\Tbb^\Lambda$ and $\xvec,\yvec\notin\Ccal$, 
we define $\{\xvec\cn{1,2}{}\yvec$ in $\Ccal^\compl\}$ to be the event that 
either $\xvec=\added[id=AS]{\yvec\in\Ccal^\compl}$ or there is an open path in 
$\Ccal^\compl$ from $\xvec$ to $\yvec$.  We omit the proviso ``\,in 
$\Ccal^\compl$\,'' if $\Ccal=\vno$ (i.e., $\Ccal^\compl=\replaced[id=YK]{\Tbb^\Lambda}{\Lambda}$).  Let
\begin{align}\label{eq:through}
\Big\{\xvec\cn{1,2}{\Ccal}\yvec\Big\}=\Big\{\xvec\cn{1,2}{}\yvec\Big\}
 \setminus\Big\{\xvec\cn{1,2}{}\yvec\text{ in }\Ccal^\compl\Big\}.
\end{align}
\end{enumerate}
\end{definition}
\end{shaded}

Unlike the random-current representation for the classical Ising model~\cite{a1982geometric},
the above connectivity is defined by the superposition of the spin configurations $\bm{\psi}^1$ and $\bm{\psi}^2$\added[id=AS]{, not by} a single spin configuration.
One of the reasons comes from compatibility with the source switching.

\added[id=YK]{In the rest of this section, we describe the source switching, which is widely used to prove various correlation inequalities.
We also use it to derive a new lace expansion for the classical Ising model in Section~\ref{s:LE}.}
Recall \eqref{eq:2pt-diff-rewr2}.  Since $\psivec^2$ is a spin configuration on 
$\Ccal^\compl$ with the source constraint $\partial\psivec^2=\ovec\vtri\xvec$, 
there must be a path in $\Ccal^\compl$ from $\ovec$ to $\xvec$ along which 
$\psivec^2$ is always $\ttl$, hence open.  On the 
other hand, an open path in \eqref{eq:2pt-diff-rewr1} does not have to be 
confined in $\Ccal^\compl$.  Therefore, we obtain the identities
\begin{align}
\Ebb^1\Ebb_{\Ccal^\compl}^2\Bigg[\sum_{\substack{\partial\psivec^1=\ovec\vtri
 \xvec\\ \partial\psivec^2=\vno}}1\Bigg]&=\tilde\Ebb\Bigg[\bigg(\frac12
 \bigg)^{{}^\#\xivec|_{\Ccal^\compl}+{}^\#\mvec|_{\Ccal^\compl}}\hskip-1.5pc
 \sum_{\substack{(\partial\psivec^1,\partial\psivec^2)=(\ovec\vtri\xvec,\,
 \vno)\\ \psivec^2\text{ on }\Ccal^\compl}}\ind{\ovec\cn{1,2}{}\xvec}\Bigg],
 \label{eq:2pt-diff-par1}\\
\Ebb^1\Ebb_{\Ccal^\compl}^2\Bigg[\sum_{\substack{\partial\psivec^1=\vno\\
 \partial\psivec^2=\ovec\vtri\xvec}}1\Bigg]&=\tilde\Ebb\Bigg[\bigg(\frac12
 \bigg)^{{}^\#\xivec|_{\Ccal^\compl}+{}^\#\mvec|_{\Ccal^\compl}}\hskip-1.5pc
 \sum_{\substack{(\partial\psivec^1,\partial\psivec^2)=(\vno,\,\ovec\vtri
 \xvec)\\ \psivec^2\text{ on }\Ccal^\compl}}\ind{\ovec\cn{1,2}{}\xvec
 \text{ in }\Ccal^\compl}\Bigg].\label{eq:2pt-diff-par2}
\end{align}

Next we swap the source constraints in \eqref{eq:2pt-diff-par2} as follows.  
Fix a bridge configuration $\xivec$ in which there are finitely many paths 
$\Pcal_1,\dots,\Pcal_\nu$ in $\Ccal^\compl$ from $\ovec$ to $\xvec$.  
Suppose that they are ordered in a certain way ($\Pcal_j$ is earlier than 
$\Pcal_k$ in that order if $j<k$) and let 
$\Ocal_j=\Ocal_j(\psivec^1,\psivec^2,\mvec)$ be the event that $\Pcal_j$ 
is the earliest path which is $(\psivec^1,\psivec^2,\mvec)$-open.  
Then, by conditioning on $\xivec$,  we have
\begin{align}\label{eq:2pt-diff-par3}
\tilde\Ebb\Bigg[\bigg(\frac12\bigg)^{{}^\#\mvec|_{\Ccal^\compl}}\hskip-1.5pc
 \sum_{\substack{(\partial\psivec^1,\partial\psivec^2)=(\vno,\,\ovec\vtri
 \xvec)\\ \psivec^2\text{ on }\Ccal^\compl}}\ind{\ovec\cn{1,2}{}\xvec
 \text{ in }\Ccal^\compl}\Bigg|\xivec\Bigg]=\sum_{j=1}^\nu\tilde\Ebb\Bigg[
 \bigg(\frac12\bigg)^{{}^\#\mvec|_{\Ccal^\compl}}\hskip-1.5pc\sum_{\substack
 {(\partial\psivec^1,\partial\psivec^2)=(\vno,\,\ovec\vtri\xvec)\\ \psivec^2
 \text{ on }\Ccal^\compl}}\indic{\Ocal_j}\Bigg|\xivec\Bigg].
\end{align}
On the event $\Ocal_j$, the following map $\Phi_{\Pcal_j}:\{(\psivec^1,\mvec^1),
(\psivec^2,\mvec^2)\}\mapsto\{(\tilde\psivec^1,\tilde\mvec^1),(\tilde\psivec^2,
\tilde\mvec^2)\}$ is a measure-preserving bijection (recall that $\psivec^1$ 
and $\psivec^2$ uniquely determines the splitting 
$\xivec=\xivec^1\cup\xivec^2$) and $\Pcal_j$ is also the earliest path which is 
$(\tilde\psivec^1,\tilde\psivec^2,\tilde\mvec^1\cup\tilde\mvec^2)$-open:
\begin{enumerate}[(i)]
\item
If $\xvec\notin\Pcal_j$, then nothing changes.
\item
If $\xvec\in\Pcal_j$ and $(\psivec^1,\psivec^2)(\xvec)=(\ttl,\ttl)$ 
(hence $\xvec\notin\mvec^1\cup\mvec^2$, 
i.e., $t_{\xvec}\notin m_x^1\cup m_x^2$), then we let 
$(\tilde\psivec^1,\tilde\psivec^2)(\xvec)=(\ttr,\ttr)$.  Likewise, if 
$\xvec\in\Pcal_j$ and $(\psivec^1,\psivec^2)(\xvec)=(\ttr,\ttr)$ 
(hence $\xvec\notin\mvec^1\cup\mvec^2$ on the event $\Ocal_j$), 
then we let $(\tilde\psivec^1,\tilde\psivec^2)(\xvec)=(\ttl,\ttl)$.
\item
If $\xvec\in\Pcal_j$ and $(\psivec^1,\psivec^2)(\xvec)=(\ttl,\ttr)$ 
(hence $\xvec\notin\mvec^1$), then we let 
$(\tilde\psivec^1,\tilde\psivec^2)(\xvec)=(\ttr,\ttl)$; in addition, 
if $\xvec\in\mvec^2$, then we let $\xvec\in\tilde\mvec^1$ and 
$\xvec\notin\tilde\mvec^2$.  Likewise, if $\xvec\in\Pcal_j$ and 
$(\psivec^1,\psivec^2)(\xvec)=(\ttr,\ttl)$ (hence $\xvec\notin\mvec^2$), 
then we let $(\tilde\psivec^1,\tilde\psivec^2)(\xvec)=(\ttl,\ttr)$; in addition, 
if $\xvec\in\mvec^1$, then we let $\xvec\notin\tilde\mvec^1$ and 
$\xvec\in\tilde\mvec^2$.  
\item
Let $\partial\tilde\psivec^1=\partial\psivec^1\vtri\ovec\vtri\xvec$ 
and $\partial\tilde\psivec^2=\partial\psivec^2\vtri\ovec\vtri\xvec$.
\end{enumerate}
As a result, we can swap the source constraints in \eqref{eq:2pt-diff-par3}, 
hence in \eqref{eq:2pt-diff-par2} as well, as
\begin{align}\label{eq:2pt-diff-par4}
\Ebb^1\Ebb_{\Ccal^\compl}^2\Bigg[\sum_{\substack{\partial\psivec^1=\vno\\
 \partial\psivec^2=\ovec\vtri\xvec}}1\Bigg]&=\tilde\Ebb\Bigg[\bigg(\frac12
 \bigg)^{{}^\#\xivec|_{\Ccal^\compl}+{}^\#\mvec|_{\Ccal^\compl}}\hskip-1.5pc
 \sum_{\substack{(\partial\psivec^1,\partial\psivec^2)=(\ovec\vtri\xvec,\,
 \vno)\\ \psivec^2\text{ on }\Ccal^\compl}}\ind{\ovec\cn{1,2}{}\xvec
 \text{ in }\Ccal^\compl}\Bigg]\nn\\
&\equiv\Ebb^1\Ebb_{\Ccal^\compl}^2\Bigg[\sum_{\substack{\partial\psivec^1=\ovec
 \vtri\xvec\\ \partial\psivec^2=\vno}}\ind{\ovec\cn{1,2}{}\xvec\text{ in }
 \Ccal^\compl}\Bigg],
\end{align}
where $\{\ovec\cn{1,2}{}\xvec\text{ in }\Ccal^\compl\}$ in the last line 
should be interpreted as the event that, in the bridge configuration 
$\xivec^1\cup\xivec^2$, there is a 
$(\psivec^1,\psivec^2,\mvec^1\cup\mvec^2)$-open path in $\Ccal^\compl$.  
Similarly, we rewrite \eqref{eq:2pt-diff-par1} as
\begin{align}\label{eq:2pt-diff-par5}
\Ebb^1\Ebb_{\Ccal^\compl}^2\Bigg[\sum_{\substack{\partial\psivec^1=\ovec
 \vtri\xvec\\ \partial\psivec^2=\vno}}1\Bigg]=\Ebb^1\Ebb_{\Ccal^\compl}^2
 \Bigg[\sum_{\substack{\partial\psivec^1=\ovec\vtri\xvec\\ \partial\psivec^2
 =\vno}}\ind{\ovec\cn{1,2}{}\xvec}\Bigg],
\end{align}
so that, by using the notation \eqref{eq:through}, we arrive at the identity
\begin{align}\lbeq{2pt-diff-par6}
\Ebb^1\Ebb_{\Ccal^\compl}^2\Bigg[\sum_{\substack{\partial\psivec^1=\ovec
 \vtri\xvec\\ \partial\psivec^2=\vno}}1\Bigg]-\Ebb^1\Ebb_{\Ccal^\compl}^2\Bigg[
 \sum_{\substack{\partial\psivec^1=\vno\\ \partial\psivec^2=\ovec\vtri\xvec}}1
 \Bigg]=\Ebb^1\Ebb_{\Ccal^\compl}^2\Bigg[\sum_{\substack{\partial\psivec^1=\ovec
 \vtri\xvec\\ \partial\psivec^2=\vno}}\ind{\ovec\cn{1,2}{\Ccal}\xvec}\Bigg],
\end{align}
where $\{\ovec\cn{1,2}{\Ccal}\xvec\}$ should be interpreted as the event 
that, in the bridge configuration $\xivec^1\cup\xivec^2$, all 
$(\psivec^1,\psivec^2,\mvec^1\cup\mvec^2)$-open paths must go through 
the set $\Ccal$.

\bigskip

The following is \added[id=AS]{a} generalization of the source switching \added[id=AS]{explained above}:

\begin{shaded}
    \begin{lemma}[{Source switching 
    \cite[Section~2]{ci2010random}}]
\label{lmm:SST}
Let $\Ccal\subset\Tbb^\Lambda$ be such that its complement 
$\Ccal^\compl=\Tbb^\Lambda\setminus\Ccal$ is a union of finite number of 
intervals, and let $A\subset\Tbb^\Lambda$ and $B\subset\Ccal^\compl$ be finite 
sets.  Let $F(\psivec^1,\psivec^2,\mvec)$ be a function that depends only on 
the connectivity properties using open paths of $(\psivec^1,\psivec^2,\mvec)$. 
Then we have
\begin{align}
&\Ebb^1\Ebb_{\Ccal^\compl}^2\Bigg[\sum_{\substack{\partial\psivec^1=A\\ \partial
 \psivec^2=B}}F(\psivec^1,\psivec^2,\mvec^1\cup\mvec^2)\,\ind{\xvec\cn{1,2}{}
 \yvec\text{ in }\Ccal^\compl}\Bigg]\nn\\
&=\Ebb^1\Ebb_{\Ccal^\compl}^2\Bigg[\sum_{\substack{\partial\psivec^1=A\vtri\xvec
 \vtri\yvec\\ \partial\psivec^2=B\vtri\xvec\vtri\yvec}}F(\psivec^1,\psivec^2,
 \mvec^1\cup\mvec^2)\,\ind{\xvec\cn{1,2}{}\yvec\text{ in }\Ccal^\compl}\Bigg].
\end{align}
\end{lemma}
\end{shaded}

\added[id=AS]{For example, if $F\equiv1$, $A=\vno$ and $B=\xvec\vtri\yvec$, 
then $A\vtri\xvec\vtri\yvec=\xvec\vtri\yvec$ and $B\vtri\xvec\vtri\yvec=\vno$, 
just as in obtaining \eqref{eq:2pt-diff-par4}.}

\section{Sketch proof of the differential inequalities for the susceptibility}
\label{s:bjornberg}
As an example of application of the stochastic-geometric representation, 
we review the proof of the differential inequalities in Lemma~\ref{lem:diff-ineq_susceptibility_coupling-coefficient_transverse-field}.  

First, by using the stochastic-geometric representation in the previous section, 
we derive a much simpler expression for 
\eqref{eq:pre_derivative_susceptibility_coupling-coefficient}. 
Let 
\begin{align}
    \Expect{A\,;B}_\Lambda=\Expect{AB}_\Lambda-\Expect{A}_\Lambda\Expect{B}_\Lambda
\end{align} 
be the truncated correlation function for operators 
$A,B:(\mathbb{C}^2)^{\otimes\Lambda}\to(\mathbb{C}^2)^{\otimes\Lambda}$. 
Then, by \eqref{eq:crawford-ioffe2}, we have the rewrite
\begin{align}
&\int_{0}^{t} \odif{s}\,\Expect*{S^{\sss(3)}_o(0)\,S^{\sss(3)}_u(s)\,
 S^{\sss(3)}_v(s)\, S^{\sss(3)}_x(t)}_\Lambda 
 +\int_{t}^{1} \odif{s}\,\Expect*{S^{\sss(3)}_o(0)\,S^{\sss(3)}_x(t)\,
 S^{\sss(3)}_u(s)\,S^{\sss(3)}_v(s)}_\Lambda\nn\\
&=\int_0^t\diff s\,\frac1Z\Ebb\Bigg[\sum_{\psivec}\ind{\partial\psivec
 =(0,o)\vtri(t,x)\vtri\{(s,u),(s,v)\}}\Bigg]
 +\int_t^1\diff s\,\frac1Z\Ebb\Bigg[\sum_{\psivec}\ind{\partial\psivec
 =(0,o)\vtri(t,x)\vtri\{(s,u),(s,v)\}}\Bigg]\nn\\
&=\int_{\Tbb}\diff s\,\frac1Z\Ebb\Bigg[\sum_{\psivec}\ind{\partial\psivec
 =(0,o)\vtri(t,x)\vtri\{(s,u),(s,v)\}}\Bigg]\nn\\
&=\int_{\Tbb}\diff s\,\Expect*{S^{\sss(3)}_o(0)\,S^{\sss(3)}_x(t)\,
 S^{\sss(3)}_u(s)\,S^{\sss(3)}_v(s)}_\Lambda,
\end{align}
where the last equality is due to the symmetry with respect to 
how those four points are arranged.  Therefore, 
\begin{align}\lbeq{derivative_susceptibility_coupling-coefficient}
\eqref{eq:pre_derivative_susceptibility_coupling-coefficient}
 =\beta\int_{\Tbb}\diff t\sum_{x\in\Lambda}\int_{\Tbb}\diff s\,
 \Expect*{S^{\sss(3)}_o(0)\,S^{\sss(3)}_x(t)\,;S^{\sss(3)}_u(s)\,
 S^{\sss(3)}_v(s)}_\Lambda.
\end{align}
Recalling the definition of $\chi_\Lambda$ (\added[id=AS]{see} resp., \eqref{eq:chidef}) 
and using translation-invariance (since $\Lambda$ is a torus), we arrive at
\begin{align}\lbeq{pre_diff-ineq_susceptibility_coupling-coefficient}
&\sum_{\{u,v\}\in\Jbb_\Lambda}\frac{J_{u,v}}{\beta}
 \pdv{\chi_\Lambda}{J_{u,v}}=\ft J(0)\,\chi_\Lambda^2\nn\\
&+\int_{\Tbb}\diff t\int_{\Tbb}\diff s\hskip-1pc
 \sum_{\substack{\xvec:t_{\xvec}=t\\ \{\uvec,\vvec\}:t_{\uvec}=t_{\vvec}=s}}
 \hskip-1pc J_{u,v}\bigg(\underbrace{\Expect*{S^{\sss(3)}_o(0)\,
 S^{\sss(3)}_x(t)\,;S^{\sss(3)}_u(s)\,S^{\sss(3)}_v(s)}_\Lambda
 -G(\ovec,\uvec)\,G(\xvec,\vvec)-G(\ovec,\vvec)\,G(\xvec,\uvec)}_{=:\,
 F_4(\ovec,\xvec,\uvec,\vvec)}\bigg),
\end{align}
where we have used $\sum_{\{u,v\}\in\Jbb_\Lambda}J_{u,v}=\frac12\hat J(0)$.
Notice that the first term on the right, $\ft J(0)\,\chi_\Lambda^2$, appears 
in both sides of \eqref{eq:diff-ineq_susceptibility_coupling-coefficient}. 
The integrand $F_4(\ovec,\xvec,\uvec,\vvec)$ is often called the fourth 
Ursell function.

Next, we simplify \Refeq{pre_derivative_susceptibility_transverse-field}.  
First, by $S^{\sss(1)}=2U^{\sss(1)}-I$ and 
\eqref{eq:representation_x-correlation}, we have the rewrite
\begin{align}
&\int_0^t\diff s\,\Expect*{S^{\sss(3)}_o(0)\,S^{\sss(1)}_y(s)\,
 S^{\sss(3)}_x(t)}_\Lambda+\int_t^1\diff s\,\Expect*{S^{\sss(3)}_o(0)\,
 S^{\sss(3)}_x(t)\,S^{\sss(1)}_y(s)}_\Lambda\nn\\
&=2\int_0^t\diff s\,\Expect*{S^{\sss(3)}_o(0)\,U^{\sss(1)}_y(s)\,
 S^{\sss(3)}_x(t)}_\Lambda+2\int_t^1\diff s\,\Expect*{S^{\sss(3)}_o(0)\,
 S^{\sss(3)}_x(t)\,U^{\sss(1)}_y(s)}_\Lambda-\Expect*{S^{\sss(3)}_o(0)\,S^{\sss(3)}_x(t)}_\Lambda\nn\\
&=\int_0^t\diff s\,\frac2Z\Ebb\bigg[\sum_{\partial\psivec=(0,o)\vtri(t,x)}
 \ind{\psi_y(s)=\ttr}\bigg]+\int_t^1\diff s\,\frac2Z\Ebb\bigg[\sum_{\partial
 \psivec=(0,o)\vtri(t,x)}\ind{\psi_y(s)=\ttr}\bigg]-
 \Expect*{S^{\sss(3)}_o(0)\,S^{\sss(3)}_x(t)}_\Lambda\nn\\
&=\int_{\Tbb}\diff s\,\frac2Z\Ebb\bigg[\sum_{\partial\psivec=(0,o)\vtri(t,x)}
 \ind{\psi_y(s)=\ttr}\bigg]-\added[id=AS]{\Expect*{S^{\sss(3)}_o(0)\,
 S^{\sss(3)}_x(t)}_\Lambda}\nn\\
&=\int_{\Tbb}\diff s\,\Expect*{S^{\sss(3)}_o(0)\,S^{\sss(3)}_x(t)\,
 S^{\sss(1)}_y(s)}_\Lambda,
\end{align}
where the last equality is due to the symmetry with respect to how those \added[id=AS]{three} points are arranged. Therefore, 
\begin{align}\label{eq:derivative_susceptibility_transverse-field}
\Refeq{pre_derivative_susceptibility_transverse-field}=\beta\int_{\Tbb}
 \diff t\sum_{x,y\in\Lambda}\int_{\Tbb}\diff s~\Expect*{S^{\sss(3)}_o(0)\,
 S^{\sss(3)}_x(t)\,;S^{\sss(1)}_y(s)}_\Lambda.
\end{align}
Again, by the definition of $\chi_\Lambda$ and using translation-invariance, we arrive at
\begin{align}\lbeq{pre_diff-ineq_susceptibility_transverse-field}
-\frac1\beta\pdv{\chi_\Lambda}{\TField}
 =2\chi_\Lambda^2-\int_{\Tbb}\odif{t}\int_{\Tbb}\odif{s}
 \sum_{\substack{\xvec:t_{\xvec}=t\\ \yvec:t_{\yvec}=s}}
 \bigg(\underbrace{\Expect*{S^{\sss(3)}_o(0)\,
 S^{\sss(3)}_x(t)\,;S^{\sss(1)}_y(s)}_\Lambda
+2G(\ovec,\yvec)\,G(\yvec,\xvec)}_{=:\,F_3(\ovec,\xvec,\yvec)}\bigg),
\end{align}
where the first term on the right, $2\chi_\Lambda^2$, appears in both 
sides of \eqref{eq:diff-ineq_susceptibility_transverse-field}.

The following lemma provides bounds on $F_4$ and $F_3$, which are the 
counterparts of the differential inequalities in \cite{ag1983renormalized}.

\begin{shaded}
\begin{lemma}[{Generalization of \cite[Inequalities~(68) and (75)]{b2013infrared}}]
  \label{lem:AG-type-ineq}
  For any spin-spin coupling that satisfies Assumption~\ref{asm:ferromagnetism},
  and for any $\bm{w}, \bm{x}, \bm{y}, \bm{z} \in \mathbb{T}\times\Lambda$,
  \begin{align}\lbeq{truncated-z-correlation_bound}
  0\ge F_4(\wvec,\xvec,\yvec,\zvec)
  &\ge-G(\wvec,\xvec)\,G(\wvec,\yvec)\,G(\wvec,\zvec)
   -G(\xvec,\wvec)\,G(\xvec,\yvec)\,G(\xvec,\zvec)\nn\\
  &\quad-\beta\int_{\Tbb}\diff s\sum_{\{\uvec,\vvec\}:t_{\uvec}=t_{\vvec}=s}
   J_{u, v}\,G(\yvec,\vvec)\,G(\zvec,\vvec)\,\Expect*{S^{\sss(3)}_w(t_{\bm{w}})\,S^{\sss(3)}_x(t_{\bm{x}})\,;S^{\sss(3)}_u(s)\,S^{\sss(3)}_v(s)}_\Lambda\nn\\
  &\quad+4\beta\TField\int_{\Tbb}\diff s\sum_{\vvec:t_{\vvec}=s}G(\yvec,\vvec)\,
   G(\zvec,\vvec)\,\Expect*{S^{\sss(3)}_w(t_{\bm{w}})\,S^{\sss(3)}_x(t_{\bm{x}})\,;
   S^{\sss(1)}_v(s)}_\Lambda,
  \end{align}
  and
  \begin{align}\lbeq{truncated-cross-correlation_bound}
  0\le F_3(\wvec,\xvec,\yvec)
  &\le G(\wvec,\yvec)^2\,G(\wvec,\xvec)+G(\xvec,\yvec)^2\,G(\xvec,\wvec)\nn\\
  &\quad+\beta\int_{\Tbb}\diff s\sum_{\{\uvec,\vvec\}:t_{\uvec}=t_{\vvec}=s}
   J_{u,v}\,G(\yvec,\vvec)^2\,\Expect*{S^{\sss(3)}_w(t_{\bm{w}})\,
   S^{\sss(3)}_x(t_{\bm{x}})\,;S^{\sss(3)}_u(s)\,S^{\sss(3)}_v(s)}_\Lambda\nn\\
  &\quad-4\beta\TField\int_{\Tbb}\diff s\sum_{\vvec:t_{\vvec}=s}G(\yvec,\vvec)^2\,
   \Expect*{S^{\sss(3)}_w(t_{\bm{w}})\,S^{\sss(3)}_x(t_{\bm{x}})\,;
   S^{\sss(1)}_v(s)}_\Lambda.    
  \end{align}
\end{lemma}
\end{shaded}

\begin{proof}[Sketch proof of Lemma~\ref{lem:diff-ineq_susceptibility_coupling-coefficient_transverse-field}]
It is readily proven by applying the above inequalities to 
\Refeq{pre_diff-ineq_susceptibility_coupling-coefficient} and 
\Refeq{pre_diff-ineq_susceptibility_transverse-field}. 
For example, the contribution to 
\Refeq{pre_diff-ineq_susceptibility_coupling-coefficient} from 
the first term on the right-hand side of 
\Refeq{truncated-z-correlation_bound} is bounded from below as follows:
\begin{align}\lbeq{schwarz}
&-\int_{\Tbb}\diff t\int_{\Tbb}\diff s\sum_{\substack{\xvec:t_{\xvec}=t\\
 \{\uvec,\vvec\}:t_{\uvec}=t_{\vvec}=s}}J_{u,v}\,G(\ovec,\xvec)\,
 G(\ovec,\uvec)\,G(\ovec,\vvec)\nn\\
&=-\chi_\Lambda\int_{\Tbb}\diff s\sum_{\{\uvec,\vvec\}:t_{\uvec}=t_{\vvec}=s}
 J_{u,v}\,G(\ovec,\uvec)\,G(\ovec,\vvec)\nn\\
&=-\chi_\Lambda\int_{\Tbb}\diff s\sum_{\substack{\xvec:t_{\xvec}=s\\
 \{\ovec,\vvec\}:t_{\vvec}=0}}J_{o,v}\,G(\xvec,\ovec)\,G(\xvec,\vvec)
 \qquad(\because~\text{translation-invariance})\nn\\
&\ge-\chi_\Lambda\,\hat J(0)\,\sup_{\vvec}\int_{\Tbb}\diff s
 \sum_{\xvec:t_{\xvec}=s}G(\xvec,\ovec)\,G(\xvec,\vvec)\nn\\
&\ge-\chi_\Lambda\hat J(0)B\qquad\qquad(\because~\text{the Schwarz inequality}).
\end{align}
The same computation applies to the contribution from the second term 
on the right-hand side of \Refeq{truncated-z-correlation_bound}.
On the other hand, the contribution from the third term on the right-hand 
side of \Refeq{truncated-z-correlation_bound} equals
\begin{align}
&-\beta\int_{\Tbb}\diff t\int_{\Tbb}\diff s\hskip-1pc
 \sum_{\substack{\xvec:t_{\xvec}=t\\ \{\uvec,\vvec\}:t_{\uvec}=t_{\vvec}=s}}
 \hskip-1pc J_{u,v}\int_{\Tbb}\diff s'\sum_{\substack{\{\uvec',\vvec'\}:\\
 t_{\uvec'}=t_{\vvec'}=s'}}J_{u',v'}\,G(\uvec,\vvec')\,G(\vvec,\vvec')\,
 \Expect*{S^{\sss(3)}_o(0)\,S^{\sss(3)}_x(t)\,;S^{\sss(3)}_{u'}(s')\,
 S^{\sss(3)}_{v'}(s')}_\Lambda\nn\\
&=-\sum_{\{u',v'\}\in\Jbb_\Lambda}J_{u',v'}\,
 \underbrace{\beta\int_{\Tbb}\diff t\sum_{\xvec:t_{\xvec}=t}\int_{\Tbb}\diff
 s'~\Expect*{S^{\sss(3)}_o(0)\,S^{\sss(3)}_x(t)\,;S^{\sss(3)}_{u'}(s')\,
 S^{\sss(3)}_{v'}(s')}_\Lambda}_{\pdv{\chi_\Lambda}/{J_{u',v'}}\quad
 (\because~\Refeq{derivative_susceptibility_coupling-coefficient})}\nn\\
&\hskip12pc\times\int_{\Tbb}\diff s\sum_{\{\uvec,\vvec\}:t_{\uvec}=t_{\vvec}=s}
 J_{u,v}\,G\big(\uvec,(s',v')\big)\,G\big(\vvec,(s',v')\big).
\end{align}
As in \Refeq{schwarz}, the last line is bounded by $\hat J(0)B$, 
resulting in the third term on the left-hand side of 
\Refeq{diff-ineq_susceptibility_coupling-coefficient}.  
Since the other terms can be computed similarly, 
we refrain from giving tedious details.
\end{proof}

In the remainder of this section, we explain the inequalities in Lemma~\ref{lem:AG-type-ineq} schematically.  
To do so, we define some notions and notation that are also used in Section~\ref{s:LE} to derive the lace expansion.

\begin{shaded}
\begin{definition}\label{def:piv}
  Fix a bridge configuration \added[id=AS]{$\xivec=\xivec^1\cup\xivec^2$}, a mark configuration \added[id=AS]{$\mvec=\mvec^1\cup\mvec^2$} and 
  two time-dependent spin configurations $\psivec^1$ and $\psivec^2$.
  \begin{enumerate}[label=(\roman*)]
  \item
    Let $\Ccal(\xvec)$ be the set of vertices connected to $\xvec$ by a 
    $(\psivec^1,\psivec^2,\mvec)$-open path:
    \begin{align}
      \Ccal(\xvec) = \Set[\Big]{\yvec \separator \xvec\cn{1,2}{}\yvec}.
    \end{align}
  \item
    We say that $\xvec$ is doubly connected to $\yvec$ by open paths, denoted 
    $\xvec\db{1,2}{}\yvec$, if $\xvec=\yvec$ or there are at least two disjoint 
    paths from $\xvec$ to $\yvec$ that are $(\psivec^1,\psivec^2,\mvec)$-open.
  \item
    For a bridge $\bvec=\{\uvec,\vvec\}$ (i.e., 
    $t_{\uvec}=t_{\vvec}$), 
    we say that $\xvec$ is connected to $\yvec$ by an open path off $\bvec$, 
    denoted $\xvec\cn{1,2}{}\yvec$ off $\bvec$, if $\xvec=\yvec$ or there is a 
    $(\psivec^1,\psivec^2,\mvec)$-open path from $\xvec$ to $\yvec$ 
    in the new bridge configuration $\xivec\setminus\{\bvec\}$.  Let
    \begin{align}
      \tilde{\Ccal}^{\bvec}(\xvec) =
        \Set[\Big]{
          \yvec
          \separator
          \xvec\cn{1,2}{}\yvec\text{ off }\bvec
        }.
    \end{align}
    We say that the oriented bridge $\vec\bvec=(\uvec,\vvec)$ is pivotal for 
    $\xvec\cn{1,2}{}\yvec$ from $\xvec$, denoted 
    $\vec\bvec\in\piv\big\{\xvec\ocn{1,2}{}\yvec\big\}$, 
    if $\xvec\cn{1,2}{}\uvec$ off $\bvec$ and $\vvec\cn{1,2}{}\yvec$ in 
    $\tilde{\Ccal}^{\bvec}(\xvec)^\compl$.
  \item
    For \added[id=AS]{a vertex} $\vvec$, \added[id=AS]{we say that $\xvec$ is connected to $\yvec$ by an open path off 
    $\vvec$, denoted $\xvec\cn{1,2}{}\yvec$ off $\vvec$, if $\xvec=\yvec\ne\vvec$ or there is a 
    $(\psivec^1,\psivec^2,\mvec)$-open path from $\xvec$ to $\yvec\in\Tbb^\Lambda\setminus\{\vvec\}$.  Let}

    \begin{equation}
      \tilde{\mathcal{C}}^{\bm{v}}(\bm{x}) =
        \Set[\Big]{
          \bm{y}
          \separator
          \bm{x} \cn{1, 2}{} \bm{y}
          \added[id=AS]{\text{ off }} \added[id=AS]{\vvec}
        }.
    \end{equation}
    We say that the \added[id=AS]{mark $\bm{v} \in \mvec$} is pivotal for $\bm{x} \cn{1, 2}{} \bm{y}$,
    if $\tilde{\mathcal{C}}^{\bm{v}}(\bm{x}) \cap \tilde{\mathcal{C}}^{\bm{v}}(\bm{y}) = \varnothing$.
  \end{enumerate}
\end{definition}
\end{shaded}
\todo[inline]{The fourth item was added by Kamijima.}

\begin{figure}[t]
  \centering
  \begin{subfigure}[t]{0.33\linewidth}
    \centering
    \includegraphics[scale=0.50]{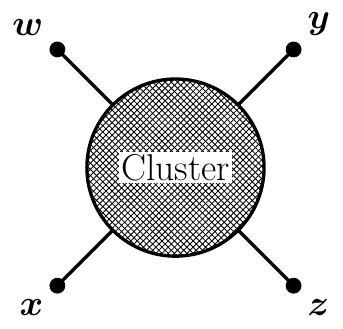}
    \caption{$F_4(\wvec,\xvec,\yvec,\zvec)$.}
    \label{fig:representation_4-point}
  \end{subfigure}%
  \begin{subfigure}[t]{0.66\linewidth}
    \centering
    \begin{tikzpicture}
      \node (plus) {$+$};
      \node[left=5mm of plus] {\includegraphics[scale=0.50]{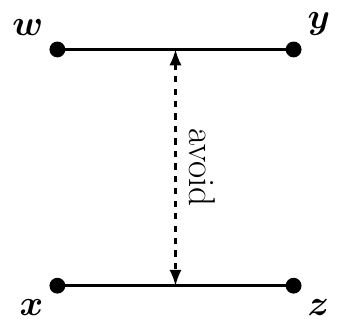}};
      \node[right=5mm of plus] {\includegraphics[scale=0.50]{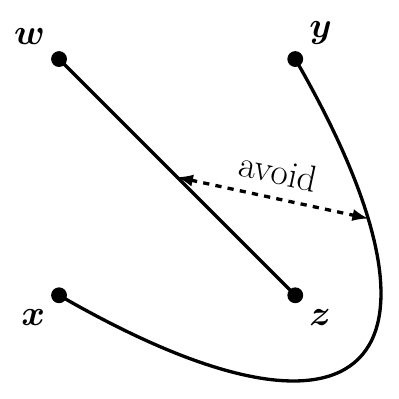}};
    \end{tikzpicture}
    \caption{$\Expect{S^{\sss(3)}_w(t_{\bm{w}}) S^{\sss(3)}_x(t_{\bm{x}})\,; S^{\sss(3)}_y(t_{\bm{y}}) S^{\sss(3)}_z(t_{\bm{z}})}_\Lambda$.}
    \label{fig:representation_truncated_4-point}
  \end{subfigure}
  \begin{subfigure}{0.49\linewidth}
    \centering
    \includegraphics[scale=0.50]{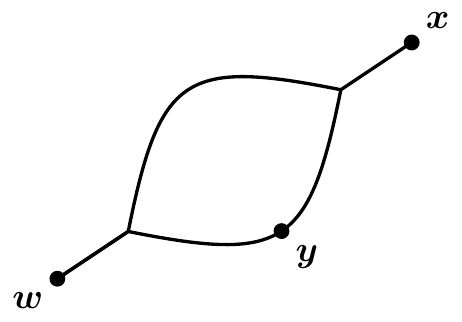}
    \caption{$\bm{y}$ is not pivotal.}
    \label{fig:representation_non-pivotal}
  \end{subfigure}%
  \begin{subfigure}{0.49\linewidth}
    \centering
    \includegraphics[scale=0.50]{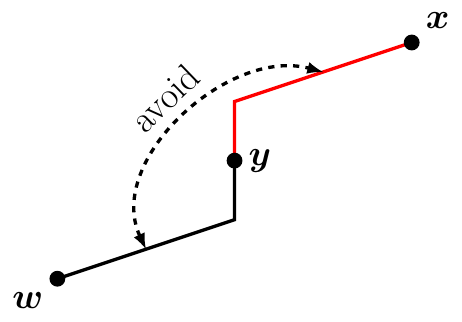}
    \caption{$\bm{y}$ is pivotal.}
    \label{fig:representation_pivotal}
  \end{subfigure}
  \caption{Schematic representations of the correlation functions.  Each dashed arrow labeled ``avoid'' means that the concerned clusters do not intersect.  
  In Figure~\subref{fig:representation_pivotal}, for 
  example, $\tilde{\mathcal{C}}^{\bm{y}}(\bm{w})$ (in black) and $\tilde{\mathcal{C}}^{\bm{y}}(\bm{x})$ (in red) do not intersect.}
  \label{fig:schematic-drawing_correlation}
\end{figure}

By the notion of connectivity (by $(\psivec_1,\psivec_2,\mvec)$-open paths), 
the fourth Ursell function $F_4(\wvec,\xvec,\yvec,\zvec)$ in \Refeq{truncated-z-correlation_bound} 
may be illustrated as a connected component containing all four points, as in Figure~\ref{fig:representation_4-point}.
Similarly, the truncated four-point function $\Expect{S^{\sss(3)}_w(t_{\bm{w}})\,S^{\sss(3)}_x(t_{\bm{x}})\,;S^{\sss(3)}_y(t_{\bm{y}})\,S^{\sss(3)}_z(t_{\bm{z}})}_\Lambda$, 
the three-point function $F_3(\wvec,\xvec,\yvec)$ and the cross-correlation function 
$\Expect{S^{\sss(3)}_w(t_{\bm{w}})\,S^{\sss(3)}_x(t_{\bm{x}})\,;S^{\sss(1)}_y(t_{\bm{y}})}_\Lambda$ in 
\Refeq{truncated-z-correlation_bound}--\Refeq{truncated-cross-correlation_bound} may be illustrated as in 
Figures~\ref{fig:representation_truncated_4-point}, \ref{fig:representation_non-pivotal} and \ref{fig:representation_pivotal}, respectively.

To explain the inequality \Refeq{truncated-z-correlation_bound}, we first 
rewrite $F_4(\wvec,\xvec,\yvec,\zvec)$.  By the stochastic-geometric 
representation and repeated use of the source switching, we obtain
\begin{align}
F_4(\wvec,\xvec,\yvec,\zvec)
&=\frac1Z\Ebb\bigg[\sum_{\partial\psivec=\wvec\vtri\xvec\vtri\yvec\vtri\zvec}1
 \bigg]-\frac1{Z^2}\Ebb^1\Ebb^2\Bigg[\sum_{\substack{\partial\psivec_1
 =\wvec\vtri\xvec\\ \partial\psivec_2=\yvec\vtri\zvec}}1\Bigg]
 -G(\wvec,\yvec)\,G(\xvec,\zvec)-G(\wvec,\zvec)\,G(\xvec,\yvec)\nn\\
&=\frac1{Z^2}\Ebb^1\Ebb^2\Bigg[\sum_{\substack{\partial\psivec_1=\wvec\vtri
 \xvec\vtri\yvec\vtri\zvec\\ \partial\psivec_2=\vno}}\ind{\yvec\ncn{1,2}{}\zvec}
 \Bigg]-G(\wvec,\yvec)\,G(\xvec,\zvec)-G(\wvec,\zvec)\,G(\xvec,\yvec)\nn\\
&=\frac1{Z^2}\Ebb^1\Ebb^2\Bigg[\sum_{\substack{\partial\psivec_1=\xvec\vtri
 \zvec\\ \partial\psivec_2=\wvec\vtri\yvec}}\ind{\yvec\ncn{1,2}{}\zvec}
 \Bigg]-G(\wvec,\yvec)\,G(\xvec,\zvec)\nn\\
&\quad+\frac1{Z^2}\Ebb^1\Ebb^2\Bigg[\sum_{\substack{\partial\psivec_1=\wvec
 \vtri\zvec\\ \partial\psivec_2=\xvec\vtri\yvec}}\ind{\yvec\ncn{1,2}{}\zvec}
 \Bigg]-G(\wvec,\zvec)\,G(\xvec,\yvec).
\end{align}
By the so-called conditioning-on-clusters argument that is heavily used in 
Section~\ref{s:LE}, we can rewrite the expectations as
\begin{align}
\frac1{Z^2}\Ebb^1\Ebb^2\Bigg[\sum_{\substack{\partial\psivec_1=\xvec\vtri
 \zvec\\ \partial\psivec_2=\wvec\vtri\yvec}}\ind{\yvec\ncn{1,2}{}\zvec}
 \Bigg]&=\frac1{Z^2}\Ebb^1\Ebb^2\Bigg[\sum_{\substack{\partial\psivec_1
 =\xvec\vtri\zvec\\ \partial\psivec_2=\vno}}G_{\Ccal_{1,2}^\compl}(\wvec,
 \yvec)\Bigg],\\
\frac1{Z^2}\Ebb^1\Ebb^2\Bigg[\sum_{\substack{\partial\psivec_1=\wvec
 \vtri\zvec\\ \partial\psivec_2=\xvec\vtri\yvec}}\ind{\yvec\ncn{1,2}{}\zvec}
 \Bigg]&=\frac1{Z^2}\Ebb^1\Ebb^2\Bigg[\sum_{\substack{\partial\psivec_1
 =\wvec\vtri\zvec\\ \partial\psivec_2=\vno}}G_{\Ccal_{1,2}^\compl}(\xvec,
 \yvec)\Bigg],
\end{align}
where $\Ccal_{1,2}\equiv\Ccal(\zvec)$ is random against the outer 
expectation, but deterministic against the inner expectation.  
Since (\added[id=AS]{see} \Refeq{2pt-difference} and \Refeq{2pt-diff-par6})
\begin{align}
G(\wvec,\yvec)-G_{\Ccal_{1,2}^\compl}(\wvec,\yvec)=\frac1{Z
 Z_{\Ccal_{1,2}^\compl}}\Ebb^3\Ebb^4_{\Ccal_{1,2}^\compl}
 \Bigg[\sum_{\substack{\partial\psivec^3=\wvec\vtri\yvec\\
 \partial\psivec^4=\vno}}\ind{\wvec\cn{3,4}{\Ccal_{1,2}}\yvec}\Bigg],
\end{align}
we obtain the rewrite
 \begin{align}\lbeq{F4-rewr}
   F_4(\bm{w}, \bm{x}, \bm{y}, \bm{z})
   = {} \MoveEqLeft[0]
     -\frac{1}{Z^2}\Ebb^1\Ebb^2\bigggl[
       \sum_{
         \substack{
           \partial\bm{\psi}^1 = \bm{x} \vtri \bm{z} \\
           \partial\bm{\psi}^2 = \varnothing
         }
       }
       \frac{1}{Z Z_{\mathcal{C}_{1, 2}^\compl}}
       \Ebb^3\Ebb^4_{\mathcal{C}_{1, 2}^\compl} \Biggl[
         \sum_{
           \substack{
             \partial\bm{\psi}^3 = \bm{w} \vtri \bm{y} \\
             \partial\bm{\psi}^4 = \varnothing
           }
         }
         \ind{
           \bm{w} \cn{3, 4}{\Ccal_{1, 2}} \bm{y}
         }
       \Biggr]
     \bigggr] \notag \\
     &-
       \frac{1}{Z^2}
       \Ebb^1 \Ebb^2 \bigggl[
         \sum_{
           \substack{
             \partial\bm{\psi}^1 = \bm{w} \vtri \bm{z} \\
             \partial\bm{\psi}^2 = \varnothing
           }
         }
         \frac{1}{Z Z_{\mathcal{C}_{1, 2}^\compl}}
         \Ebb^3 \Ebb^4_{\mathcal{C}_{1, 2}^\compl} \Biggl[
           \sum_{
             \substack{
               \partial\bm{\psi}^3 = \bm{x} \vtri \bm{y} \\
               \partial\bm{\psi}^4 = \varnothing
             }
           }
           \ind{
             \bm{x} \cn{3, 4}{\Ccal_{1, 2}} \bm{y}
           }
         \Biggr]
       \bigggr]
 \end{align}

\begin{figure}[t]
  \centering
  \begin{subfigure}{0.33\linewidth}
    \centering
    \includegraphics[scale=0.50]{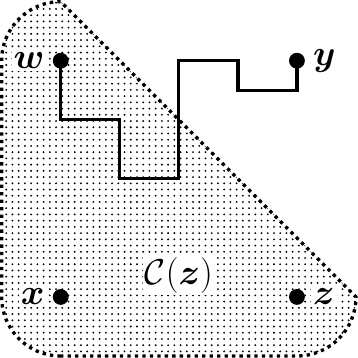}
    \caption{$\bm{w} \in \mathcal{C}(\bm{z})$.}
    \label{fig:split-event_four-point-correlation_belonging-to-cluster}
  \end{subfigure}%
  \begin{subfigure}{0.33\linewidth}
    \centering
    \includegraphics[scale=0.50]{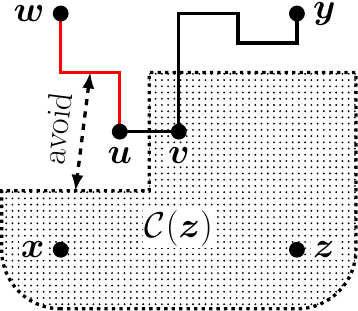}
    \caption{The bridge $(\bm{u}, \bm{v})$ is pivotal.}
    \label{fig:split-event_four-point-correlation_spatial-pivotal}
  \end{subfigure}%
  \begin{subfigure}{0.33\linewidth}
    \centering
    \includegraphics[scale=0.50]{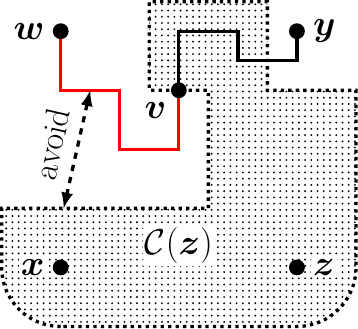}
    \caption{The mark $\bm{v}$ is pivotal.}
    \label{fig:split-event_four-point-correlation_temporal-pivotal}
  \end{subfigure}
  \caption{Decomposition of the event $\{\wvec\cn{3,4}{\Ccal(\zvec)}\yvec\}$, 
  assuming $\Ccal(\zvec)$ is a fixed set.  
  In Figure~\subref{fig:split-event_four-point-correlation_spatial-pivotal}, $\tilde{\Ccal}^{\{\bm{u},\bm{v}\}}(\bm{w})$ (in red) and 
  $\Ccal(\zvec)$ do not intersect.  In Figure~\subref{fig:split-event_four-point-correlation_temporal-pivotal}, $\tilde{\mathcal{C}}^{\bm{v}}(\bm{w})$ (in red) and $\mathcal{C}(\bm{z})$ do not intersect.}
  \label{fig:split-event_four-point-correlation}
\end{figure}

Take the first term on the right, for example, where all 
$(\psivec^3,\psivec^4,\mvec')$-open paths from $\wvec$ to $\yvec$ 
must go through $\Ccal_{1,2}$.  This is realized in three disjoint 
cases: (a)~$\wvec\in\Ccal_{1,2}$ (\added[id=AS]{see} 
Figure~\ref{fig:split-event_four-point-correlation_belonging-to-cluster}), 
(b)~$\exists(\uvec,\vvec)\in\piv\{\wvec\cn{3,4}{}\yvec\}$ such that 
$\tilde{\Ccal}_{3,4}^{\{\uvec,\vvec\}}(\wvec)\cap\Ccal_{1,2}=\vno$
(\added[id=AS]{see} Figure~\ref{fig:split-event_four-point-correlation_spatial-pivotal}) 
and (c)~$\exists$a pivotal mark $\vvec$ such that 
$\tilde{\Ccal}_{3,4}^{\sss\vvec}(\wvec)\cap\Ccal_{1,2}=\vno$ (\added[id=AS]{see} 
Figure~\ref{fig:split-event_four-point-correlation_temporal-pivotal}). 
The case~(a) corresponds to the first term on the right-hand side of 
\eqref{eq:truncated-z-correlation_bound}, while the case~(b) and the 
case~(c) correspond to (parts of) the third and fourth terms on the 
right-hand side of \eqref{eq:truncated-z-correlation_bound}, respectively.

\begin{figure}[t]
  \centering
  \begin{subfigure}{0.33\linewidth}
    \centering
    \includegraphics[scale=0.50]{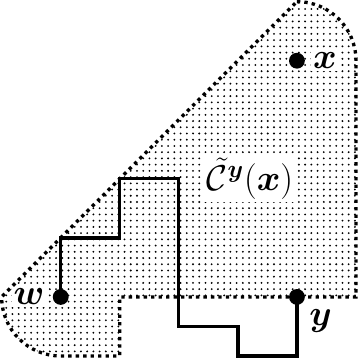}
    \caption{$\bm{w} \in \tilde{\mathcal{C}}^{\bm{y}}(\bm{x})$.}
    \label{fig:split-event_non-pivotal_belonging-to-cluster}
  \end{subfigure}%
  \begin{subfigure}{0.33\linewidth}
    \centering
    \includegraphics[scale=0.50]{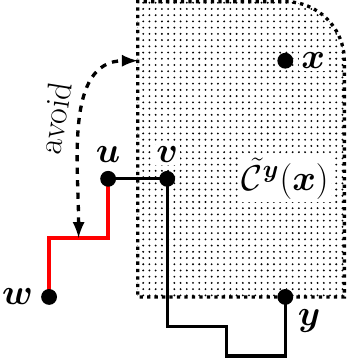}
    \caption{The bridge $(\bm{u}, \bm{v})$ is pivotal.}
    \label{fig:split-event_non-pivotal_spatial-pivotal}
  \end{subfigure}%
  \begin{subfigure}{0.33\linewidth}
    \centering
    \includegraphics[scale=0.50]{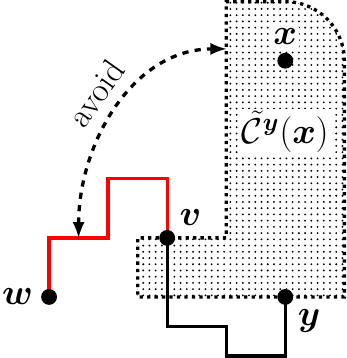}
    \caption{The mark $\bm{v}$ is pivotal.}
    \label{fig:split-event_non-pivotal_temporal-pivotal}
  \end{subfigure}
  \caption{Decomposition of the event $\{\wvec\cn{3,4}{\tilde{\Ccal}^{\yvec}(\xvec)}\yvec\}$, assuming $\tilde{\Ccal}^{\yvec}(\xvec)$ is a fixed set. 
  In Figure~\subref{fig:split-event_non-pivotal_spatial-pivotal}, $\tilde{\mathcal{C}}^{\{\bm{u}, \bm{v}\}}(\bm{w})$ (in red) and $\tilde{\mathcal{C}}^{\bm{y}}(\bm{x})$ do not intersect.  In Figure~\subref{fig:split-event_non-pivotal_temporal-pivotal}, $\tilde{\mathcal{C}}^{\bm{v}}(\bm{w})$ (in red) and $\tilde{\mathcal{C}}^{\bm{y}}(\bm{x})$ do not intersect.}
  \label{fig:split-event_non-pivotal}
\end{figure}

We can follow the same strategy to obtain the 
inequality \Refeq{truncated-cross-correlation_bound}.  
Similarly to the above rewrite for $F_4(\wvec,\xvec,\yvec,\zvec)$, 
we can also obtain the rewrite 
 \begin{align}
   F_3(\bm{w}, \bm{x}, \bm{y})
   = {} \MoveEqLeft[0]
     \frac{1}{Z^2}\Ebb^1 \Ebb^2 \bigggl[
       \sum_{
         \substack{
           \partial\bm{\psi}^1 = \bm{y} \vtri \bm{x} \\
           \partial\bm{\psi}^2 = \varnothing
         }
       }
       \frac{1}{Z Z_{\Ccal_{1, 2}^\compl}}
       \Ebb^3 \Ebb^4_{\Ccal_{1, 2}^\compl} \Biggl[
         \sum_{
           \substack{
             \partial\bm{\psi}^3 = \bm{w} \vtri \bm{y} \\
             \partial\bm{\psi}^4 = \varnothing
           }
         }
         \ind{
           \bm{w} \cn{3, 4}{\Ccal_{1, 2}} \bm{y}
         }
       \Biggr]
     \bigggr] \notag \\
     &+
     \frac{1}{Z^2}\Ebb^1 \Ebb^2 \bigggl[
       \sum_{
         \substack{
           \partial\bm{\psi}^1 = \bm{y} \vtri \bm{w} \\
           \partial\bm{\psi}^2 = \varnothing
         }
       }
       \frac{1}{Z Z_{\Ccal_{1, 2}^{\prime\compl}}}
       \Ebb^3 \Ebb^4_{\Ccal_{1, 2}^{\prime\compl}} \Biggl[
         \sum_{
           \substack{
             \partial\bm{\psi}^3 = \bm{x} \vtri \bm{y} \\
             \partial\bm{\psi}^4 = \varnothing
           }
         }
         \ind{
           \bm{x} \cn{3, 4}{\Ccal_{1, 2}'} \bm{y}
         }
       \Biggr]
     \bigggr].
 \end{align}
where $\Ccal_{1,2}=\tilde{\Ccal}^{\yvec}(\xvec)$ for this case and $\Ccal_{1,2}'=\tilde{\Ccal}^{\yvec}(\wvec)$.
Then we decompose the event $\{\wvec\cn{3,4}{\Ccal_{1,2}}\yvec\}$ 
into three disjoint cases, as depicted in 
Figure~\ref{fig:split-event_non-pivotal}.

\section{New lace expansion for the classical Ising model}\label{s:LE}
As another example of application of the stochastic-geometric representation 
and the source switching explained in Section~\ref{s:SGRSW}, 
we will \added[id=AS]{prove Theorem~\ref{thm:LEintro} (see also} Theorem~\ref{thm:LE} below), which is a new lace expansion for the classical Ising model \added[id=AS]{(i.e., $q=0$).  
From the perspective of difficulty, there is not much difference between the new expansion and the previous one in~\cite{s2007lace,s2022correct} in deriving the expansion and bounding the expansion coefficients.
However, the new expansion has advantages in its extension to the quantum case $q>0$, which is far more involved due to the introduction of pivotal vertices (see Definition~\ref{def:piv}(iii) for  pivotal bridges). 
It} will be reported separately in the \added[id=AS]{forthcoming paper}~\cite{ks2025lace}.

The lace expansion is one of the few methods to rigorously prove critical 
behavior in high dimensions for various models, such as self-avoiding walk 
\cite{bs1985selfavoiding,hs1992selfavoiding}, lattice trees and lattice animals \cite{hs1990upper}, 
percolation \cite{hs1990meanfield}, the classical Ising model \cite{s2007lace,s2022correct} and 
the lattice $\varphi^4$ model \cite{s2015application,bhh2021continuoustime}.  David Brydges, who established 
the methodology of the lace expansion for the first time in 1985 with Thomas 
Spencer, was awarded the Henri Poincar\'e Prize in 2024.  One of the 
implications of the lace expansion is an infrared bound on the two-point 
function, which is uniform in the subcritical regime, without assuming 
reflection positivity.  In the forthcoming paper~\cite{ks2025lace}, we will show an infrared 
bound for the quantum Ising model without assuming reflection positivity, 
which was assumed in the previous section to prove mean-field divergence 
of the susceptibility.

From now on, we fix $q=0$.  Let us roughly explain how to derive an infrared bound \added[id=AS]{from the lace expansion \Refeq{LEintro} for the classical Ising model:} 
\begin{align}\lbeq{LE1stshow}
G(\ovec,\xvec)=\pi^{\sss(\le j)}(\ovec,\xvec)+\int_{\Tbb}\diff t\sum_{\substack{
 (\yvec,\zvec):\\ t_{\yvec}=t_{\zvec}=t}}\beta J_{y,z}\,\pi^{\sss(\le j)}(\ovec,
 \yvec)\,G(\zvec,\xvec)+(-1)^{j+1}R^{\sss(j+1)}(\ovec,\xvec).
\end{align}
Let $\hat G(\omega,k)$ be the Fourier transform of $G(\ovec,\xvec)$: 
for $\omega\in2\pi\Z$ and \added[id=AS]{$k\in\frac\pi{L}\Lambda$},
\begin{align}
\hat G(\omega,k)=\int_{\Tbb}\diff t\sum_{x\in\Lambda}e^{\im\omega t+\im k\cdot
 x}G\big(\ovec,(t,x)\big).
\end{align}
Similarly define $\hat\pi^{\sss(\le j)}(\omega,k)$ and $\hat R^{\sss(j)}(\omega,k)$.  
\added[id=AS]{Suppose that the limit $\hat\pi(\omega,k)=\lim_{j\uparrow\infty}\hat\pi^{\sss(\le j)}(\omega,k)$ exists 
and $\lim_{j\uparrow\infty}\hat R^{\sss(j)}(\omega,k)=0$ for $\beta<\betac$}, which can be verified eventually in sufficiently high dimensions or for sufficiently spread-out models with $d>4$.  
Then we \added[id=AS]{formally} obtain
\begin{align}
\hat G(\omega,k)=\hat\pi(\omega,k)+\beta\hat J(k)\,\hat\pi(\omega,k)\,\hat
 G(\omega,k)=\frac{\hat\pi(\omega,k)}{1-\beta\hat J(k)\,\hat\pi(\omega,k)},
\end{align}
where \added[id=AS]{we recall} $\hat J(k)=\sum_xe^{\im k\cdot x}J_{o,x}$.  
\added[id=AS]{Since $\chi_\Lambda=\hat G(0,0)$, we can rewrite the above as}
\begin{align}
&\hat G(\omega,k)=\frac1{\hat\pi(\omega,k)^{-1}-\beta\hat J(k)}
 =\frac1{\added[id=AS]{\chi_\Lambda^{-1}}-\hat\pi(0,0)^{-1}+\beta \hat J(0)+\hat\pi(\omega,k)^{-1}
 -\beta\hat J(k)}\nn\\
&=\Bigg(\added[id=AS]{\chi_\Lambda^{-1}}-\hat\pi(0,0)^{-1}+\hat\pi(\omega,0)^{-1}+\bigg(\beta
 -\frac{\hat\pi(\omega,0)^{-1}-\hat\pi(\omega,k)^{-1}}{\hat J(0)-\hat J(k)}
 \bigg)\big(\hat J(0)-\hat J(k)\big)\Bigg)^{-1},
\end{align}
which may imply an infrared bound (\added[id=AS]{compare this with} \Refeq{infrared-bound})
\begin{align}\lbeq{IRbd}
|\hat G(\omega,k)|\lesssim\Bigg(\frac{-\partial_\omega^2\hat\pi(0,0)^{-1}}2
 \omega^2+\bigg(\beta-\frac{\Delta\hat\pi(0,0)^{-1}}{\Delta\hat J(0)}\bigg)
 \big(\hat J(0)-\hat J(k)\big)\Bigg)^{-1},
\end{align}
assuming existence of \added[id=AS]{those derivatives, which can also be verified in sufficiently high dimensions or for sufficiently spread-out models with $d>4$}.  
Notice that the above argument does 
not require reflection positivity.  Letting $\omega=0$, in particular, yields 
an infrared bound on the classical Ising 
two-point function. 

Next we derive the lace expansion \Refeq{LE1stshow} for $q=0$.

\subsection{Derivation of the expansion}\label{ss:derivation}
\paragraph{The 1st stage:}
By \Refeq{crawford-ioffe2} and \Refeq{2pt-diff-par5} 
with $\Ccal=\vno$, we have 
\begin{align}
G(\ovec,\xvec)=\frac1{Z^2}\Ebb^1\Ebb^2\Bigg[\sum_{\substack{\partial\psivec^1
 =\ovec\vtri\xvec\\ \partial\psivec^2=\vno}}\ind{\ovec\cn{1,2}{}\xvec}\Bigg].
\end{align}
The indicator $\ind{\ovec\cn{1,2}{}\xvec}$ can be split into two, depending 
on whether $\ovec$ is doubly connected to $\xvec$ or is not:
\begin{align}\lbeq{G-dec1}
G(\ovec,\xvec)=\pi^{\sss(0)}(\ovec,\xvec)+\frac1{Z^2}\Ebb^1\Ebb^2\Bigg[
 \sum_{\substack{\partial\psivec^1=\ovec\vtri\xvec\\ \partial\psivec^2=\vno}}
 \ind{\ovec\cn{1,2}{}\xvec\}\setminus\{\ovec\db{1,2}{}\xvec}\Bigg],
\end{align}
where
\begin{align}\lbeq{pi0-def}
\pi^{\sss(0)}(\ovec,\xvec)=\frac1{Z^2}\Ebb^1\Ebb^2\Bigg[\sum_{\substack{\partial
 \psivec^1=\ovec\vtri\xvec\\ \partial\psivec^2=\vno}}\ind{\ovec\db{1,2}{}
 \xvec}\Bigg].
\end{align}

On the event $\{\ovec\cn{1,2}{}\xvec\}\setminus\{\ovec\db{1,2}{}\xvec\}$, 
there is at least one pivotal bridge for $\ovec\cn{1,2}{}\xvec$ \added[id=AS]{(recall Definition~\ref{def:piv}; there are no pivotal marks when $\TField=0$)}.  
Taking the first among those pivotal bridges yields the decomposition
\begin{align}\lbeq{piv-dec}
\ind{\ovec\cn{1,2}{}\xvec\}\setminus\{\ovec\db{1,2}{}\xvec}=\sum_{(\uvec,
 \vvec):\{\uvec,\vvec\}\in\xivec}\ind{\ovec\db{1,2}{}\uvec\text{ off }\{\uvec,
 \vvec\}}\,\ind{\vvec\cn{1,2}{}\xvec\text{ in }\Ccal_{1,2}^\compl},
\end{align}
where $\xivec=\xivec^1\cup\xivec^2$ and 
$\Ccal_{1,2}=\tilde\Ccal^{\{\uvec,\vvec\}}(\ovec)$.  Since 
$\partial\psivec^1=\ovec\vtri\xvec$ and $\partial\psivec^2=\vno$, we can 
replace the sum over $\{\uvec,\vvec\}\in\xivec$ with that over 
$\{\uvec,\vvec\}\in\xivec^1$.  
By the Mecke equation\footnote{In this paper, we need the following form of the 
Mecke equation~\cite{lp2017Lectures}: let $\xi$ be the $\Tbb$-valued Poisson 
point process with intensity $\lambda$, whose expectation is denoted by 
$\Ebb_\lambda$, and let $f$ be a measurable 
function of $\xi$ and $t\in\xi$.  Then we have
\begin{align}
\Ebb_\lambda\bigg[\sum_{t\in\xi}f(\xi,t)\bigg]=\lambda\int_{\Tbb}\Ebb_\lambda
 [f(\xi^t,t)]\,\diff t,
\end{align}
where $\xi^t$ be the Poisson point process \added[id=AS]{augmented by} $t\in\Tbb$.} 
for Poisson point processes, we obtain the rewrite
\begin{align}\lbeq{mecke1}
&\Ebb^1\Ebb^2\Bigg[\sum_{\substack{\partial\psivec^1=\ovec\vtri\xvec\\ \partial
 \psivec^2=\vno}}\sum_{\substack{(\uvec,\vvec):\\ \{\uvec,\vvec\}\in\xivec^1}}
 \ind{\ovec\db{1,2}{}\uvec\text{ off }\{\uvec,\vvec\}}\,\ind{\vvec\cn{1,2}{}
 \xvec\text{ in }\Ccal_{1,2}^\compl}\Bigg]\nn\\
&=\int_{\Tbb}\diff t\sum_{\substack{(\uvec,\vvec):\\ t_{\uvec}=t_{\vvec}=t}}
 \beta J_{u,v}\,\Ebb^1\Ebb^2\Bigg[\sum_{\substack{\partial\psivec^1=\ovec\vtri
 \xvec\vtri\{\uvec,\vvec\}\\ \partial\psivec^2=\vno}}\ind{\ovec\db{1,2}{}\uvec
 \text{ off }\{\uvec,\vvec\}}\,\ind{\vvec\cn{1,2}{}\xvec\text{ in }
 \Ccal_{1,2}^\compl}\Bigg].
\end{align}
Conditioning on the cluster $\Ccal_{1,2}$ and splitting each $\psivec^j$ into 
$\psivec^{j'}\equiv\psivec^j|_{\Ccal_{1,2}}$ and 
$\psivec^{j''}\equiv\psivec^j|_{\Ccal_{1,2}^\compl}$, we can further rewrite 
the expectation on the right-hand side as 
\begin{align}\lbeq{nested}
&\Ebb^1\Ebb^2\Bigg[\sum_{\substack{\partial\psivec^{1'}=\ovec\vtri\uvec\\
 \partial\psivec^{2'}=\vno}}\ind{\ovec\db{1',2'}{}\uvec\text{ off }\{\uvec,
 \vvec\}}\sum_{\substack{\partial\psivec^{1''}=\vvec\vtri\xvec\\ \partial
 \psivec^{2''}=\vno}}\ind{\vvec\cn{1'',2''}{}\xvec\text{ in }\Ccal_{1,2}^\compl}
 \Bigg]\nn\\
&=\Ebb^{1'}\Ebb^{2'}\Bigg[\sum_{\substack{\partial\psivec^{1'}=\ovec\vtri\uvec\\
 \partial\psivec^{2'}=\vno}}\ind{\ovec\db{1',2'}{}\uvec\text{ off }\{\uvec,
 \vvec\}}~\Ebb_{\Ccal_{1',2'}^\compl}^{1''}\Ebb_{\Ccal_{1',2'}^\compl}^{2''}
 \Bigg[\sum_{\substack{\partial\psivec^{1''}=\vvec\vtri\xvec\\ \partial
 \psivec^{2''}=\vno}}\ind{\vvec\cn{1'',2''}{}\xvec}\Bigg]\Bigg],
\end{align}
where $\Ccal_{1',2'}$ is random against the outer expectation 
$\Ebb^{1'}\Ebb^{2'}$.  Notice that $\ind{\vvec\cn{1'',2''}{}\xvec}=1$ under 
the source constraint $\partial\psivec^{1''}=\vvec\vtri\xvec$.  Therefore,
\begin{align}
\Ebb_{\Ccal_{1',2'}^\compl}^{1''}\Ebb_{\Ccal_{1',2'}^\compl}^{2''}\Bigg[
 \sum_{\substack{\partial\psivec^{1''}=\vvec\vtri\xvec\\ \partial\psivec^{2''}
 =\vno}}\ind{\vvec\cn{1'',2''}{}\xvec}\Bigg]=Z_{\Ccal_{1',2'}^\compl}\,
 \Ebb_{\Ccal_{1',2'}^\compl}^{1''}\Bigg[\sum_{\partial\psivec^{1''}=\vvec\vtri
 \xvec}1\Bigg]=Z_{\Ccal_{1',2'}^\compl}^2\,G_{\Ccal_{1',2'}^\compl}(\vvec,
 \xvec)\nn\\
=\Ebb_{\Ccal_{1',2'}^\compl}^{1''}\Ebb_{\Ccal_{1',2'}^\compl}^{2''}\Bigg[
 \sum_{\partial\psivec^{1''}=\partial\psivec^{2''}=\vno}1\Bigg]\,
 G_{\Ccal_{1',2'}^\compl}(\vvec,\xvec).
\end{align}
Substituting this back into \Refeq{nested}, we arrive at
\begin{align}\lbeq{inclexcl1}
&\frac1{Z^2}\Ebb^1\Ebb^2\Bigg[\sum_{\substack{\partial\psivec^1=\ovec\vtri\xvec
 \vtri\{\uvec,\vvec\}\\ \partial\psivec^2=\vno}}\ind{\ovec\db{1,2}{}\uvec
 \text{ off }\{\uvec,\vvec\}}\,\ind{\vvec\cn{1,2}{}\xvec\text{ in }\Ccal_{1,
 2}^\compl}\Bigg]\nn\\
&=\frac1{Z^2}\Ebb^1\Ebb^2\Bigg[\sum_{\substack{\partial\psivec^1=\ovec\vtri
 \uvec\\ \partial\psivec^2=\vno}}\ind{\ovec\db{1,2}{}\uvec}\,G_{\Ccal_{1,
 2}^\compl}(\vvec,\xvec)\Bigg]\nn\\
&=\pi^{\sss(0)}(\ovec,\uvec)\,G(\vvec,\xvec)-\frac1{Z^2}\Ebb^1\Ebb^2\Bigg[
 \sum_{\substack{\partial\psivec^1=\ovec\vtri\uvec\\ \partial\psivec^2=\vno}}
 \ind{\ovec\db{1,2}{}\uvec}\,\Big(G(\vvec,\xvec)-G_{\Ccal_{1,2}^\compl}(\vvec,\xvec)\Big)\Bigg],
\end{align}
where, in the second line, we have dropped ``off $\{\uvec,\vvec\}$'' since 
$G_{\Ccal_{1,2}^\compl}(\vvec,\xvec)=0$ on the event 
$\{\vvec\in\Ccal_{1,2}\}\supset\{\ovec\db{1,2}{}
\uvec\}\setminus\{\ovec\db{1,2}{}\uvec\text{ off }\{\uvec,\vvec\}\}$.

Summarizing the above, we obtain
\begin{align}
G(\ovec,\xvec)&=\pi^{\sss(0)}(\ovec,\xvec)+\int_{\Tbb}\diff t\sum_{\substack{
 (\uvec,\vvec):\\ t_{\uvec}=t_{\vvec}=t}}\beta J_{u,v}\,\pi^{\sss(0)}(\ovec,
 \uvec)\,G(\vvec,\xvec)-R^{\sss(1)}(\ovec,\xvec),
\end{align}
where 
\begin{align}\lbeq{R1def}
R^{\sss(1)}(\ovec,\xvec)=\int_{\Tbb}\diff t\sum_{\substack{(\uvec,\vvec):\\
 t_{\uvec}=t_{\vvec}=t}}\frac{\beta J_{u,v}}{Z^2}~\Ebb^1\Ebb^2\Bigg[
 \sum_{\substack{\partial\psivec^1=\ovec\vtri\uvec\\ \partial\psivec^2=\vno}}
 \ind{\ovec\db{1,2}{}\uvec}\,\Big(G(\vvec,\xvec)-G_{\Ccal_{1,2}^\compl}(\vvec,
 \xvec)\Big)\Bigg].
\end{align}
This completes the first stage of the expansion.

\paragraph{The 2nd stage:}
Next we expand the remainder $R^{\sss(1)}(\ovec,\xvec)$.  
By \Refeq{2pt-difference} and \Refeq{2pt-diff-par6}, the expectation in 
\Refeq{R1def} can be written as 
\begin{align}
\Ebb^1\Ebb^2\Bigg[\sum_{\substack{\partial\psivec^1=\ovec\vtri\uvec\\ \partial
 \psivec^2=\vno}}\ind{\ovec\db{1,2}{}\uvec}\,\frac1{ZZ_{\Ccal_{1,2}^\compl}}
 \Ebb^3\Ebb_{\Ccal_{1,2}^\compl}^4\Bigg[\sum_{\substack{\partial\psivec^3=\vvec
 \vtri\xvec\\ \partial\psivec^4=\vno}}\ind{\vvec\cn{3,4}{\Ccal_{1,2}}\xvec}
 \Bigg]\Bigg].
\end{align}
We split the indicator 
$\ind{\vvec\cn{3,4}{\Ccal_{1,2}}\xvec}$ into two, depending on whether the 
event $E_{3,4}(\vvec,\xvec;\Ccal_{1,2})$ occurs or does not, where 
\begin{align}
E_{3,4}(\vvec,\xvec;\Ccal_{1,2})=\big\{\vvec\cn{3,4}{\Ccal_{1,2}}\xvec\big\}
 \setminus\Big\{\exists(\yvec,\zvec)\in\piv\big\{\vvec\ocn{3,4}{}\xvec\big\}
 \text{ s.t. }\vvec\cn{3,4}{\Ccal_{1,2}}\yvec\Big\}.
\end{align}
We define the contribution to $R^{\sss(1)}(\ovec,\xvec)$ from 
$E_{3,4}(\vvec,\xvec;\Ccal_{1,2})$ as $\pi^{\sss(1)}(\ovec,\xvec)$:
\begin{align}\lbeq{pi1def}
&\pi^{\sss(1)}(\ovec,\xvec)\nn\\
&=\int_{\Tbb}\diff t\sum_{\substack{(\uvec,\vvec):\\
 t_{\uvec}=t_{\vvec}=t}}\frac{\beta J_{u,v}}{Z^2}~\Ebb^1\Ebb^2\Bigg[
 \sum_{\substack{\partial\psivec^1=\ovec\vtri\uvec\\ \partial\psivec^2=\vno}}
 \ind{\ovec\db{1,2}{}\uvec}\,\frac1{ZZ_{\Ccal_{1,2}^\compl}}\Ebb^3
 \Ebb_{\Ccal_{1,2}^\compl}^4\Bigg[\sum_{\substack{\partial\psivec^3=\vvec
 \vtri\xvec\\ \partial\psivec^4=\vno}}\indic{E_{3,4}(\vvec,\xvec;\Ccal_{1,2})}
 \Bigg]\Bigg].
\end{align}

On the event $\{\vvec\cn{3,4}{\Ccal_{1,2}}\xvec\}\setminus E_{3,4}(\vvec,
\xvec;\Ccal_{1,2})$, which equals 
\begin{align}
\big\{\vvec\cn{3,4}{\Ccal_{1,2}}\xvec\big\}\cap\Big\{\exists(\yvec,\zvec)\in
 \piv\big\{\vvec\ocn{3,4}{}\xvec\big\}\text{ s.t. }\vvec\cn{3,4}{\Ccal_{1,
 2}}\yvec\Big\},
\end{align}
we take the first bridge $(\yvec,\zvec)\in\piv\{\vvec\ocn{3,4}{}\xvec\}$ that 
satisfies $\vvec\cn{3,4}{\Ccal_{1, 2}}\yvec$.  Then we obtain the 
decomposition
\begin{align}
\indic{\{\vvec\cn{3,4}{\Ccal_{1,2}}\xvec\}\setminus E_{3,4}(\vvec,\xvec;
 \Ccal_{1,2})}=\sum_{(\yvec,\zvec):\{\yvec,\zvec\}\in\xivec}\ind{E_{3,4}
 (\vvec,\yvec;\Ccal_{1,2})\text{ off }\{\yvec,\zvec\}}\,\ind{\zvec\cn{3,4}
 {}\xvec\text{ in }\Ccal_{3,4}^\compl},
\end{align}
where $\xivec=\xivec^3\cup\xivec^4$ and 
$\Ccal_{3,4}=\tilde\Ccal^{\{\yvec,\zvec\}}(\vvec)$.  Since 
$\partial\psivec^3=\vvec\vtri\xvec$ and $\partial\psivec^4=\vno$, 
the sum over $\{\yvec,\zvec\}\in\xivec$ can be replaced by that over 
$\{\yvec,\zvec\}\in\xivec^3$.  By the Mecke equation again, we obtain
\begin{align}\lbeq{Mecke-appl2}
&\Ebb^3\Ebb_{\Ccal_{1,2}^\compl}^4\Bigg[\sum_{\substack{\partial\psivec^3=\vvec
 \vtri\xvec\\ \partial\psivec^4=\vno}}\indic{\{\vvec\cn{3,4}{\Ccal_{1,2}}\xvec\}
 \setminus E_{3,4}(\vvec,\xvec;\Ccal_{1,2})}\Bigg]\nn\\
&=\Ebb^3\Ebb_{\Ccal_{1,2}^\compl}^4\Bigg[\sum_{\substack{\partial\psivec^3=\vvec
 \vtri\xvec\\ \partial\psivec^4=\vno}}\sum_{(\yvec,\zvec):\{\yvec,\zvec\}\in
 \xivec^3}\ind{E_{3,4}(\vvec,\yvec;\Ccal_{1,2})\text{ off }\{\yvec,\zvec\}}\,
 \ind{\zvec\cn{3,4}{}\xvec\text{ in }\Ccal_{3,4}^\compl}\Bigg]\nn\\
&=\int_{\Tbb}\diff s\sum_{\substack{(\yvec,\zvec):\\ t_{\yvec}=t_{\zvec}=s}}
 \beta J_{y,z}\,\Ebb^3\Ebb_{\Ccal_{1,2}^\compl}^4\Bigg[\sum_{\substack{\partial
 \psivec^3=\vvec\vtri\xvec\vtri\{\yvec,\zvec\}\\ \partial\psivec^4=\vno}}
 \ind{E_{3,4}(\vvec,\yvec;\Ccal_{1,2})\text{ off }\{\yvec,\zvec\}}\,\ind{\zvec
 \cn{3,4}{}\xvec\text{ in }\Ccal_{3,4}^\compl}\Bigg].
\end{align}
Then, as done in \Refeq{nested}, we condition on the cluster $\Ccal_{3,4}$ and 
split each $\psivec^j$ into $\psivec^{j'}\equiv\psivec^j|_{\Ccal_{3,4}}$ and 
$\psivec^{j''}\equiv\psivec^j|_{\Ccal_{3,4}^\compl}$, so that we obtain the 
rewrite of the expectation on the right-hand side as 
\begin{align}\lbeq{nested2}
&\Ebb^3\Ebb_{\Ccal_{1,2}^\compl}^4\Bigg[\sum_{\substack{\partial\psivec^{3'}
 =\vvec\vtri\yvec\\ \partial\psivec^{4'}=\vno}}\ind{E_{3',4'}(\vvec,\yvec;
 \Ccal_{1,2})\text{ off }\{\yvec,\zvec\}}\sum_{\substack{\partial\psivec^{3''}
 =\zvec\vtri\xvec\\ \partial\psivec^{4''}=\vno}}\ind{\zvec\cn{3'',4''}{}\xvec
 \text{ in }\Ccal_{3,4}^\compl}\Bigg]\nn\\
&=\Ebb^{3'}\Ebb_{\Ccal_{1,2}^\compl}^{4'}\Bigg[\sum_{\substack{\partial
 \psivec^{3'}=\vvec\vtri\yvec\\ \partial\psivec^{4'}=\vno}}\ind{E_{3',4'}
 (\vvec,\yvec;\Ccal_{1,2})\text{ off }\{\yvec,\zvec\}}~\Ebb_{\Ccal_{3',
 4'}^\compl}^{3''}\Ebb_{\Ccal_{1,2}^\compl\cap\Ccal_{3',4'}^\compl}^{4''}\Bigg[
 \sum_{\substack{\partial\psivec^{3''}=\zvec\vtri\xvec\\ \partial\psivec^{4''}
 =\vno}}\ind{\zvec\cn{3'',4''}{}\xvec}\Bigg]\Bigg],
\end{align}
where $\Ccal_{3',4'}$ is random against the outer expectation 
$\Ebb^{3'}\Ebb_{\Ccal_{1,2}^\compl}^{4'}$.  
Since $\ind{\zvec\cn{3'',4''}{}\xvec}=1$ under 
the source constraint $\partial\psivec^{3''}=\zvec\vtri\xvec$, we obtain
\begin{align}
\Ebb_{\Ccal_{3',4'}^\compl}^{3''}\Ebb_{\Ccal_{1,2}^\compl\cap\Ccal_{3',
 4'}^\compl}^{4''}\Bigg[\sum_{\substack{\partial\psivec^{3''}=\zvec\vtri\xvec\\
 \partial\psivec^{4''}=\vno}}\ind{\zvec\cn{3'',4''}{}\xvec}\Bigg]
 =Z_{\Ccal_{3',4'}^\compl}Z_{\Ccal_{1,2}^\compl\cap\Ccal_{3',4'}^\compl}
 G_{\Ccal_{3',4'}^\compl}(\zvec,\xvec)\nn\\
=\Ebb_{\Ccal_{3',4'}^\compl}^{3''}\Ebb_{\Ccal_{1,2}^\compl\cap\Ccal_{3',
 4'}^\compl}^{4''}\Bigg[\sum_{\partial\psivec^{3''}=\partial\psivec^{4''}
 =\vno}1\Bigg]\,G_{\Ccal_{3',4'}^\compl}(\zvec,\xvec),
\end{align}
so that
\begin{align}\lbeq{inclexcl2}
\Refeq{nested2}&=\Ebb^3\Ebb_{\Ccal_{1,2}^\compl}^4\Bigg[\sum_{\substack{\partial
 \psivec^3=\vvec\vtri\yvec\\ \partial\psivec^4=\vno}}\indic{E_{3,4}(\vvec,\yvec;
 \Ccal_{1,2})}\,G_{\Ccal_{3,4}^\compl}(\zvec,\xvec)\Bigg]\nn\\
&=\Ebb^3\Ebb_{\Ccal_{1,2}^\compl}^4\Bigg[\sum_{\substack{\partial\psivec^3
 =\vvec\vtri\yvec\\ \partial\psivec^4=\vno}}\indic{E_{3,4}(\vvec,\yvec;\Ccal_{1,
 2})}\Bigg]\,G(\zvec,\xvec)\nn\\
&\quad-\Ebb^3\Ebb_{\Ccal_{1,2}^\compl}^4\Bigg[\sum_{\substack{\partial\psivec^3
 =\vvec\vtri\yvec\\ \partial\psivec^4=\vno}}\indic{E_{3,4}(\vvec,\yvec;\Ccal_{1,
 2})}\,\Big(G(\zvec,\xvec)-G_{\Ccal_{3,4}^\compl}(\zvec,\xvec)\Big)\Bigg],
\end{align}
where, in the first line, we have dropped ``off $\{\yvec,\zvec\}$'' since 
$G_{\Ccal_{3,4}^\compl}(\zvec,\xvec)=0$ on the event 
$\{\zvec\in\Ccal_{3,4}\}\supset E_{3,4}(\vvec,\yvec;\Ccal_{1,2})\setminus
\{E_{3,4}(\vvec,\yvec;\Ccal_{1,2})\text{ off }\{\yvec,\zvec\}\}$.  
Substituting this back into \Refeq{Mecke-appl2} and using \Refeq{R1def} and 
\Refeq{pi1def}, we arrive at
\begin{align}
R^{\sss(1)}(\ovec,\xvec)=\pi^{\sss(1)}(\ovec,\xvec)+\int_{\Tbb}\diff s
 \sum_{\substack{(\yvec,\zvec):\\ t_{\yvec}=t_{\zvec}=s}}\beta J_{y,z}\,
 \pi^{\sss(1)}(\ovec,\yvec)\,G(\zvec,\xvec)-R^{\sss(2)}(\ovec,\xvec),
\end{align}
where
\begin{align}\lbeq{R2def}
R^{\sss(2)}(\ovec,\xvec)&=\int_{\Tbb}\diff t\sum_{\substack{(\uvec,\vvec):\\
 t_{\uvec}=t_{\vvec}=t}}\beta J_{u,v}\int_{\Tbb}\diff s\sum_{\substack{(\yvec,
 \zvec):\\ t_{\yvec}=t_{\zvec}=s}}\beta J_{y,z}~\frac1{Z^2}\Ebb^1\Ebb^2\Bigg[
 \sum_{\substack{\partial\psivec^1=\ovec\vtri\uvec\\ \partial\psivec^2=\vno}}
 \ind{\ovec\db{1,2}{}\uvec}\nn\\
&\qquad\times\frac1{ZZ_{\Ccal_{1,2}^\compl}}\Ebb^3\Ebb_{\Ccal_{1,2}^\compl}^4
 \Bigg[\sum_{\substack{\partial\psivec^3=\vvec\vtri\yvec\\ \partial\psivec^4
 =\vno}}\indic{E_{3,4}(\vvec,\yvec;\Ccal_{1,2})}\,\Big(G(\zvec,\xvec)
 -G_{\Ccal_{3,4}^\compl}(\zvec,\xvec)\Big)\Bigg]\Bigg].
\end{align}
Notice that the difference of the two-point function and its restricted version 
shows up again.  By repeated use of \Refeq{2pt-difference} and 
\Refeq{2pt-diff-par6} and following the same argument as above, 
we obtain the lace expansion for $q=0$ as follows:

\begin{shaded}
\begin{theorem}[The lace expansion for $q=0$]\label{thm:LE}
For $j\ge0$, we let
\begin{align}
\pi^{\sss(j)}(\ovec,\xvec)&=\int_{\Tbb}\diff t_1\sum_{\substack{(\yvec_1,
 \zvec_1):\\ t_{\yvec_1}=t_{\zvec_1}=t_1}}\beta J_{y_1,z_1}\cdots\int_{\Tbb}
 \diff t_j\sum_{\substack{(\yvec_j,\zvec_j):\\ t_{\yvec_j}=t_{\zvec_j}=t_j}}
 \beta J_{y_j,z_j}\frac1{Z^2}\Ebb^1\Ebb^2\Bigg[\sum_{\substack{\partial\psivec^1
 =\ovec\vtri\yvec_1\\ \partial\psivec^2=\vno}}\ind{\ovec\db{1,2}{}\yvec_1}\nn\\
&\quad\times\cdots\frac1{ZZ_{\Ccal_{2j-1,2j}^\compl}}\Ebb^{2j+1}\Ebb_{\Ccal_{2j
 -1,2j}^\compl}^{2j+2}\Bigg[\sum_{\substack{\partial\psivec^{2j+1}=\zvec_{j}
 \vtri\xvec\\ \partial\psivec^{2j+2}=\vno}}\indic{E_{2j+1,2j+2}(\zvec_j,\xvec;
 \Ccal_{2j-1,2j})}\Bigg]\cdots\Bigg],
\end{align}
where $\Ccal_{2j-1,2j}=\tilde\Ccal^{\{\yvec_j,\zvec_j\}}(\zvec_{j-1})$ (with 
$\zvec_0=\ovec$), which is random for $\Ebb^1\Ebb^2$ when $j=1$ and for 
$\Ebb^{2j-1}\Ebb_{\Ccal_{2j-3,2j-2}^\compl}^{2j}$ when $j\ge2$.  Let 
$\pi^{\sss(\le j)}(\ovec,\xvec)=\sum_{i=0}^j(-1)^i\pi^{\sss(i)}(\ovec,\xvec)$. 
Then, for any $j\ge0$, the two-point function $G(\ovec,\xvec)$ satisfies the 
recursion equation
\begin{align}
G(\ovec,\xvec)=\pi^{\sss(\le j)}(\ovec,\xvec)+\int_{\Tbb}\diff t\sum_{\substack{
 (\yvec,\zvec):\\ t_{\yvec}=t_{\zvec}=t}}\beta J_{y,z}\,\pi^{\sss(\le j)}(\ovec,
 \yvec)\,G(\zvec,\xvec)+(-1)^{j+1}R^{\sss(j+1)}(\ovec,\xvec),
\end{align}
where the remainder $R^{\sss(j+1)}(\ovec,\xvec)$ is bounded as
\begin{align}
0\le R^{\sss(j+1)}(\ovec,\xvec)\le\int_{\Tbb}\diff t\sum_{\substack{(\yvec,
 \zvec):\\ t_{\yvec}=t_{\zvec}=t}}\beta J_{y,z}\pi^{\sss(j)}(\ovec,\yvec)\,
 G(\zvec,\xvec).
\end{align}
\end{theorem}
\end{shaded}

\subsection{Diagrammatic bounds on the expansion coefficients}\label{ss:DiagBds}
The lace expansion in the previous subsection is quite similar in spirit to that for the classical Ising model~\cite{s2007lace}\added[id=AS]{; correct bounds on the expansion coefficients 
are  proven in \cite{s2022correct}} by using a ``double expansion'', that is, a lace expansion for the expansion coefficients. 

For example, the $0^\text{th}$ expansion coefficient in \cite{s2007lace}, which corresponds to $\pi^{\sss(0)}(\ovec,\xvec)$ in this paper, is defined in terms of the event that there are two disjoint paths of bonds with positive currents between two sources.  
Since the sum of the currents on the bonds incident on each source is odd, there must be a path of bonds with odd currents joining the two sources.  Then we use the ``earliest'' among such paths as a time line for the double expansion \cite[Step~2 in Section~4.1]{s2022correct}. 

In the present setting, since Poisson bridges do not share their end vertices almost surely, there is a unique path ($\subset\ttl(\psivec^1)\equiv\{(t,z):\psi^1_z(t)=\ttl\}$) from $\ovec$ to $\xvec$ in the bridge configuration $\xivec^1$ almost surely.  
We call \added[id=AS]{the unique path} a backbone, denoted $\Scal_1$, and use it as a ``time line'' for the double expansion.  
\added[id=AS]{In this respect, the new lace expansion looks simpler than the previous one in bounding the expansion coefficients.  However, due to the anisotropy of (discrete) space and (continuous) time, we have to deal with two types of lace edges separately (type-B and type-I, introduced in the proof of Lemma~\ref{lmm:pi02bd} below), depending on how they land on the backbone (see also Figure~\ref{fig:lacegraph}).  
This introduces new complexity, resulting in nearly the same level of difficulty in both lace extensions.}

\added[id=AS]{In this subsection, we demonstrate the double expansion to show diagrammatic bounds (in terms of two-point functions) on a part of the expansion coefficient} $\pi^{\sss(0)}(\ovec,\xvec)$. 
A complete proof of diagrammatic bounds on the other expansion coefficients, including the quantum case $q>0$, will be reported in the \added[id=AS]{forthcoming paper}~\cite{ks2025lace}.

\begin{figure}[t]
\begin{center}
\includegraphics[scale=0.6]{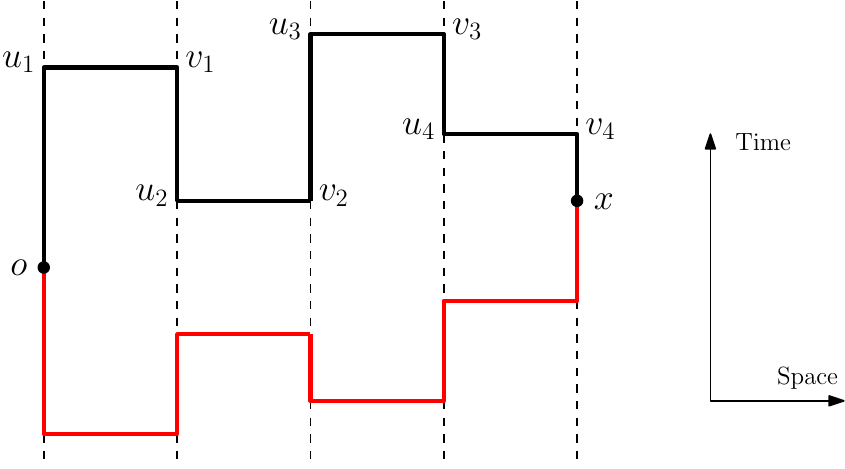}
\quad
\includegraphics[scale=0.6]{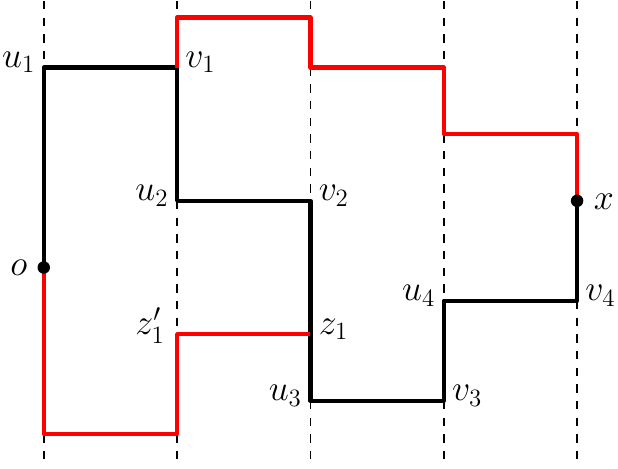}
\end{center}
\caption{Two examples of $\ovec\db{\Scal_1,1',2}{}\xvec$, where $\Scal_1$ is 
depicted as a path from $\ovec$ to $\xvec$ (\added[id=AS]{thicker} in black).  On the left, 
$\ovec$ and $\xvec$ are still connected, even after removal of $\Scal_1$, by 
a $(\psivec^{1'},\psivec^2,\zerovec)$-open path (\added[id=AS]{thinner} in red), while on the right, 
they are not.  Since $\zvec_1$ is located in the middle of an interval, there 
must be a bridge $\{\zvec'_1,\zvec_1\}\in\xivec^2$ such that 
$\ovec\cn{1',2}{}\zvec'_1$ in $\Scal_1^\compl$.}
\label{fig:doubly}
\end{figure}

First, by conditioning on the backbone $\Scal_1$, we can rewrite 
$\pi^{\sss(0)}(\ovec,\xvec)$ as
\begin{align}\lbeq{pi0-rewr}
\pi^{\sss(0)}(\ovec,\xvec)
&=\frac1{Z^2}\Ebb^1\Bigg[\sum_{\partial\psivec^1=\ovec\vtri\xvec}\Ebb^2
 \bigg[\sum_{\partial\psivec^2=\vno}\ind{\ovec\db{1,2}{}\xvec}\Bigg]\Bigg]\nn\\
&=\frac1{Z^2}\Ebb^1\Bigg[\sum_{\partial\psivec^1=\ovec\vtri\xvec}
 \frac1{Z_{\Scal_1^\compl}}\Ebb_{\Scal_1^\compl}^{1'}\Ebb^2\Bigg[\sum_{\partial
 \psivec^{1'}=\partial\psivec^2=\vno}\ind{\ovec\db{\Scal_1,1',2}{}\xvec}\Bigg]
 \Bigg],
\end{align}
where $\ovec\db{\Scal_1,1',2}{}\xvec$ is an abuse of notation meaning that 
$\ovec$ is doubly connected to $\xvec$ in the superposition of 
$(\xivec^{1'},\psivec^{1'})$, $(\xivec^2,\psivec^2)$ and $\Scal_1$.  
We note that $\Scal_1$ is random against the outer expectation $\Ebb^1$, but 
deterministic against the inner expectation $\Ebb_{\Scal_1^\compl}^{1'}\Ebb^2$ 
(\added[id=AS]{see} Figure~\ref{fig:doubly}). 

Now we use a double expansion.  
Denote $\yvec\stackrel{\sss\Scal_1}<\zvec$ if $\yvec$ is closer to $\ovec$ 
along $\Scal_1$ than $\zvec$; we will use below $\max$ and $\min$ as defined 
by this relationship.  Given $(\xivec^{1'},\psivec^{1'})$ and 
$(\xivec^2,\psivec^2)$ satisfying $\partial\psivec^{1'}=\partial\psivec^2=\vno$, 
we define a lace $\lace_{1',2}=\{\yvec_j\zvec_j\}_{j=1}^N$ as follows 
(\added[id=AS]{see} Figure~\ref{fig:lacegraph}):
\begin{itemize}
\item
First we define
\begin{align}
\yvec_1=\ovec,&&
\zvec_1=\max\Big\{\wvec\in\Scal_1:\ovec\cn{1',2}{}\wvec\text{ in }
 \Scal_1^\compl\Big\},
\end{align}
where ``in $\Scal_1^\compl$'' means that an open path from $\ovec$ to $\wvec$ 
does not intersect $\Scal_1$ except for the endvertices $\ovec$ and $\wvec$.  
If $\zvec_1=\xvec$, then it is done with $\lace_{1',2}=\{\ovec\xvec\}$ and 
$N=1$.
\item
If $\zvec_1\stackrel{\sss\Scal_1}<\xvec$, then there is almost surely a unique 
$\yvec_2\zvec_2$ defined as
\begin{align}
\zvec_2&=\max\Big\{\wvec\in\Scal_1:\exists\wvec'\stackrel{\sss\Scal_1}<\zvec_1
 \stackrel{\sss\Scal_1}<\wvec\text{ such that }\wvec'\cn{1',2}{}\wvec\text{ in }
 \Scal_1^\compl\Big\},\\
\yvec_2&=\min\Big\{\wvec'\in\Scal_1:\wvec'\cn{1',2}{}\zvec_2\text{ in }
 \Scal_1^\compl\Big\}.
\end{align}
If $\zvec_2=\xvec$, then it is done with 
$\lace_{1',2}=\{\ovec\zvec_1,\yvec_2\xvec\}$ and $N=2$.
\item
Repeat this procedure until it reaches $\zvec_N=\xvec$ with 
$\lace_{1',2}=\{\yvec_j\zvec_j\}_{j=1}^N$.
\end{itemize}
Notice that, due to the above construction, the lace edges 
$\{\yvec_j\zvec_j\}_{j=1}^N$ are mutually avoiding.  Then we can rewrite 
\Refeq{pi0-rewr} as
\begin{align}
\pi^{\sss(0)}(\ovec,\xvec)=\sum_{N=1}^\infty\pi_N^{\sss(0)}(\ovec,\xvec),
\end{align}
where
\begin{align}
\pi_N^{\sss(0)}(\ovec,\xvec)=\frac1{Z^2}\Ebb^1\Bigg[\sum_{\partial\psivec^1
 =\ovec\vtri\xvec}\frac1{Z_{\Scal_1^\compl}}\Ebb_{\Scal_1^\compl}^{1'}\Ebb^2\Bigg[
 \sum_{\partial\psivec^{1'}=\partial\psivec^2=\vno}\sum_{\{
 \yvec_j\zvec_j\}_{j=1}^N}\ind{\lace_{1',2}=\{\yvec_j\zvec_j\}_{j=1}^N}\Bigg]
 \Bigg].
\end{align}

\begin{figure}[t]
\begin{center}
\includegraphics[scale=0.75]{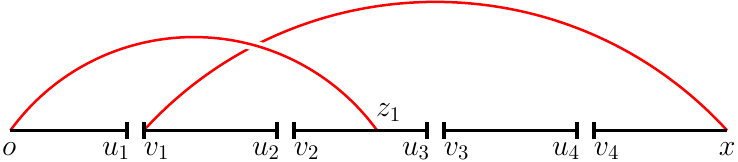}
\end{center}
\caption{A lace graph corresponding to the right figure in 
Figure~\ref{fig:doubly}.  Each line segment (e.g., from $\ovec$ to $\uvec_1$) 
is an oriented interval in $\Scal_1$, while each gap (e.g., between 
$\uvec_1$ and $\vvec_1$) is a bridge in $\Scal_1$.  Since the first lace edge 
$\ovec\zvec_1$ (in red) lands on the middle of the interval $(\vvec_2,\uvec_3)$, 
there must be a bridge $\{\zvec'_1,\zvec_1\}\in\xivec^2$ that is not explicitly 
shown in this figure.}
\label{fig:lacegraph}
\end{figure}

As a first attempt, we investigate the contribution from the case of $N=1$.  

\begin{shaded}
\begin{lemma}\label{lmm:pi00bd}
\begin{align}
\pi_1^{\sss(0)}(\ovec,\xvec)\equiv\frac1{Z^2}\Ebb^1\Bigg[\sum_{\partial\psivec^1
 =\ovec\vtri\xvec}\frac1{Z_{\Scal_1^\compl}}
 \Ebb_{\Scal_1^\compl}^{1'}\Ebb^2\Bigg[\sum_{\partial\psivec^{1'}
 =\partial\psivec^2=\vno}\ind{\lace_{1',2}=\{\ovec\xvec\}}\Bigg]\Bigg]
 \le G(\ovec,\xvec)^3.
\end{align}
\end{lemma}
\end{shaded}

\begin{proof}
By the source switching (\added[id=AS]{see} Lemma~\ref{lmm:SST}), we have 
\begin{align}
\frac1{Z_{\Scal_1^\compl}Z}\Ebb_{\Scal_1^\compl}^{1'}\Ebb^2\Bigg[\sum_{\partial
 \psivec^{1'}=\partial\psivec^2=\vno}\ind{\lace_{1',2}=\{\ovec\xvec\}}\Bigg]
&=\frac1{Z_{\Scal_1^\compl}Z}\Ebb_{\Scal_1^\compl}^{1'}\Ebb^2\Bigg[
 \sum_{\partial\psivec^{1'}=\partial\psivec^2=\vno}\ind{\ovec\cn{1',2}{}\xvec
 \text{ in }\Scal_1^\compl}\Bigg]\nn\\
&=\frac1{Z_{\Scal_1^\compl}Z}\Ebb_{\Scal_1^\compl}^{1'}\Ebb^2\Bigg[
 \sum_{\partial\psivec^{1'}=\partial\psivec^2=\ovec\vtri\xvec}1\Bigg]\nn\\
&=G_{\Scal_1^\compl}(\ovec,\xvec)\,G(\ovec,\xvec),
\end{align}
where the second equality is due to the fact that 
$\ind{\ovec\cn{1',2}{}\xvec\text{ in }\Scal_1^\compl}=1$ under the 
constraint $\partial\psivec^{1'}=\partial\psivec^2=\ovec\vtri\xvec$.  By 
monotonicity in terms of the volume, i.e., 
$G_{\Scal_1^\compl}(\ovec,\xvec)\le G(\ovec,\xvec)$, we obtain
\begin{align}\lbeq{pi0-N1}
\pi_1^{\sss(0)}(\ovec,\xvec)\le\frac1Z\Ebb^1\Bigg[\sum_{\partial\psivec^1
 =\ovec\vtri\xvec}1\Bigg]\,G(\ovec,\xvec)^2=G(\ovec,\xvec)^3,
\end{align}
as required.
\end{proof}

Next we investigate the contribution from the case of $N=2$.  Let
\begin{align}\lbeq{betaJG}
(\beta J*G)(\zvec_1,\xvec)=\sum_{\zvec'_1:t_{\zvec'_1}=t_{\zvec_1}}\beta J_{z_1,
 z'_1}\,G(\zvec'_1,\xvec),
\end{align}
and define $(G*\beta J)(\zvec_1,\xvec)$ and 
$(\beta J*G*\beta J)(\zvec_1,\xvec)$ similarly.

\begin{shaded}
\begin{lemma}\label{lmm:pi02bd}
\begin{align}\lbeq{pi0-N2}
\pi_2^{\sss(0)}(\ovec,\xvec)&\equiv\frac1{Z^2}\Ebb^1\Bigg[\sum_{\partial
 \psivec^1=\ovec\vtri\xvec}\:\sum_{\zvec_1,\yvec_2}\indic{\big\{\yvec_2
 \stackrel{\Scal_1}<\zvec_1\big\}}\,
 \frac1{Z_{\Scal_1^\compl}}
 \Ebb_{\Scal_1^\compl}^{1'}\Ebb^2\Bigg[\sum_{\partial\psivec^{1'}=\partial
 \psivec^2=\vno}\ind{\lace_{1',2}=\{\ovec\zvec_1,\yvec_2\xvec\}}\Bigg]\Bigg]\nn\\
&\le\iint_{\Tbb\times\Tbb}\diff s\,\diff t\sum_{\substack{\zvec_1:t_{\zvec_1}=s\\
 \yvec_2:t_{\yvec_2}=t}}\Big(G(\ovec,\zvec_1)^2\,G(\yvec_2,\xvec)^2\,G(\ovec,\yvec_2)
 \,(\beta J*G)(\yvec_2,\zvec_1)\,(\beta J*G)(\zvec_1,\xvec)\nn\\
&\hskip9pc+[\text{8 other combinations}]\Big).
\end{align}
\end{lemma}
\end{shaded}

\begin{proof}
Given $\Scal_1$, we say that $\yvec$ is type-B if $\yvec$ is an endvertex of 
a bridge in $\Scal_1$ (e.g., $\yvec_2\equiv\vvec_1$ in 
Figure~\ref{fig:lacegraph}), or type-I if $\yvec$ is in the middle of an 
interval in $\Scal_1$ (e.g., $\zvec_1\in(\vvec_2,\uvec_3)$ in 
Figure~\ref{fig:lacegraph}).  
Then, we can rewrite $\pi_2^{\sss(0)}(\ovec,\xvec)$ as
\begin{align}\lbeq{pi0-N2-rewr}
\frac1Z\Ebb^1\Bigg[\sum_{\partial\psivec^1=\ovec\vtri\xvec}\:&\sum_{\zvec_1,
 \yvec_2}\indic{\big\{\yvec_2\stackrel{\Scal_1}<\zvec_1\big\}}\Big(\ind{\zvec_1\text{ type-B}}+\ind{\zvec_1\text{ type-I}}\Big)\Big(
 \ind{\yvec_2\text{ type-B}}+\ind{\yvec_2\text{ type-I}}\Big)\nn\\
&\times\frac1{Z_{\Scal_1^\compl}Z}\Ebb_{\Scal_1^\compl}^{1'}\Ebb^2\Bigg[
 \sum_{\partial\psivec^{1'}=\partial\psivec^2=\vno}\ind{\ovec\cn{1',2}{}
 \zvec_1\text{ in }\Scal_1^\compl\}\circ\{\yvec_2\cn{1',2}{}\xvec\text{ in }
 \Scal_1^\compl}\Bigg]\Bigg],
\end{align}
where we have used ``$\circ$'' to mean that the two events occur ``without 
touching each other'' i.e.,
\begin{align}
\ind{\ovec\cn{1',2}{}\zvec_1\text{ in }\Scal_1^\compl\}\circ\{\yvec_2\cn{1',2}{}
 \xvec\text{ in }\Scal_1^\compl}&=\ind{\ovec\cn{1',2}{}\zvec_1\text{ in }
 \Scal_1^\compl}\,\ind{\yvec_2\cn{1',2}{}\xvec\text{ in }\Scal_1^\compl}\,
 \ind{\tilde\Ccal_{1',2}^{\Scal_1}(\ovec)\cap\tilde\Ccal_{1',2}^{\Scal_1}(\xvec)
 =\vno},
\end{align}
where $\tilde\Ccal_{1',2}^{\Scal_1}(\ovec)=\{\wvec\in\Scal_1^\compl:\ovec
\cn{1',2}{}\wvec$ in $\Scal_1^\compl\}$.  Now we show how to bound the 
contributions from (i)~$\ind{\zvec_1\text{ type-B}}\,\ind{\yvec_2\text{ type-B}}$ 
and (ii)~$\ind{\zvec_1\text{ type-I}}\,\ind{\yvec_2\text{ type-B}}$.

\bigskip

(i)~ To bound the contribution from 
$\ind{\zvec_1\text{ type-B}}\ind{\yvec_2\text{ type-B}}$, we rewrite the 
inner expectation in \Refeq{pi0-N2-rewr} by conditioning on 
$\tilde\Ccal_{1',2}^{\Scal_1}(\ovec)$ (\added[id=AS]{see} \Refeq{nested}--\Refeq{inclexcl1} 
or \Refeq{nested2}--\Refeq{inclexcl2}) as
\begin{align}\lbeq{pi0-N2-cal1}
&\frac1{Z_{\Scal_1^\compl}Z}\Ebb_{\Scal_1^\compl}^{1'}\Ebb^2\Bigg[\sum_{\partial
 \psivec^{1'}=\partial\psivec^2=\vno}\ind{\ovec\cn{1',2}{}\zvec_1\text{ in }
 \Scal_1^\compl\}\circ\{\yvec_2\cn{1',2}{}\xvec\text{ in }\Scal_1^\compl}
 \Bigg]\nn\\
&=\frac1{Z_{\Scal_1^\compl}Z}\Ebb_{\Scal_1^\compl}^{1'}\Ebb^2\Bigg[\sum_{
 \partial\psivec^{1'}=\partial\psivec^2=\vno}\ind{\ovec\cn{1',2}{}\zvec_1
 \text{ in }\Scal_1^\compl}\,\frac1{Z_{\Scal_1^\compl\cap\tilde\Ccal_{1',
 2}^{\Scal_1}(\ovec)^\compl}Z_{\tilde\Ccal_{1',2}^{\Scal_1}(\ovec)^\compl}}\nn\\
&\hskip7pc\times\Ebb^{1''}_{\Scal_1^\compl\cap\tilde\Ccal_{1',2}^{\Scal_1}
 (\ovec)^\compl}\Ebb^{2'}_{\tilde\Ccal_{1',2}^{\Scal_1}(\ovec)^\compl}\Bigg[
 \sum_{\partial\psivec^{1''}=\partial\psivec^{2'}=\vno}\ind{\yvec_2\cn{1'',2'}{}
 \xvec\text{ in }\Scal_1^\compl\cap\tilde\Ccal_{1',2}^{\Scal_1}(\ovec)^\compl}
 \Bigg]\Bigg],
\end{align}
which is bounded, by using the source switching and monotonicity, by
\begin{align}\lbeq{pi0-N2-cal2}
\frac1{Z_{\Scal_1^\compl}Z}\Ebb_{\Scal_1^\compl}^{1'}\Ebb^2\Bigg[\sum_{\partial
 \psivec^{1'}=\partial\psivec^2=\vno}\ind{\ovec\cn{1',2}{}\zvec_1\text{ in }
 \Scal_1^\compl}\Bigg]\,G(\yvec_2,\xvec)^2\le G(\ovec,\zvec_1)^2\,G(\yvec_2,
 \xvec)^2.
\end{align}
Therefore, the contribution to \Refeq{pi0-N2-rewr} from this case is bounded by
\begin{align}\lbeq{pi0-N2-cal3}
\frac1Z\Ebb^1\Bigg[\sum_{\partial\psivec^1=\ovec\vtri\xvec}\:\sum_{\zvec_1,
 \yvec_2}\indic{\big\{\yvec_2\stackrel{\Scal_1}<\zvec_1\big\}}\,\ind{\zvec_1\text{ type-B}}\,\ind{\yvec_2\text{ type-B}}\,G(\ovec,
 \zvec_1)^2\,G(\yvec_2,\xvec)^2\Bigg].
\end{align}
Since $\ind{\zvec_1\text{ type-B}}=1$ implies existence of a bridge 
$\{\zvec_1,\zvec'_1\}\in\xivec^1$ that satisfies 
$\zvec_1\stackrel{\sss\Scal_1}<\zvec'_1$ or vice versa, we have the rewrite
\begin{align}\lbeq{spine-decomp}
&\sum_{\zvec_1,\yvec_2}\ind{\Scal_1=(\ovec,\yvec_2)\cdot(\yvec_2,\zvec_1)\cdot
 (\zvec_1,\xvec)}\,\ind{\zvec_1\text{ type-B}}\,\ind{\yvec_2\text{ type-B}}\nn\\
&=\sum_{\{\zvec_1,\zvec'_1\},\{\yvec_2,\yvec'_2\}\in\xivec^1}\Big(\indic{\big\{
 \yvec_2\stackrel{\Scal_1}<\yvec'_2\stackrel{\Scal_1}<\zvec_1\stackrel{\Scal_1}
 <\zvec'_1\big\}}+\indic{\big\{\yvec_2\stackrel{\Scal_1}<\yvec'_2
 \stackrel{\Scal_1}<\zvec'_1\stackrel{\Scal_1}<\zvec_1\big\}}\nn\\
&\hskip10pc+\indic{\big\{\yvec'_2\stackrel{\Scal_1}<\yvec_2\stackrel{\Scal_1}
 <\zvec_1\stackrel{\Scal_1}<\zvec'_1\big\}}+\indic{\big\{\yvec'_2
 \stackrel{\Scal_1}<\yvec_2\stackrel{\Scal_1}<\zvec'_1\stackrel{\Scal_1}<\zvec_1
 \big\}}\Big).
\end{align}
Then, by the Mecke equation (\added[id=AS]{see} \Refeq{mecke1}), we can rewrite the 
contribution from the first indicator as
\begin{align}\lbeq{pi0-N2-cal4}
\int_{\Tbb}\diff s\sum_{\substack{\{\zvec_1,\zvec'_1\}:\\ t_{\zvec_1}
 =t_{\zvec'_1}=s}}\beta J_{z_1,z'_1}\int_{\Tbb}\diff t\sum_{\substack{\{\yvec_2,
 \yvec'_2\}:\\ t_{\yvec_2}=t_{\yvec'_2}=t}}\beta J_{y_2,y'_2}\,G(\ovec,
 \zvec_1)^2\,G(\yvec_2,\xvec)^2\nn\\
\times\frac1Z\Ebb^1\Bigg[\sum_{\partial\psivec^1=\ovec\vtri\xvec\vtri\{\zvec_1,
 \zvec'_1\}\vtri\{\yvec_2,\yvec'_2\}}\indic{\big\{
 \yvec_2\stackrel{\Scal_1}<\yvec'_2\stackrel{\Scal_1}<\zvec_1\stackrel{\Scal_1}
 <\zvec'_1\big\}}\Bigg].
\end{align}
Finally, by conditioning on parts of $\Scal_1$ (\added[id=AS]{see} \Refeq{pi0-rewr}) 
and monotonicity, the last line is bounded as
\begin{align}\lbeq{spinebd}
&\frac1Z\Ebb^1\Bigg[\sum_{\partial\psivec^1=\ovec\vtri\yvec_2\vtri\yvec'_2\vtri
 \zvec_1}\indic{\big\{\yvec_2\stackrel{\Scal_1}<\yvec'_2\stackrel{\Scal_1}<
 \zvec_1\big\}}\underbrace{\frac1{Z_{\Scal_1^\compl}}\Ebb^{1'}_{\Scal_1^\compl}
 \Bigg[\sum_{\partial\psivec^{1'}=\zvec'_1\vtri\xvec}\ind{\Scal_1\cap\Scal_{1'}
 =\vno}\Bigg]}_{\le\,G(\zvec'_1,\xvec)}\Bigg]\nn\\
&\le\frac1Z\Ebb^1\Bigg[\sum_{\partial\psivec^1=\ovec\vtri\yvec_2}
\underbrace{\frac1{Z_{\Scal_1^\compl}}\Ebb^{1'}_{\Scal_1^\compl}
 \Bigg[\sum_{\partial\psivec^{1'}=\yvec'_2\vtri
 \zvec_1}\ind{\Scal_1\cap\Scal_{1'}
 =\vno}\Bigg]}_{\le\,G(\yvec'_2,\zvec_1)}\Bigg]\,G(\zvec'_1,\xvec)\nn\\
&\le G(\ovec,\yvec_2)\,G(\yvec'_2,\zvec_1)\,G(\zvec'_1,\xvec),
\end{align}
where, in the leftmost expression, $\Scal_1$ is the backbone from $\ovec$ to 
$\zvec_1$, which is random against the outer expectation $\Ebb^1$, and 
$\Scal_{1'}$ is the backbone from $\zvec'_1$ to $\xvec$, which is random against 
the inner expectation $\Ebb^{1'}_{\Scal_1^\compl}$; in the middle expression, 
$\Scal_1$ is the backbone from $\ovec$ to $\yvec_2$, which is random against 
the outer expectation $\Ebb^1$, and $\Scal_{1'}$ is the backbone from $\yvec'_2$ 
to $\zvec_1$, which is random against the inner expectation 
$\Ebb^{1'}_{\Scal_1^\compl}$.  Therefore,
\begin{align}\lbeq{bound(i)}
\Refeq{pi0-N2-cal4}\le\iint_{\Tbb\times\Tbb}\diff s\,\diff t\sum_{\substack{\zvec_1:
 t_{\zvec_1}=s\\ \yvec_2:t_{\yvec_2}=t}}G(\ovec,\zvec_1)^2\,G(\yvec_2,\xvec)^2\,
 G(\ovec,\yvec_2)\,(\beta J*G)(\yvec_2,\zvec_1)\,(\beta J*G)(\zvec_1,\xvec),
\end{align}
which already appeared in \Refeq{pi0-N2}.

\begin{figure}[t]
\begin{center}
\includegraphics[scale=0.4]{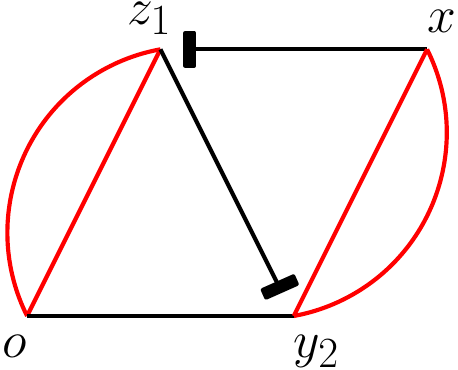}\qquad
\includegraphics[scale=0.4]{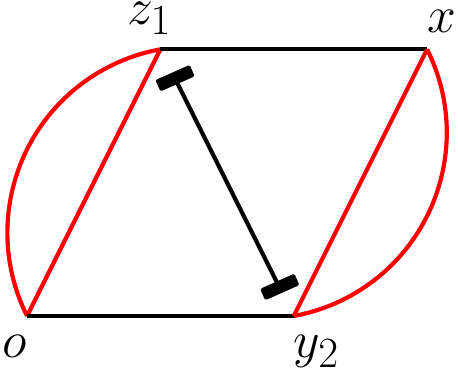}\qquad
\includegraphics[scale=0.4]{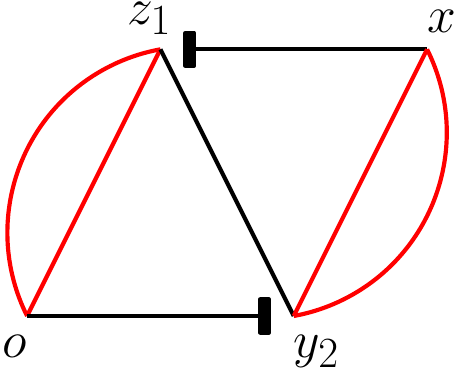}\qquad
\includegraphics[scale=0.4]{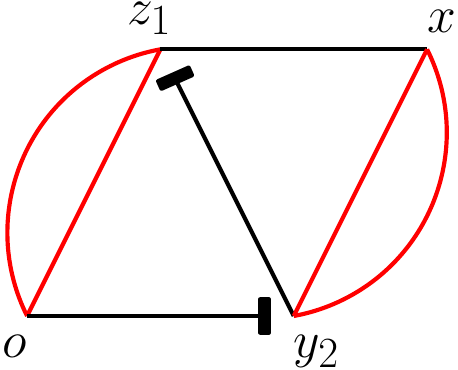}
\end{center}
\caption{Schematic representation for \Refeq{bound(i)} (leftmost) and 
diagrammatic bounds on the contributions from the other three indicators 
in \Refeq{spine-decomp}.  Each gap represents a bridge, and the four red 
segments in each figure represent $G(\ovec,\zvec_1)^2\,G(\yvec_2,\xvec)^2$.}
\label{fig:indic-decomp}
\end{figure}

The other three cases in \Refeq{spine-decomp} are bounded similarly by 
replacing the last three terms in \Refeq{bound(i)} by 
$G(\ovec,\yvec_2)\,(\beta J*G*\beta J)(\yvec_2,\zvec_1)\,G(\zvec_1,\xvec)$ 
(for the $2^\text{nd}$ indicator in \Refeq{spine-decomp}), 
$(G*\beta J)(\ovec,\yvec_2)\,G(\yvec_2,\zvec_1)\,(\beta J*G)(\zvec_1,\xvec)$ 
(for the $3^\text{rd}$ indicator in \Refeq{spine-decomp}) and 
$(G*\beta J)(\ovec,\yvec_2)\,(G*\beta J)(\yvec_2,\zvec_1)\,G(\zvec_1,\xvec)$ 
(for the $4^\text{th}$ indicator in \Refeq{spine-decomp}), respectively (\added[id=AS]{see} 
Figure~\ref{fig:indic-decomp}).

\bigskip

(ii)~ To bound the contribution to \Refeq{pi0-N2-rewr} from 
$\ind{\zvec_1\text{ type-I}}\,\ind{\yvec_2\text{ type-B}}$, we follow 
the same strategy as in (i) to bound the inner expectation in 
\Refeq{pi0-N2-rewr} by the left-hand side of \Refeq{pi0-N2-cal2}.  
Then, we obtain the counterpart to \Refeq{pi0-N2-cal3}:
\begin{align}\lbeq{pi0-N2-cal5}
\frac1Z\Ebb^1\Bigg[&\sum_{\partial\psivec^1=\ovec\vtri\xvec}\:\sum_{\zvec_1,
 \yvec_2}\indic{\big\{\yvec_2\stackrel{\Scal_1}<\zvec_1\big\}}\,\ind{\zvec_1\text{ type-I}}\,\ind{\yvec_2\text{ type-B}}\nn\\
&\times\frac1{Z_{\Scal_1^\compl}Z}\Ebb_{\Scal_1^\compl}^{1'}\Ebb^2\Bigg[
 \sum_{\partial\psivec^{1'}=\partial\psivec^2=\vno}\ind{\ovec\cn{1',2}{}\zvec_1\text{ in }
 \Scal_1^\compl}\Bigg]\,G(\yvec_2,\xvec)^2\Bigg].
\end{align}
Since $\ind{\zvec_1\text{ type-I}}=1$ implies existence of a bridge 
$\{\zvec'_1,\zvec_1\}\in\xivec^2$ that satisfies $\ovec\cn{1',2}{}\zvec'_1$ in 
$\Scal_1^\compl$ (as explained in 
Figures~\ref{fig:doubly}--\ref{fig:lacegraph}), 
we can bound \Refeq{pi0-N2-cal5} by
\begin{align}\lbeq{pi0-N2-cal6}
\frac1Z\Ebb^1\Bigg[&\sum_{\partial\psivec^1=\ovec\vtri\xvec}\sum_{\zvec_1}
 \sum_{\{\yvec_2,\yvec'_2\}\in\xivec^1}\Big(\indic{\big\{\yvec_2
 \stackrel{\Scal_1}<\yvec'_2\stackrel{\Scal_1}<\zvec_1\big\}}+\indic{\big\{
 \yvec'_2\stackrel{\Scal_1}<\yvec_2\stackrel{\Scal_1}<\zvec_1\big\}}\Big)\nn\\
&\times\frac1{Z_{\Scal_1^\compl}Z}\Ebb_{\Scal_1^\compl}^{1'}\Ebb^2\Bigg[
 \sum_{\partial\psivec^{1'}=\partial\psivec^2=\vno}\sum_{\substack{\zvec'_1:\\
 \{\zvec'_1,\zvec_1\}\in\xivec^2}}\ind{\ovec\cn{1',2}{}\zvec'_1
 \text{ in }\Scal_1^\compl}\Bigg]\,G(\yvec_2,\xvec)^2\Bigg].
\end{align}
Then, by the Mecke equation (\added[id=AS]{see} \Refeq{mecke1}) and the source switching 
(\added[id=AS]{see} Lemma~\ref{lmm:SST}), we can further bound it by
\begin{align}\lbeq{pi0-N2-cal7}
&\int_{\Tbb}\diff s\sum_{\substack{\{\zvec'_1,\zvec_1\}:\\ t_{\zvec'_1}
 =t_{\zvec_1}=s}}\beta J_{z'_1,z_1}\int_{\Tbb}\diff t\sum_{\substack{\{\yvec_2,
 \yvec'_2\}:\\ t_{\yvec_2}=t_{\yvec'_2}=t}}\beta J_{y_2,y'_2}\,\frac1Z\Ebb^1
 \Bigg[\sum_{\partial\psivec^1=\ovec\vtri\xvec\vtri\{\yvec_2,\yvec'_2\}}\Big(
 \indic{\big\{\yvec_2\stackrel{\Scal_1}<\yvec'_2\stackrel{\Scal_1}<\zvec_1
 \big\}}\nn\\
&\qquad\qquad+\indic{\big\{\yvec'_2\stackrel{\Scal_1}<\yvec_2\stackrel{\Scal_1}<
 \zvec_1\big\}}\Big)\underbrace{\frac1{Z_{\Scal_1^\compl}Z}\Ebb_{\Scal_1^\compl}^{1'}\Ebb^2
 \Bigg[\sum_{\substack{\partial\psivec^{1'}=\vno\\ \partial\psivec^2=\zvec'_1
 \vtri\zvec_1}}\ind{\ovec\cn{1',2}{}\zvec'_1\text{ in }\Scal_1^\compl}
 \Bigg]}_{G_{\Scal_1^\compl}(\ovec,\zvec'_1)\,G(\ovec,\zvec_1)~(\because\,
 \text{Lemma~\ref{lmm:SST}})}G(\yvec_2,\xvec)^2\Bigg]\nn\\
&\le\iint_{\Tbb\times\Tbb}\diff s\,\diff t\sum_{\substack{\zvec_1:t_{\zvec_1}=s\\ \yvec_2:
 t_{\yvec_2}=t}}(G*\beta J)(\ovec,\zvec_1)\,G(\ovec,\zvec_1)\,G(\yvec_2,\xvec)^2
 \sum_{\yvec'_2:t_{\yvec'_2}=t}\beta J_{y_2,y'_2}\nn\\
&\qquad\times\frac1Z\Ebb^1\Bigg[\sum_{\partial\psivec^1=\ovec\vtri\xvec\vtri
 \{\yvec_2,\yvec'_2\}}\Big(\indic{\big\{\yvec_2\stackrel{\Scal_1}<\yvec'_2
 \stackrel{\Scal_1}<\zvec_1\big\}}+\indic{\big\{\yvec'_2\stackrel{\Scal_1}<
 \yvec_2\stackrel{\Scal_1}<\zvec_1\big\}}\Big)\Bigg].
\end{align}
Finally, by conditioning on parts of the backbone and monotonicity (\added[id=AS]{see} 
\Refeq{spinebd}), 
the last 
line is bounded by $(G(\ovec,\yvec_2)\,G(\yvec'_2,\zvec_1)+G(\ovec,\yvec'_2)\,
G(\yvec_2,\zvec_1))\,G(\zvec_1,\xvec)$.  As a result, the contribution to 
\Refeq{pi0-N2-rewr} from 
$\ind{\zvec_1\text{ type-I}}\,\ind{\yvec_2\text{ type-B}}$ is bounded by 
(\added[id=AS]{see} Figure~\ref{fig:indic-decomp2})
\begin{align}\lbeq{bound(ii)}
&\iint_{\Tbb\times\Tbb}\diff s\,\diff t\sum_{\substack{\zvec_1:t_{\zvec_1}=s\\ \yvec_2:
 t_{\yvec_2}=t}}(G*\beta J)(\ovec,\zvec_1)\,G(\ovec,\zvec_1)\,G(\yvec_2,\xvec)^2\nn\\
&\qquad\times\Big(G(\ovec,\yvec_2)\,(\beta J*G)(\yvec_2,\zvec_1)+(G*\beta J)
 (\ovec,\yvec_2)\,G(\yvec_2,\zvec_1)\Big)\,G(\zvec_1,\xvec).
\end{align}

\begin{figure}[t]
\begin{center}
\includegraphics[scale=0.4]{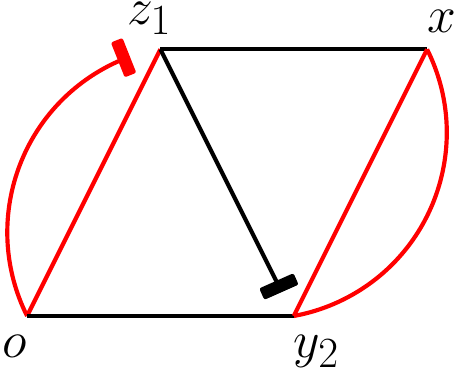}\hskip5pc
\includegraphics[scale=0.4]{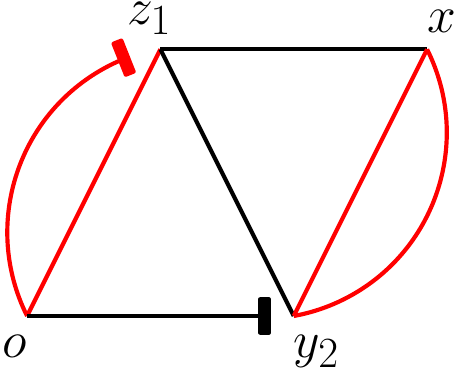}
\end{center}
\caption{Schematic representation for the two terms in \Refeq{bound(ii)}.}
\label{fig:indic-decomp2}
\end{figure}

\bigskip

In the same way, we can bound the contribution to \Refeq{pi0-N2-rewr} from 
$\ind{\zvec_1\text{ type-B}}\,\ind{\yvec_2\text{ type-I}}$ by an expression 
similar to \Refeq{bound(ii)}, 
and the contribution from 
$\ind{\zvec_1\text{ type-I}}\,\ind{\yvec_2\text{ type-I}}$ by an expression 
consisting of a single term. 
Therefore, $\pi_2^{\sss(0)}(\ovec,\xvec)$ has a diagrammatic bound consisiting 
of $3\times3=9$ terms, where 3 is the number of bridge embeddings at each 
internal vertex.
\end{proof}

\begin{remark}
Similar structure appears in the higher-$N$ case (with slight modification in 
the number of bridge embeddings at internal vertices).  Thanks to those bridges, we can 
show that, by the bootstrapping argument used in the lace-expansion literature, 
nonzero space-time bubbles made of $\beta J*G$ are small uniformly in 
$\beta<\betac$ (for the nearest-neighbor model with $d\gg4$ and for sufficiently spread-out models with $d>4$), so that $\pi_N^{\sss(0)}$ decays 
exponentially in $N$, hence convergence of the series.  

The full details of diagrammatic bounds on the expansion coefficients as well 
as the bootstrapping argument, for $q\ge0$, will be reported in the forthcoming paper~\cite{ks2025lace}.
\end{remark}

\section*{Statements and Declarations}
On behalf of all authors, the corresponding author states that there is no
conflict of interest and data sharing is not applicable to this article as no datasets were generated or analysed during the current study.

\section*{Acknowledgements}
\added[id=AS]{We would like to thank two anonymous referees for many valuable comments and encouragement.}
\added[id=YK]{The work of YK was supported in part by the Postdoctoral Fellow of NCTS.}
The work of AS was supported by JSPS KAKENHI Grant Number 23K03143.  
This work is dedicated to Toyoji Sakai, who continued to support AS until 
he passed away a month before the completion of the draft.

\printbibliography

\end{document}